\documentclass[journal]{IEEEtran}
\usepackage{ifsym}
\ifCLASSINFOpdf
\else
\fi
\usepackage{cite}
\usepackage{amsmath,amssymb,amsfonts}
\usepackage{algorithmic}
\usepackage{graphicx}
\usepackage{textcomp}
\usepackage{makecell}  
\usepackage{multirow}
\usepackage{hyperref}
\usepackage{placeins}
\usepackage[justification=centering]{caption} 
\begin{document}
%
\title{Deep Learning-Powered Electrical Brain Signals\\ Analysis: Advancing Neurological Diagnostics}

\author{Jiahe Li, Xin Chen, Fanqi Shen, Junru Chen, Yuxin Liu, Daoze Zhang, Zhizhang Yuan, Fang Zhao, Meng Li$^{\dagger}$ and Yang Yang$^{\dagger}$
\thanks{$^{\dagger}$ Corresponding author.
Manuscript received 24 February 2025; revised 08 August 2025; accepted 21 October 2025. 
This work is supported by NSFC (62322606) and Zhejiang NSF (LR22F020005). }
\thanks{Jiahe Li, Xin Chen, Fanqi Shen, Junru Chen, Yuxin Liu, Daoze Zhang, Zhizhang Yuan and Yang Yang are with the College of Computer Science and Technology, Zhejiang University, Hangzhou, Zhejiang 310027, China (e-mail: jiaheli@zju.edu.cn, xin.21@intl.zju.edu.cn, fanqishen@zju.edu.cn, jrchen\_cali@zju.edu.cn, yuxin.liu@zju.edu.cn, zhangdz@zju.edu.cn, zhizhangyuan@zju.edu.cn, yangya@zju.edu.cn).}
\thanks{Fang Zhao is with the DiFint Technology (Shanghai) Co, Shanghai, 201210, China (email: zhaofang@difint.cn).}
\thanks{Meng Li is with the Shanghai Institute of Microsystem and Information Technology, Chinese Academy of Sciences, Shanghai, China, the School of Graduate Study, University of Chinese Academy of Sciences, Beijing, 100049, China, and The INSIDE Institute for Biological and Artificial Intelligence, Shanghai, 201210, China (email: limeng.braindecoder@gmail.com). }}

\markboth{Journal of \LaTeX\ Class Files,~Vol.~14, No.~8, August~2015}%
{Shell \MakeLowercase{\textit{et al.}}: Bare Demo of IEEEtran.cls for IEEE Journals}

\maketitle

\begin{abstract}
Neurological disorders pose major global health challenges, driving advances in brain signal analysis.
Scalp electroencephalography (EEG) and intracranial EEG (iEEG) are widely used for diagnosis and monitoring.
However, dataset heterogeneity and task variations hinder the development of robust deep learning solutions.
This review systematically examines recent advances in deep learning approaches for EEG/iEEG-based neurological diagnostics, focusing on applications across 7 neurological conditions using 46 datasets. 
For each condition, we review representative methods and their quantitative results, integrating performance comparisons with analyses of data usage, model design, and task-specific adaptations, while highlighting the role of pre-trained multi-task models in achieving scalable, generalizable solutions.
Finally, we propose a standardized benchmark to evaluate models across diverse datasets and improve reproducibility, emphasizing how recent innovations are transforming neurological diagnostics toward intelligent, adaptable healthcare systems.
\end{abstract}

\section{Introduction}
\label{sec:introduction}
Neurological disorders are among the most significant global health challenges, with profound consequences for healthcare systems.
According to the World Health Organization (WHO), neurological disorders affect over one-third of the global population, making them a leading cause of illness and disability worldwide~\cite{WHO2024}.
Dementia, affecting 47.5 million people, is a primary concern, with Alzheimer's disease being the most common form~\cite{better2024alzheimer}. 
Seizures impact more than 50 million individuals~\cite{WHO_epilepsy}, while sleep disorders are widespread yet underdiagnosed~\cite{recoveryvillage_sleep_statistics_2023}. 
Other significant disorders, including Parkinson's disease~\cite{who2023parkinson}, schizophrenia~\cite{WHO_SZ}, depression~\cite{WHO_depression}, and ADHD~\cite{who_adolescent_mental_health}, further exacerbate the burden, placing strain on healthcare systems~\cite{WHO_MentalHealth}.
In low-income countries, where resources limit access to care, the situation is particularly dire.

Practical diagnostic tools are essential to alleviate growing global burden of neurological disorders, and electrical brain signals are indispensable.
Specifically, electroencephalography is critical for understanding and diagnosing neurological disorders.
Electroencephalography evaluates electrical activity in the brain and is categorized into scalp electroencephalography (EEG) and intracranial electroencephalography (iEEG). 
EEG is non-invasive, recording brain activity from electrodes on the scalp~\cite{berger1929elektroenkephalogramm}. 
iEEG places electrodes into the brain (stereo-electroencephalography, SEEG) or onto brain's surface (electrocorticography, ECoG), providing localized information~\cite{ramantani2016correlation}.

\begin{figure*}[t] 
    \centering
    \includegraphics[width=0.85\textwidth]{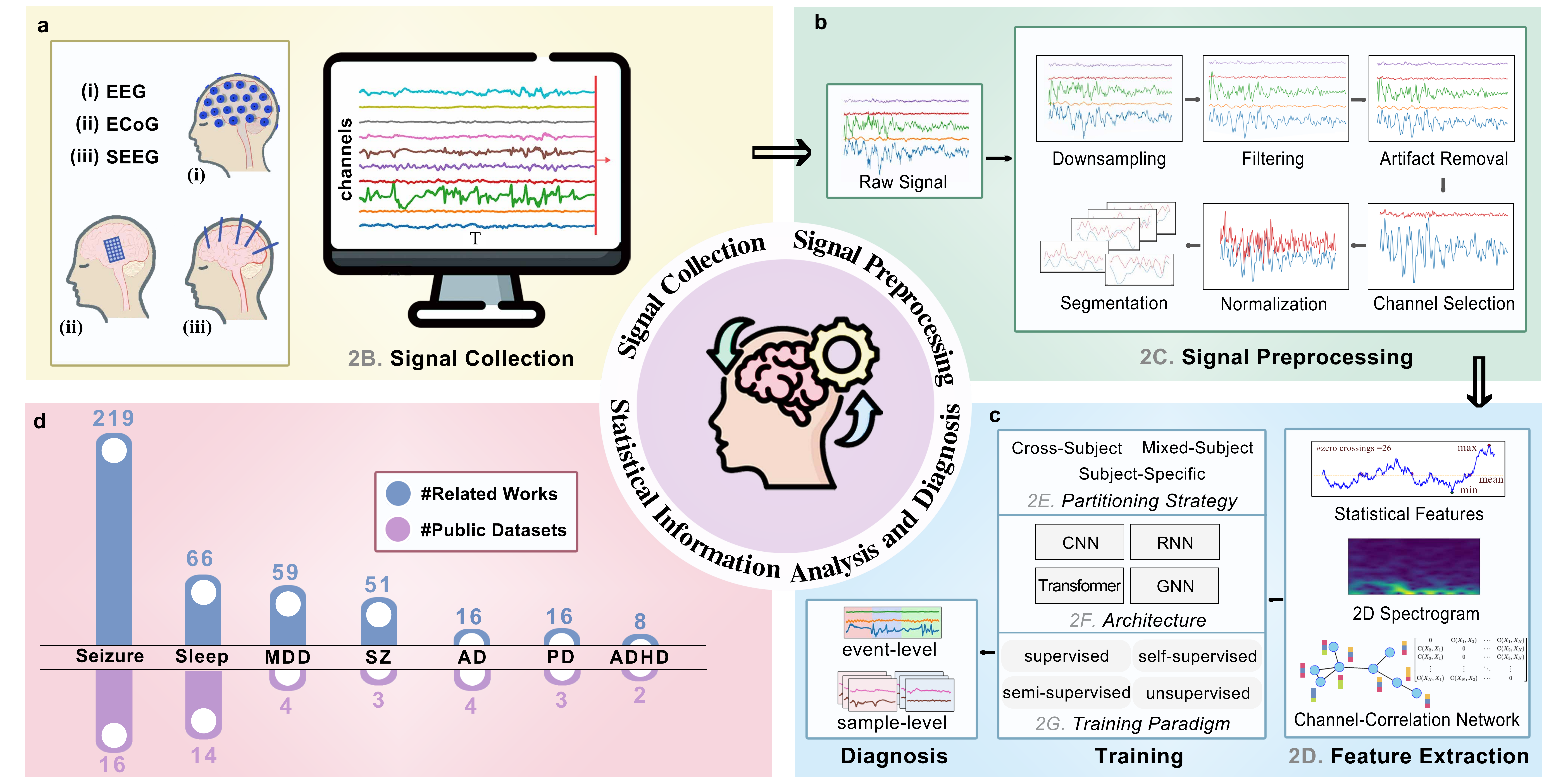} %
  \caption{\textbf{General Workflow of Electrical Brain Signals Analysis in Neurological Diagnostics.} \newline \textbf{a. Signal Collection:} Acquisition of EEG/iEEG signals from patients, capturing brain electrical activity for clinical purposes. \textbf{b. Signal Preprocessing:} A feasible workflow to process raw signals, ensuring their suitability for subsequent analysis.  \newline \textbf{c. Analysis and Diagnosis:} Feature extraction and deep learning-based training for neurological classification. \newline \textbf{d. Statistical Information:} Statistical summary of resources for neurological conditions, including related work and datasets.}
    \label{fig:main}
\end{figure*}

The analysis of EEG/iEEG signals poses significant challenges for traditional machine learning (ML) approaches. These methods typically rely on manually engineered features that may not fully capture complex patterns in neurophysiological data, while their performance is often compromised by inherent noise and artifacts in raw recordings~\cite{ACHARYA2018270, acharya2018automated}. Deep learning (DL) addresses these limitations by automatically extracting features, modeling temporal dependencies, and improving robustness against signal variability~\cite{lecun1995convolutional, vaswani2017attention}. The ability of DL to detect and classify neurological disorders with high accuracy has driven widespread adoption in brain signal analysis~\cite{turk2019epilepsy, Kostas2021BENDR}. This survey systematically examines the workflow of DL models in brain signal analysis, focusing on applications in diagnosing neurological disorders.

\subsection{General Workflow}

The general workflow of EEG/iEEG analysis in neurological diagnostics is shown in Fig.~\ref{fig:main}, including three stages: signal collection, signal preprocessing, and analysis and diagnosis.

In the signal collection stage (Fig.~\ref{fig:main}.a), electrical brain activity is recorded using EEG/iEEG systems, typically across multiple channels at specific sampling rates with task-related labels.
In the preprocessing stage (Fig.~\ref{fig:main}.b), techniques including denoising, filtering and normalization reduce noise and structure the data for feature extraction.
In the analysis and diagnosis stage (Fig.\ref{fig:main}c), preprocessed signals undergo feature extraction and classification. Traditional methods rely on manually designed features, whereas DL automatically learns diagnostically relevant patterns.
Finally, the extracted features are applied to downstream tasks. Fig.~\ref{fig:main}.d highlights the distribution of related research efforts and publicly available datasets across various neurological conditions, including seizure, sleep disorders, major depressive disorder (MDD), schizophrenia (SZ), Alzheimer's disease (AD), Parkinson's disease (PD), and attention deficit hyperactivity disorder (ADHD).

\subsection{Related Studies and Our Contributions}

Existing brain signal analysis surveys exhibit diverse scopes and focuses.
Some focus on EEG, emphasizing their wide availability~\cite{roy2019deep, amer2023eeg}. 
Others broaden the scope to include brain signals like magnetic resonance imaging (MRI)~\cite{zhang2021survey, khan2021machine}, which differ from EEG/iEEG in temporal resolution and preprocessing requirements.
From a task perspective, some reviews focus specifically on diseases such as seizure~\cite{shoeibi2021epileptic, rahul2024systematic}, providing in-depth insights into disease-specific applications. 
Others take a broader view, covering brain-computer interface (BCI) applications~\cite{hossain2023status, weng2024self}, which focus on interaction and control, differing from neurological diagnostic tasks. 

To provide a systematic perspective, we conducted a structured literature search in \textit{PubMed}, \textit{Science Direct}, and \textit{Google Scholar} over the past ten years. 
Combinations of the terms \textit{EEG/iEEG}, \textit{deep learning}, and disease-related tasks were used. 
Studies focusing on clinically relevant diagnostics were retained, while those relying on traditional ML or non-healthcare applications were excluded.
This process resulted in 450 manuscripts, from which information on publicly available datasets was extracted and cross-checked against repositories like \textit{PhysioNet}, \textit{Zenodo}, and \textit{OpenNeuro}, yielding 46 open datasets that form the empirical foundation of this survey.

Building on this systematic basis, our work establishes three contributions to advance deep learning–driven neurodiagnosis: 
First, we curate and analyze 46 public EEG/iEEG datasets across seven neurological conditions, establishing the most comprehensive data landscape to date while unifying fragmented methodologies by standardizing data processing, model architectures, and evaluation protocols.
Besides, we identify self-supervised learning as the optimal paradigm for developing multi-task diagnostic frameworks, offering a comprehensive overview of pre-trained multi-task frameworks and their advancements.
Additionally, we propose a benchmarking methodology to evaluate brain signal models across tasks, providing a foundation for scalable and versatile solutions in EEG/iEEG-based neurological diagnostics.
\section{Methods}

\subsection{Problem Definition}
\label{sec:pdef}
In this survey, we classify neurological diagnostic tasks into sample-level and event-level classification, both under the broader framework of classification problems. 
Sample-level classification involves assigning a single label to an entire signal, typically representing a specific subject or sample (e.g., Alzheimer’s disease diagnosis). 
Event-level classification focuses on identifying and classifying distinct temporal segments within a longer signal, thereby introducing an implicit segmentation process by associating each segment with a specific event or state (e.g., seizure detection or sleep staging).

Electrical brain signals, which capture the brain's electrical activity over time, can be modeled as multivariate time series.
Specifically, let  $\mathbf{X} \in \mathbb{R}^{C \times T}$  represent the EEG/iEEG time series, where $C$ is the number of channels, and $T$ is the number of sampling points. Each channel  $\mathbf{x}^c = \{x^c_1, x^c_2, \dots, x^c_T\}$  corresponds to the measurements from a specific source, such as an EEG electrode or a contact of an iEEG electrode.

\subsubsection{Sample-Level Classification}
In sample-level classification, the objective is to assign a single label $y \in \mathcal{Y}$ to the entire signal $\mathbf{X}$. This can be formulated as:
\[
y = \Phi_{\text{sample}}(\mathbf{X}; \boldsymbol{\theta}), \quad y \in \mathcal{Y},
\]
where $\Phi_{\text{sample}}$ represents the deep learning model parameterized by $\boldsymbol{\theta}$, and $\mathcal{Y}$ denotes the set of possible classes. Here, $\mathbf{X}$ is treated as a unified entity, capturing sample-level or subject-level characteristics.

\subsubsection{Event-Level Classification}
In event-level classification, the goal is to classify smaller temporal segments of the signal. The signal $\mathbf{X}$ is divided into $K$ segments $\mathbf{X}_1, \mathbf{X}_2, \dots, \mathbf{X}_K$, where $\mathbf{X}_k \in \mathbb{R}^{C \times T_k}$ and $T_k$ is the duration of the $k$-th segment. A classification model is applied to each segment to produce a sequence of labels $\mathbf{Y} = \{y_1, y_2, \dots, y_K\}, \, y_k \in \mathcal{Y}$:
\[
y_k = \Phi_{\text{segment}}(\mathbf{X}_k; \boldsymbol{\theta}), \quad \mathbf{Y} = \bigcup_{k=1}^K \{y_k\},
\]
where $\Phi_{\text{segment}}$ denotes the deep learning model parameterized by $\boldsymbol{\theta}$. This process associates each segment $\mathbf{X}_k$ with a specific label $y_k$, allowing the temporal localization of events within the signal.
Event-level classification captures natural temporal dependencies between consecutive segments, reflecting the continuity of events in time~\cite{chen2024con4m}.

\begin{table}[t]
\renewcommand{\arraystretch}{1.2}
\caption{Signal Preprocessing Techniques}
\label{tab:lowlevel}
\footnotesize
\centering
\begin{tabular}{p{80pt}p{90pt}p{30pt}}
\hline
\textbf{Techniques}     & \textbf{Details}             & \textbf{Reference}             \\
\hline
\multirow{4}{80pt}{Noise Reduction \& Filtering} 
                            & FIR Filter           & ~\cite{banville2021uncovering}  \\
                            & IIR Filter           & ~\cite{oh2020deep} \\
                            & Adaptive Filters     & ~\cite{9353630}\\
                            & Manual \& Custom     & ~\cite{ay2019automated} \\
\hline
\multirow{2}{80pt}{Artifact Removal} 
                            & Blind Source Separation  & ~\cite{10023506} \\
                            & Artifact Correction  & ~\cite{MOGHADDARI2020105738} \\
\hline
\multirow{3}{80pt}{Baseline Correction \& Detrending} 
                            & Baseline Correction  & ~\cite{sun2021hybrid} \\
                            & Baseline Removal     & ~\cite{nouri2024detection} \\
                            & Detrending           & ~\cite{Seizure49} \\
\hline
\multirow{3}{80pt}{Channel Processing} 
                            & Channel Selection    & ~\cite{wen2018deep} \\
                            & Channel Mapping      & ~\cite{Kostas2021BENDR} \\
                            & Re-Referencing       & ~\cite{nouri2024detection} \\
\hline
\multirow{3}{80pt}{Normalization \& Scaling} 
                            & Z-Normalization      & ~\cite{ACHARYA2018270} \\
                            & Quantile Normalization & ~\cite{ko2022eeg} \\
                            & Scaling \& Shifting  & ~\cite{Kostas2021BENDR} \\
\hline
\multirow{4}{80pt}{Sampling Adjustment} 
                            & Downsampling         & ~\cite{mousavi2019deep} \\
                            & Resampling           & ~\cite{ZHANG2020105089} \\
                            & Interpolation        & ~\cite{SZ27} \\
                            & Imputation           & ~\cite{sharma2021dephnn} \\
\hline
\multirow{1}{80pt}{Segmentation} 
                            & Windowing            & ~\cite{seal2021deprnet} \\
\hline
\multirow{2}{80pt}{Signal Alignment \& Synchronization} 
                            & Time Synchronization & ~\cite{iwama2023two} \\
                            & Temporal Alignment   & ~\cite{iwama2023two} \\
\hline
\end{tabular}
\end{table}

\subsection{Signal Collection}
EEG has evolved significantly since Hans Berger first recorded signals from the human scalp in 1924~\cite{berger1929elektroenkephalogramm}.
While EEG is typically collected non-invasively with scalp electrodes placed according to the 10-20 system~\cite{jasper1958ten},
recent studies employ higher-density configurations for enhanced spatial resolution.
EEG captures oscillations across frequency bands linked to neural states: delta (deep sleep), theta (light sleep), alpha (relaxation), beta (focus), and gamma (higher cognition)~\cite{buzsaki2004neuronal}.
Depending on the study, participants may perform tasks or rest to elicit relevant brain activity. Resting-state EEG evaluates baseline activity, while specific tasks can highlight disease-related abnormalities~\cite{jeong2004eeg}.

iEEG involves implanting electrodes within deep or superficial structures via burr holes (SEEG) or placing grids on the cortical surface (ECoG).
Compared to EEG, iEEG offers higher spatial resolution and reduced susceptibility to artifacts from scalp and eye movements.
SEEG allows recording from deep and distributed regions with minimal invasiveness, while ECoG provides detailed cortical surface mapping with dense grids.
However, iEEG remains affected by cardiac artifacts, electrode shifts, and other noise, making rigorous preprocessing essential for reliable clinical and research use.

\begin{table}[t]
\renewcommand{\arraystretch}{1.2}
\caption{Feature Extraction Techniques}
\label{tab:highlevel}
\footnotesize
\centering
\begin{tabular}{p{80pt}p{90pt}p{30pt}} 
\hline
\textbf{Techniques}     & \textbf{Details}                 & \textbf{Reference}             \\
\hline
\multirow{2}{80pt}{Data Augmentation} & Oversampling                     & ~\cite{ZHANG2020105089}         \\
                                       & ELM-AE                           & ~\cite{9713847}                 \\
\hline
\multirow{2}{80pt}{Signal Decomposition \& Transformation} 
                                       & Time-Frequency Analysis          & ~\cite{PD2} \\
                                       & Empirical Decomposition          & ~\cite{zulfikar2022empirical}   \\
\hline
\multirow{3}{80pt}{Spectral \& Power Analysis}  
                                       & Power Spectrum                   & ~\cite{li2019eeg}               \\
                                       & Spectral Density                 & ~\cite{Seizure214}        \\
                                       & Partial Directed Coherence       & ~\cite{khan2021automated}       \\
\hline
\multirow{3}{80pt}{Time-Domain Features Extraction}  
                                       & Statistical Measures             & ~\cite{zhu2019multimodal}       \\
                                       & Amplitude \& Range               & ~\cite{Seizure7}    \\
                                       & Hjorth Parameters                & ~\cite{li2019depression}             \\
\hline
\multirow{2}{80pt}{Frequency-Domain Features Extraction}  
                                       & Band Power Features              & ~\cite{Seizure67}        \\
                                       & Spectral Measures                & ~\cite{tosun2021effects}        \\
\hline
\multirow{3}{80pt}{Time-Frequency Features Extraction}  
                                       & Wavelet Coefficients             & ~\cite{aslan2022deep}           \\
                                       & STFT Features                    & ~\cite{choi2019novel}           \\
                                       & Multitaper Spectral              & ~\cite{vilamala2017deep}        \\
\hline
\multirow{3}{80pt}{Other Features Extraction}  
                                       & Nonlinear Features               & ~\cite{Seizure109}        \\
                                       & Spatial Features                 & ~\cite{phang2019multi} \\
                                       & Transform-Based Features         & ~\cite{electronics11142265} \\
\hline
\multirow{2}{80pt}{Source Imaging}                                                        & Conventional Methods & ~\cite{sohrabpour2020noninvasive} \\
                                       & Deep Learning-based & ~\cite{sun2022deep} \\
\hline
\multirow{2}{80pt}{Graph Analysis}                                                         & Clustering Coefficient           & ~\cite{Zhan2020EpilepsyDetection}       \\
                                       & Other Graph Metrics              & ~\cite{ho2023self}   \\
\hline
\end{tabular}
\end{table}

\begin{table*}[t]
\renewcommand{\arraystretch}{1.15}
\caption{Summary of subject-level data partitioning strategies for EEG/iEEG.}
\label{tab:partitioning}
\footnotesize
\begin{tabular}{m{2.5cm}p{5.5cm}p{3.5cm}p{4.5cm}}  
\hline
\textbf{Strategy} & \textbf{Formal Definition} & \textbf{Advantages} & \textbf{Limitations} \\
\hline
\textbf{Subject-Specific} &
$\mathcal{X}_{tr} \cup \mathcal{X}_{val} \cup \mathcal{X}_{te} 
= \{\mathbf{X}^{(i)}_k\}_{k=1}^{K^{(i)}}$ &
Rapid prototyping \newline Useful for personalization &
Restricted clinical applicability \newline Poor transferability across individuals \\
\hline
\textbf{Mixed-Subject} &
\(
\begin{array}{l}
\mathcal{X}_{set} \subset \bigcup_{i \in \mathcal{P}} \bigcup_{k=1}^{K^{(i)}} \{\mathbf{X}^{(i)}_k\} \\
|\mathcal{X}_{set}| = \alpha_{set} \sum_{i=1}^{N} K^{(i)}
\end{array}
\) &
Maximizes training data \newline Robust to variability  &
Potential risk of data leakage \newline Reduced clinical relevance \\
\hline
\textbf{Cross-Subject} &
\(
\begin{array}{l}
\mathcal{P} = \mathcal{P}_{tr} \cup \mathcal{P}_{val} \cup \mathcal{P}_{te} \\
\mathcal{X}_{set} = \bigcup_{i \in \mathcal{P}_{set}} \bigcup_{k=1}^{K^{(i)}} \{\mathbf{X}^{(i)}_k\}
\end{array}
\) &
Clinically relevant \newline Realistic deployment &
High data demand \newline Computational burden \\
\hline
\end{tabular}
\end{table*}

\subsection{Signal Preprocessing}

EEG/iEEG signals require low-level preprocessing to address challenges such as noise and artifact removal, normalization for consistency, and segmentation into analyzable time windows.
These steps refine raw data, ensuring it accurately reflects brain activity and provides a robust foundation for analysis.
Representative methods are summarized in Table~\ref{tab:lowlevel}.

Noise reduction is central to this process: classical FIR/IIR filtering~\cite{banville2021uncovering,oh2020deep} efficiently removes narrow-band artifacts like power-line interference, whereas Blind Source Separation (e.g., ICA, PCA~\cite{9047940}) targets ocular and muscular noise but may also suppress neural components if applied indiscriminately. Wavelet decomposition~\cite{akut2019wavelet} offers multiscale handling of nonstationary noise, though at higher computational cost. 
Normalization techniques such as Z-score scaling~\cite{amer2023eeg} standardize channel amplitudes, improving model stability but potentially masking inter-individual variability. Segmentation and resampling further balance efficiency and fidelity: downsampling can reduce computational load~\cite{mousavi2019deep}, while shorter epochs facilitate localized analysis but risk fragmenting long-range dependencies. Finally, baseline correction~\cite{amer2023eeg}, channel selection~\cite{wen2018deep}, and alignment~\cite{iwama2023two} enhance interpretability and multimodal synchronization, though each relies on assumptions that may not hold uniformly across datasets.

\subsection{Feature Extraction}

Feature extraction techniques transform raw signals into structured representations by isolating salient features or reconstructing core components essential for modeling. Representative methods are summarized in Table~\ref{tab:highlevel}.

Time-domain features are straightforward and interpretable (e.g., statistical moments, Hjorth parameters~\cite{oh2014novel}), but insufficient to capture complex spectral dynamics. Frequency-domain features such as power spectral density and band power~\cite{li2019eeg} reveal oscillatory activity, yet assume stationarity. Time–frequency approaches address this by linking temporal and spectral information, making them effective for transient, nonstationary patterns in seizure detection and cognitive monitoring~\cite{akut2019wavelet, choi2019novel}, though at higher computational cost.

At a higher level, electrophysiological source imaging (ESI) improves spatial specificity by projecting EEG into cortical source space~\cite{sohrabpour2020noninvasive}, but depends on accurate head models. Graph analysis instead quantifies network-level organization~\cite{phang2019multi}, offering system-wide insights while remaining sensitive to noise and thresholding. Together, these methods extend analysis from local dynamics to global connectivity, supporting applications from seizure focus localization to network alterations in Alzheimer’s disease.

\subsection{Data Partitioning Strategies}

Building on the definition of \(\mathbf{X}^{(i)} \in \mathbb{R}^{C \times T}\) in Section~\ref{sec:pdef}, where \(\mathbf{X}^{(i)}\) represents the EEG/iEEG signal of subject \(i\), we define notations to formalize data partitioning strategies:

\begin{itemize}
    \item \(\mathcal{P} = \{1, 2, \dots, N\}\): The set of \(N\) subjects in the dataset.
    \item \(\mathcal{X}_\text{train}, \mathcal{X}_\text{val}, \mathcal{X}_\text{test}\): The training, validation, and testing sets, respectively.
    \item \(\alpha_\text{train}, \alpha_\text{val}, \alpha_\text{text}  \in (0, 1)\): The proportion of data used for training, validation and test, and \(\alpha_\text{train} + \alpha_\text{val} + \alpha_\text{test} = 1.\)
    \item \(K^{(i)}\): The total number of temporal segments or events derived from subject \(i\)'s data.
\end{itemize}

Previous studies have examined data partitioning strategies; for instance, Zancanaro et al.~\cite{zancanaro2021cnn} compared leave-one-subject-out, fixed subject splits, and pooled training in motor imagery classification.
Building on these insights, we introduce a formal taxonomy encompassing subject-specific, mixed-subject, and cross-subject strategies, with mathematical definitions and mapping to practical EEG/iEEG applications (Table~\ref{tab:partitioning}). 
\textbf{Subject-specific} methods are typically adopted in personalized or closed-loop systems where individual calibration is critical.  
\textbf{Mixed-subject} methods are widely used in early studies for efficient training, though they risk data leakage across sets.  
\textbf{Cross-subject} methods are clinically most relevant, ensuring evaluation on unseen patients and reflecting real-world deployment.

Extending subject-level partitioning strategies, dataset-level partitioning includes three approaches: \textbf{dataset-specific} (independent partitioning per dataset), \textbf{mixed-dataset} (pooling data across datasets), and \textbf{cross-dataset} (disjoint datasets for training, validation, and testing). 
While subject-based partitions remain the standard for evaluating patient-level clinical relevance, dataset-based strategies have become increasingly common—particularly in self-supervised learning to mitigate data scarcity and in multi-domain models to demonstrate cross-dataset transferability. In practice, some studies combine both paradigms, using subject-based partitioning to assess patient-level performance and dataset-based partitioning to evaluate broader generalization, thereby testing whether methods can achieve both specialization and generalizability.

\subsection{Deep Learning Architectures}
Neurological data processing relies on several key architectures:
\textbf{Convolutional Neural Networks (CNNs)}~\cite{lecun1995convolutional} excel at extracting spatial/spectral features through hierarchical convolutions.
\textbf{Recurrent Neural Networks (RNNs)}~\cite{elman1990finding} capture temporal dependencies via recurrent connections.
\textbf{Transformers}~\cite{vaswani2017attention} model long-range spatiotemporal relationships using self-attention.
\textbf{Graph Neural Networks (GNNs)}~\cite{4700287} analyze functional connectivity in graph-structured data.
\textbf{Autoencoders (AEs)}~\cite{hinton1993autoencoders} learn compressed representations through encoder-decoder structures.
\textbf{Generative Adversarial Networks (GANs)}~\cite{goodfellow2014generative} synthesize signals through adversarial training.
\textbf{Spiking Neural Networks (SNNs)}~\cite{maass1997networks} leverage spike-based computation for temporal dynamics.

\begin{table*}[t]
\centering
\renewcommand{\arraystretch}{1.2}
\caption{Public EEG/iEEG datasets for seizure detection, with \textbf{Seizures} indicating the number of episodes, \textbf{Length} the duration of each record, and \textbf{Size} the total duration of recording.}
\label{tab:ep}
\footnotesize
\begin{tabular}{lccccccc}
\hline
\textbf{Dataset} & \textbf{Type} &  \textbf{Subjects} &  \textbf{Seizures} & \textbf{Length} & \textbf{Size} &  \textbf{Frequency (Hz)} &  \textbf{Channels} \\ \hline
\href{https://www.ukbonn.de/epileptologie/arbeitsgruppen/ag-lehnertz-neurophysik/downloads/}{Bonn}~\cite{andrzejak2001indications}             & EEG & {10}                                 & {-}       & {23.6 sec}          & {$\approx$ 3.3 hours}                           & {173.61}                   & {1}               \\
\href{https://epilepsy.uni-freiburg.de/freiburg-seizure-prediction-project/eeg-database}{Freiburg}~\cite{ihle2012epilepsiae}         & iEEG & {21}            & {87}         & {4 sec}         & {$\approx$ 504 hours }         & {256}                      & {128}               \\
\href{https://www.kaggle.com/c/seizure-detection}{Mayo-UPenn}~\cite{seizure-detection}       & iEEG & {2}                                  & {48}                    & {1 sec}           & {583 min}                         & {500-5000}                 & {16-76}             \\
\href{https://physionet.org/content/chbmit/1.0.0/}{CHB-MIT}~\cite{guttag2010chb,shoeb2009application,goldberger2000physiobank}          & EEG & {22}                                 & {198}          & {1 hour}                    & {$\approx$ 686 hours}                    & {256}                      & {23 / 24 / 26}         \\ 
\href{https://www.upf.edu/web/ntsa/downloads/-/asset_publisher/xvT6E4pczrBw/content/2012-nonrandomness-nonlinear-dependence-and-nonstationarity-of-electroencephalographic-recordings-from-epilepsy-patients}{Bern-Barcelona}~\cite{andrzejak2012nonrandomness}   & iEEG & {5}                                  & {3750}              & {20 sec}               & {57 hours}                        & {512}                      & {64}                \\
\href{https://www.researchgate.net/publication/308719109_EEG_Epilepsy_Datasets}{Hauz Khas}~\cite{hauz}        & EEG & {10}                                 & {-}       & {5.12 sec}          & {87 min}                           & {200}                      & {50}                \\
\href{https://www.epilepsyecosystem.org/}{Melbourne}~\cite{melbourne}        & iEEG & {3}                                  & {-}      & {10 min}             & {81.25 hours}                            & {400}                      & {184}               \\
\href{https://isip.piconepress.com/projects/nedc/html/tuh_eeg/#c_tusz}{TUSZ}~\cite{shah2018temple}             & EEG & {642}                                & {3050}               & {-}              & {700 hours}                         & {250}                      & {19}                \\
\href{http://ieeg-swez.ethz.ch/}{SWEC-ETHZ}~\cite{burrello2018oneshot,burrello2019hdc} 
        & iEEG           & 18 / 16 & 244 / 100 & 1 hour / 3 min & 2656 hours / 48 min & 512 / 1024 & 24-128 / 36-100 \\
\href{https://zenodo.org/records/2547147#.Y7eU5uxBwlI}{Zenodo}~\cite{stevenson2019dataset}         & EEG & {79}                                 & {1379}                  & {74 min}           & {$\approx$ 97 hours}                    & {256}                      & {21}                \\
\href{https://www.kaggle.com/datasets/nejedlypetr/multicenter-intracranial-eeg-dataset}{Mayo-Clinic}~\cite{Nejedly2020}      & iEEG & {25}                                 & {-}           & {3 sec}      & {50 hours}                          & {5000}                     & {1}                 \\
\href{https://www.kaggle.com/datasets/nejedlypetr/multicenter-intracranial-eeg-dataset}{FNUSA}~\cite{Nejedly2020}           & iEEG & {14}                                 & {-}        & {3 sec}         & {7 hours}                           & {5000}                     & {1}                 \\
\href{https://physionet.org/content/siena-scalp-eeg/1.0.0/}{Siena}~\cite{detti2020eeg}            & EEG & {14}                                 & {47}                  & {145-1408 min}             & {$\approx$ 128  hours}                         & {512}                      & {27}                \\
\href{https://data.mendeley.com/datasets/5pc2j46cbc/1}{Beirut}~\cite{nasreddine2021epileptic}           & EEG & {6}                                  & {35}        & {1 sec}                       & {130 min}                          & {512}                      & {19}                \\
\href{https://openneuro.org/datasets/ds004100/versions/1.1.1}{HUP}~\cite{HUP}              & iEEG & {58}                                 & {208}                       & {300 sec}       & {$\approx$ 27 hours}                    & {500}                      & {52-232}            \\
\href{https://openneuro.org/datasets/ds004080/versions/1.2.4} {CCEP}~\cite{ds004080:1.2.4} & iEEG & {74} & {-} & {-} & {89 hours} & {2048} & {48-116} \\  \hline
\end{tabular}
\end{table*}

\subsection{Deep Learning Paradigms}
Deep learning applications in neurological diagnostics fall into four paradigms: supervised, self-supervised, unsupervised, and semi-supervised learning.
Each paradigm addresses specific challenges in processing brain signals by leveraging architectures tailored to data availability and task requirements.

\subsubsection{Supervised Learning}
Supervised learning is the dominant paradigm for neurological diagnostics tasks, training models to map signals $ \mathbf{X} \in \mathbb{R}^{C \times T} $ to labels $y \in \mathcal{Y} $.

\subsubsection{Unsupervised Learning}
Unsupervised learning is essential for uncovering intrinsic data structures in signals $\mathbf{X}$, enabling representation learning without relying on labels. 

\subsubsection{Semi-Supervised Learning}
Semi-supervised learning combines a small set of labeled examples $\{(x_i, \hat{y}_i)\}_{i=1}^l$, where $\hat{y}_i$ denotes the provided labels, with a larger set of unlabeled examples $\{x_j\}_{j=l+1}^{l+u}$ to learn a mapping from $\mathbf{X}$ to $\mathcal{Y}$. 

\subsubsection{Self-Supervised Learning}
Self-supervised learning (SSL) leverages unlabeled EEG/iEEG data by constructing pretext tasks that generate pseudo-labels $\hat{y}$ from intrinsic properties of raw signals $\mathbf{X}$. 
SSL methods fall into three main categories: contrastive, predictive, and reconstruction-based learning.
\textbf{Contrastive-based methods}, such as Contrastive Predictive Coding (CPC)~\cite{banville2021uncovering} and Transformation Contrastive Learning~\cite{mohsenvand2020contrastive}, learns by maximizing similarity between related views while minimizing it between unrelated ones, capturing distinguishing signal features.
\textbf{Predictive-based learning} employs pretext tasks such as Relative Positioning and Temporal Shuffling to extract structural patterns across temporal, frequency, and spatial domains~\cite{banville2019self, oord2018representation}. By predicting transformations applied to the data, it enhances domain-specific feature learning.
\textbf{Reconstruction-based learning} trains models to reconstruct masked signal segments. Methods like Masked Autoencoders (MAE) reconstruct temporal or spectral components, learning intrinsic patterns in the process~\cite{Kostas2021BENDR, wu2022neuro2vec}.
Studies have also explored hybrid methods, which combine elements from contrastive, predictive, and reconstruction-based approaches~\cite{cai2023mbrain, banville2021uncovering}.

\section{Applications}
\label{sec:app}

This section reviews neurological disease diagnosis methodologies. Each subsection introduces the disease, its diagnostic tasks, and related public datasets, followed by representative studies highlighting key deep learning aspects such as data types, frequency bands, and brain regions.
Summary tables report representative studies with their reported metrics (e.g., accuracy, AUC) and dataset chance levels. These metrics are for reference only, as evaluation protocols and data selection vary across studies, which may also cause slight differences in chance levels.
Technical details on preprocessing, network architectures, and training are compiled in the Appendix, covering all 450 reviewed studies.

\subsection{Seizure Disorder}

\begin{table*}[t]
\centering
\renewcommand{\arraystretch}{1.2}
\caption{Summary of related studies on EEG-based seizure detection with different learning paradigms, feature extraction methods, and backbones. The chance level is omitted due to inconsistent data selection criteria across studies on TUSZ.}
\label{tab:sz}
\footnotesize
\begin{tabular}{ccccccccccc}
\hline
\textbf{Dataset} & \textbf{Task} & \textbf{Paradigm} & \textbf{Feature} & \textbf{Backbone} &\textbf{Splitting} & \textbf{Accuracy} & \textbf{AUC} &  \\ 
\hline

\multirow{6}{*}{\centering Bonn} 
 & \multirow{6}{*}{\centering ternary} 
 & \multirow{4}{*}{\centering Supervised Learning} & raw & CNN & generalized & 0.8867 & -&\cite{ACHARYA2018270} \\ 
 &  &  & raw & CNN &generalized & 0.9900 & -& \cite{ULLAH201861} \\ 
 &  &  & Wavelet Coefficients & CNN &generalized & \textbf{0.9940} &- & \cite{akut2019wavelet} \\ 
 &  &  & Scalograms & 2D-CNN & generalized & 0.9900 & -& \cite{turk2019epilepsy} \\ 
 &  & Unsupervised Learning & AE-based & CNN &cross-subject& 0.9933  & -& \cite{abdelhameed2018epileptic} \\ 
 \hline
 &   \textit{chance level} &&  &  & & 0.4000 &  \\ 
\hline\hline
\multirow{5}{*}{\centering CHB-MIT} 
 & \multirow{5}{*}{\centering binary} 
 & \multirow{3}{*}{\centering Supervised Learning} & Spectrogram & 2D-CNN&subject-specific & 0.9750 & -& \cite{zhou2018epileptic} \\ 
 &  &  & raw & CNN-LSTM&cross-subject & \textbf{0.9771} &- & \cite{dutta2024deep} \\ 
 &  &  & Correlation Matrix & GAT-Transformer &cross-subject& 0.7315 & 0.72 & \cite{zhao2023hybrid} \\ 
 &  &  \multirow{2}{*}{\centering Self-supervised Learning} & raw & Transformer &cross-subject& 0.9707 & \textbf{0.97} & \cite{XIAO2024105464} \\ 
  &  &  & raw & CNN & cross-subject & - & 0.88 &\cite{zheng2022task} \\
\hline
 &   \textit{chance level} &&  &  & & 0.5000 &  \\ 
 \hline\hline
\multirow{5}{*}{\centering TUSZ} 
 & 8-class & Supervised Learning & Spectrogram & 2D-CNN& cross-subject & \textbf{0.8890} & - & \cite{raghu2020eeg} \\
 & binary & \multirow{4}{*}{Self-supervised Learning} & Correlation Matrix & GNN & cross-subject & - & \textbf{0.88} & \cite{tang2021self} \\
  & 4-class &  & Correlation Matrix & GNN & cross-subject & - & 0.75 & \cite{tang2021self} \\
& 4-class &  & Wavelets & Transformer & cross-subject & 0.7300 & - & \cite{peng2023wavelet2vec} \\
& binary &  & raw & CNN-GCN & cross-subject & - & 0.78 & \cite{cai2023mbrain} \\

\hline
\end{tabular}
\end{table*}

\subsubsection{Task Description}
Epilepsy, a neurological disorder affecting 50 million people, is characterized by recurrent seizures caused by abnormal brain activity. 
Seizures range from brief confusion or blanking out to severe convulsions and loss of consciousness. According to WHO, up to 70\% of cases can be effectively treated with proper care. However, in low-income regions, limited resources and stigma hinder access to treatment, increasing the risk of premature death~\cite{WHO_epilepsy}.

Seizure detection primarily relies on standardized EEG/iEEG datasets, summarized in Table~\ref{tab:ep}. The key challenge is distinguishing seizure events from background activity, typically framed as binary classification where $y_k \in \{0, 1\}$. 
Most approaches segment EEG sequences into short windows for classification, then aggregate predictions to form event-level outcomes as $\mathbf{Y} = \bigcup_{k=1}^K \{y_k\}$~\cite{xu2023patient,peng2023wavelet2vec}. 
Alternatively, some methods detect cut points in continuous recordings to define segment boundaries $\{\mathbf{X}_k\}_{k=1}^K$, each classified independently~\cite{Zhan2020EpilepsyDetection}. 
Final event-level predictions are obtained by combining labels $\mathbf{Y} = \bigcup_{k=1}^K \{\Phi_{\text{segment}}(\mathbf{X}_k; \boldsymbol{\theta})\}$.

Beyond binary tasks, more fine-grained classification has been explored. Three-class settings distinguish interictal (A, between seizures), preictal (D, before onset), and ictal (E, seizure) states~\cite{zhou2018epileptic}, while five-class tasks further subdivide the preictal phase into early, middle, and late stages~\cite{turk2019epilepsy}. 
The Temple University Seizure Corpus (TUSZ)~\cite{shah2018temple} supports such studies, providing detailed annotations of pathological events (e.g., epileptiform discharges, seizure types) and non-pathological signals (e.g., background activity, artifacts).

In addition, epileptic focus localization identifies the cortical origin of pathological discharges, formulated as classifying iEEG contacts inside versus outside the epileptogenic zone~\cite{sui2019localization}, or reconstructing source-level activity from scalp EEG via ESI~\cite{worrell2000localization}. This task is clinically critical, as accurate localization guides surgical resection in drug-resistant epilepsy.

\subsubsection{Supervised Methods}
Supervised seizure detection has advanced with public datasets and progress in deep learning.
Early studies relied on subject-specific or mixed-subject evaluations using short, pre-segmented EEG clips.
For instance, the Bonn dataset~\cite{andrzejak2001indications} contains manually labeled seizure and non-seizure segments, enabling models to operate on fixed-length inputs.
Approaches based on raw signals employ CNNs or RNNs to learn spatiotemporal features from these standardized segments~\cite{ACHARYA2018270,ULLAH201861}, while feature-based methods transform signals into handcrafted or derived representations, such as scalograms~\cite{turk2019epilepsy} and wavelet-based features~\cite{akut2019wavelet}, which are then used by shallow classifiers.
These techniques inherently assume limited temporal context and circumvent the challenges of segmenting continuous EEG.
As shown in Table~\ref{tab:sz}, the Bonn dataset is relatively simple and prone to overfitting, making it insufficient to represent real-world clinical scenarios.

With the adoption of long-term recordings like CHB-MIT~\cite{shoeb2009application}, the focus shifts toward cross-subject paradigms. 
Unlike Bonn, CHB-MIT provides continuous recordings with multiple seizure episodes per patient, requiring models to handle variable-length inputs and detect seizure onsets in unsegmented streams.
Approaches integrate temporal modeling through sliding windows~\cite{xu2023patient}, sequence-aware architectures such as Transformers~\cite{lih2023epilepsynet}, or hybrid fusion techniques~\cite{dutta2024deep}. Cross-subject validation becomes standard, reflecting clinical requirements that generalize across diverse conditions.

The necessity of cross-subject modeling in seizure detection stems from its critical role in ensuring clinical generalization.
The invasive nature of iEEG differentiates its modeling requirements from EEG through distinct acquisition paradigms and neurophysiological characteristics, its patient-specific recording conditions and electrode configurations lead to substantial inter-subject heterogeneity in temporal features and spatial sampling, unlike EEG's standardized scalp placement~\cite{zhang2024brant}. Balancing high-resolution spatiotemporal capture with robustness across patients, iEEG requires specialized methodologies to enhance generalizability while addressing its inherent complexities.
Spatial modeling is essential for capturing 3D epileptogenic networks with depth electrodes. Graph-based methods model inter-channel dependencies via neuroanatomical~\cite{9345750} or dynamic functional connections~\cite{rahmani2023meta}, while Transformers use attention mechanisms to adapt to varying electrode configurations~\cite{sun2022continuous}.
DMNet~\cite{tudmnet} improves domain generalization through self-comparison mechanisms.

\subsubsection{Semi- and Unsupervised Methods}
Semi-supervised and unsupervised techniques are increasingly applied in deep learning for seizure detection, particularly when labeled data is limited. 
A common approach incorporates clustering for event-level segmentation, allowing the model to identify and segment seizure events~\cite{Zhan2020EpilepsyDetection}.
Another application involves using models such as Autoencoders, DBNs and GANs to automatically extract relevant features or augment datasets, thereby enhancing the model’s robustness and generalizability~\cite{abdelhameed2018epileptic,you2020unsupervised}.

\subsubsection{Self-supervised Methods}
Self-supervised learning has emerged as an effective approach for seizure detection. 
Contrastive learning methods form positive and negative pairs to capture seizure-related patterns.
For instance, SLAM~\cite{XIAO2024105464} pairs an anchor with a window from a distant time point as a negative sample, while SPP-EEGNET~\cite{li2022spp} uses the absolute difference between two windows to determine pair similarity.
Predictive-based methods design pretext tasks to simulate epileptic features, such as augmenting signals with amplitude or frequency changes~\cite{zheng2022task} or predicting the next segment using graph-based modeling~\cite{tang2021self}.
Reconstruction-based methods focus on preserving context during learning.
EpilepsyNet~\cite{lih2023epilepsynet} uses Pearson Correlation Coefficients to capture spatial-temporal embeddings, while Wavelet2Vec~\cite{peng2023wavelet2vec} reconstructs wavelet-transformed EEG patches to exploit seizure-specific discharge patterns across frequency bands.
EEG-CGS\cite{ho2023self} adopts a hybrid graph-based approach, framing seizure detection as anomaly detection and integrating subgraph sampling with contrastive and reconstruction learning.
As shown in Table~\ref{tab:sz}, SSL methods exhibit considerable performance variations across datasets.
On more challenging datasets like TUSZ, performance approaches that of supervised methods, underscoring the need for larger-scale pretraining and stronger representation learning.
Furthermore, four-class seizure type classification remains more difficult than detection, highlighting persistent bottlenecks in distinguishing subtypes.

SSL paradigm is also common in iEEG-based modeling. 
BrainNet~\cite{chen2022brainnet} employs bidirectional contrastive predictive coding to capture temporal correlation in SEEG signals.
MBrain~\cite{cai2023mbrain} models time-varying propagation patterns and inter-channel phase delays of epileptic activity through a multivariant contrastive-predictive learning framework, leveraging graph-based representations for spatial-temporal correlations across EEG and SEEG channels.
PPi~\cite{yuan2024ppi} accounts for regional seizure variability, employing a channel discrimination task to ensure the model captures distinct pathological patterns across brain regions rather than treating all channels uniformly.

\begin{table}[t]
\renewcommand{\arraystretch}{1.2}
\caption{Public Sleep EEG Datasets, where \textbf{Recordings} denotes the number of whole-night PSG recordings.}
\label{tab:sleep}
\footnotesize
\centering
\begin{tabular}{lccc}
\hline
\textbf{Dataset}      & \textbf{Recordings}                     & \textbf{Frequency (Hz)} & \textbf{Channels} \\
\hline
\href{http://www.physionet.org/physiobank/
database/sleep-edfx/}{Sleep-EDF}~\cite{kemp2000analysis,goldberger2000physiobank}             & 197                 & 100          & 2                 \\
\href{http://ceams-carsm.ca/en/MASS/}{MASS}~\cite{oreilly2014montreal}                  & 200                 & 256          & 4-20         \\
\href{https://sleepdata.org/datasets/shhs}{SHHS}~\cite{quan1997sleep,zhang2018national}                  & 8362                     & 125          & 2                 \\
\href{https://physionet.org/content/ucddb/1.0.0/}{SVUH\_UCD}~\cite{ucddb2007sleep,goldberger2000physiobank}              & 25              & 128          & 3                 \\
\href{https://physionet.org/content/hmc-sleep-staging/1.1/}{HMC}~\cite{Alvarez-Estevez2022, goldberger2000physiobank} & 151 & 256 & 4\\
\href{https://physionet.org/content/challenge-2018/1.0.0/#files}{PC18}~\cite{ghassemi2018you,goldberger2000physiobank}                  & 1985                     & 200          & 6                 \\
\href{https://physionet.org/content/slpdb/1.0.0/}{MIT-BIH}~\cite{ichimaru1999development,goldberger2000physiobank}              & 16                   & 250          & 1                 \\
\href{https://zenodo.org/records/15900394}{DOD-O}~\cite{dod_dataset}                   & 55                   & 250          & 8                 \\
\href{https://zenodo.org/records/15900394}{DOD-H}~\cite{dod_dataset}                   & 25                   & 250          & 12                 \\
\href{https://sleeptight.isr.uc.pt/}{ISRUC}~\cite{khalighi2016isruc}              & 126                      & 200          & 6                 \\
\href{https://bdsp.io/content/hsp/2.0/}{MGH}~\cite{biswal2018expert}                  & 25941                   & 200          & 6                 \\
\href{http://stat.case.edu/ayp2/EEGdat}{Piryatinska}~\cite{piryatinska2009automated}           & 37         & 64           & 1                 \\
\href{https://zenodo.org/records/2650142}{DRM-SUB}~\cite{devuyst2005dreams} & 20 & 200 & 3 \\
\href{https://openneuro.org/datasets/ds004902/versions/1.0.5}{SD-71}~\cite{xiang2023resting} & 142 & 500 & 61 \\
\hline
\end{tabular}
\end{table}

\begin{table}[t]
\renewcommand{\arraystretch}{1.1}
\caption{Reported accuracies on Sleep-EDF datasets using representative models (grouped by learning paradigm).}
\label{tab:sleep-acc}
\footnotesize
\centering
\begin{tabular}{>{\centering\arraybackslash}m{0.9cm}
                >{\centering\arraybackslash}m{1.1cm}
                >{\centering\arraybackslash}m{2.0cm}
                >{\centering\arraybackslash}m{0.8cm}
                >{\centering\arraybackslash}m{0.8cm}
                >{\centering\arraybackslash}m{0.7cm}}
\hline
\textbf{Learning Paradigm}  & \textbf{Modality} & \textbf{CL Strategy} & \textbf{Sleep-EDF} & \textbf{Sleep-EDFx}& \textbf{} \\
\hline
\multirow{2}{*}{\textbf{SL}} 
   & EEG      & --                         & 0.8440 & 0.8130& \cite{eldele2021attention} \\
     & EEG+EOG  & --                         &   --    & 0.8390& \cite{Sleep28} \\
\hline
\multirow{4}{*}{\textbf{SSL}} 
            & EEG      & Global Reference       &   --    & \textbf{0.8690}& \cite{yang2023self} \\
                     & EEG      & Time–spectrogram Multi-view     &   --    & 0.7806& \cite{kumar2022muleeg} \\
                   & EEG      & Time–frequency Multi-view       & 0.7160 &   --    & \cite{ye2021cosleep}\\
                    & EEG+EOG  & Contrastive Alignment & \textbf{0.8458} & 0.8284 & \cite{zhang2024brantx}\\
\hline
\multicolumn{3}{l}{\textit{Chance Level}} & 0.4207 & 0.3537& \\
\hline
\end{tabular}
\end{table}

\subsection{Sleep Staging}

\subsubsection{Task Description}
Sleep staging is critical to understanding sleep disorders like insomnia and sleep apnea, as well as the impact on overall health.
It is estimated that 20\% to 41\% of the global population is affected by sleep disorders, which are linked to an increased risk of obesity, cardiovascular diseases, and mental health issues~\cite{recoveryvillage_sleep_statistics_2023}. 
Therefore, accurately identifying sleep stages is essential for addressing these concerns.

Sleep staging involves segmenting signals into 30-second epochs and classifying them into stages: awake (W), rapid eye movement (REM), and three non-REM (NREM) stages (N1, N2, N3).
Wake is characterized by high-frequency $\beta$ and $\alpha$ waves. In N1, the transition to sleep, low-amplitude $\theta$ waves appear. N2, light sleep, is marked by sleep spindles and K-complexes associated with sensory processing and memory consolidation. N3, or deep sleep, features slow-wave $\delta$ activity. REM sleep, essential for emotional regulation and dreaming, is characterized by rapid, low-voltage brain activities.

Multimodal modeling is fundamental for sleep analysis, as polysomnography (PSG) integrates EEG (e.g., Fpz-Cz, Pz-Oz), Electrooculography (EOG), and Electromyography (EMG) to enhance staging accuracy.
The public datasets in Table~\ref{tab:sleep} provide a comprehensive view of resourcess.


\subsubsection{Supervised methods}

Selecting biosignal modalities is critical for designing supervised learning frameworks in PSG-based sleep staging.
Two primary paradigms are widely used. 
Single-channel EEG methods, preferred in resource-constrained settings, offer hardware simplicity, reduced cross-modal interference, and enhanced computational efficiency~\cite{tsinalis2016automatic}. 
However, relying solely on EEG limits the detection of complementary cues—such as ocular and muscular activities—essential for identifying ambiguous sleep stages.
Hybrid EEG-EOG models provide a balance between diagnostic accuracy and computational efficiency, while full multimodal designs integrating EEG, EOG, and EMG most closely emulate clinical scoring protocols~\cite{Sleep28}.

\begin{table}[t]
\renewcommand{\arraystretch}{1.2}
\caption{Public EEG Datasets for Depression Detection, where \textbf{Exp (n)} represents the number of depressed individuals and \textbf{Ctrl (n)} represents the healthy control group.}
\label{tab:dep}
\footnotesize
\centering
\begin{tabular}{lcccc}
\hline
\textbf{Dataset}     & \textbf{Exp (n)}                  & \textbf{Ctrl (n)} & \textbf{Frequency (Hz)} & \textbf{Channels} \\
\hline
\href{https://figshare.com/articles/dataset/EEG_Data_New/4244171}{HUSM}~\cite{Mumtaz2016}      & 34                            & 30               & 256              & 22                \\
\href{http://predict.cs.unm.edu/downloads.php}{PRED+CT}~\cite{cavanagh2017patient} & 46                            & 75               & 500              & 64                \\
\href{https://github.com/EllieYLJ/EEG-GA-LASSO}{EDRA}~\cite{yang2023automatic} & 26                            & 24               & 500              & 63                \\
\href{https://modma.lzu.edu.cn/data/index/}{MODMA}~\cite{cai2022multi}    & \begin{tabular}[c]{@{}l@{}}24 \\ 26 \end{tabular} & \begin{tabular}[c]{@{}l@{}}29\\ 29\end{tabular} & 250              & \begin{tabular}[c]{@{}l@{}}128\\ 3\end{tabular} \\
\hline
\end{tabular}
\end{table}

\begin{table}[t]
\renewcommand{\arraystretch}{1.2}
\caption{Reported accuracies on three MDD datasets using representative backbone architectures. }
\label{tab:backbone-acc}
\footnotesize
\centering
\begin{tabular}{cccc}
\hline
\textbf{Backbone Architecture} & \textbf{HUSM} & \textbf{PRED+CT} & \textbf{MODMA} \\
\hline
CNN    & 0.9832~\cite{MUMTAZ2019103983}  & 0.9393~\cite{li2024eeg} &  0.7400~\cite{wang2023depression}  \\
CNN-RNN   & 0.9597~\cite{MUMTAZ2019103983}  & \textbf{0.9907}~\cite{thoduparambil2020eeg}  & 0.9756~\cite{yang2023gated}  \\
GCN       & \textbf{0.9844}~\cite{sun2023multi}  & 0.8317~\cite{zhang2024novel}  & \textbf{0.9968}~\cite{sun2023multi}       \\
SNN       & -  & 0.9800~\cite{sam2023depression} & -       \\
\hline
\textit{Chance Level} & 0.5313  & 0.6198  & 0.5472  \\
\hline
\end{tabular}
\end{table}

\subsubsection{Self-supervised methods}
Self-supervised contrastive methods are gradually replacing traditional supervised learning, especially on large-scale EEG datasets where they demonstrate stronger generalization and robustness (Table~\ref{tab:sleep-acc}).
Early works explore tasks like relative positioning and temporal shuffling to extract temporal structures from multivariate signals~\cite{banville2019self, oord2018representation}.
ContraWR~\cite{yang2023self} constructs contrastive pairs from distinct time windows to capture temporal dependencies, reporting notably high accuracy on Sleep-EDFx.
mulEEG~\cite{kumar2022muleeg} and CoSleep~\cite{ye2021cosleep} introduce multi-view contrastive strategies, with mulEEG focusing on cross-view consistency and CoSleep capturing temporal and spectral patterns through a dual time-frequency framework.
Multimodal modeling enhances sleep staging by integrating complementary EEG, EOG, and EMG signals.
Brant-X~\cite{zhang2024brantx} tackles alignment challenges with EEG foundation models and contrastive learning, aligning EEG and EOG at local and global levels to bridge modality gaps and achieve superior performance.

\subsection{Depression Identification}

\subsubsection{Task Description}
Depression, particularly Major Depressive Disorder (MDD), is a psychological condition affecting 5\% of individuals worldwide, with a higher prevalence among women. In low- and middle-income countries, up to 75\% of individuals lack adequate care due to limited resources and stigma, despite effective treatments being available~\cite{WHO_depression}.

Depression severity is quantified using standardized scales like the Beck Depression Inventory (BDI) to differentiate clinical depression from normal mood variations. 
Existing studies adopt heterogeneous classification criteria: some focus on binary discrimination (e.g., patients vs. healthy controls), while others stratify cohorts by treatment status (medicated vs. non-medicated) or severity levels (mild vs. moderate/severe).
Table~\ref{tab:dep} summarizes datasets used in MDD research.

\begin{table}[t]
\renewcommand{\arraystretch}{1.2}
\caption{Public EEG Datasets for Schizophrenia, where \textbf{Exp (n)} represents the number of schizophrenia patients and \textbf{Ctrl (n)} represents the control group.}
\label{tab:sz}
\footnotesize
\centering
\begin{tabular}{lcccc}
\hline
\textbf{Dataset}      & \textbf{Exp (n)} & \textbf{Ctrl (n)} & \textbf{Frequency (Hz)} & \textbf{Channels} \\
\hline
\href{https://repod.icm.edu.pl/dataset.xhtml?persistentId=doi:10.18150/repod.0107441
        
        } {CeonRepod}~\cite{olejarczyk2017graph}   & 14              & 14               & 250              & 19                \\
\href{https://www.kaggle.com/datasets/broach/button-tone-sz}{NIMH}~\cite{ford2014did}                & 49              & 32               & 1024             & 64                \\
\href{http://brain.bio.msu.ru/eeg_schizophrenia.htm}{MHRC}~\cite{borisov2005analysis}        & 45              & 39               & 128              & 16                \\
\hline
\end{tabular}
\end{table}

\begin{table}[t]
\renewcommand{\arraystretch}{1.1}
\caption{Reported accuracies on three SZ datasets using representative learning strategies.}
\label{tab:sz-acc}
\centering
\begin{tabular}{cccc}
\hline
\textbf{Modeling Strategy} & \textbf{CeonRepod}~\cite{olejarczyk2017graph} & \textbf{NIMH}~\cite{ford2014did} & \textbf{MHRC}~\cite{borisov2005analysis}   \\
\hline
Timeseries-based    & 0.9807~\cite{oh2019deep}  & 0.9200~\cite{guo2021deep}& \textbf{0.9800}~\cite{supakar2022deep} \\
2D Representation & \textbf{0.9974}~\cite{khare2023schizonet}  & \textbf{0.9635}~\cite{khare2023schizonet}& 0.9740~\cite{khare2023schizonet} \\
Transfer Learning   & 0.9900~\cite{SZ22}  & 0.9336~\cite{khare2021spwvd}& 0.9773~\cite{9713847} \\
\hline
\textit{Chance Level} & 0.5492 & 0.5962 & 0.5357  \\
\hline
\end{tabular}
\end{table}

\subsubsection{Approach overview}
Depression impacts both superficial and deeper brain structures, challenging traditional handcrafted features.
Acharya introduced the first end-to-end DL model for EEG-based depression detection, showing right-hemisphere signals to be more distinctive than left, consistent with clinical findings~\cite{acharya2018automated}. 
Sun et al.~\cite{sun2023multi} further reported that with increasing granularity, MDD patients exhibited weakened connectivity between RF–RT and LT–LP regions; by embedding these patterns into the Multi-Granularity Graph Convolutional Network (MGGCN), clinically relevant disruptions were captured, yielding superior accuracy (Table~\ref{tab:backbone-acc}). 

Spiking neural networks (SNNs) offer another direction:
Shah et al.\cite{shah2019deep} used the NeuCube SNN to map EEG into a 3D reservoir aligned with the Talairach atlas, modeling spatiotemporal dynamics via STDP with interpretable connectivity visualization
Sam et al.~\cite{sam2023depression} integrates a 3D brain-inspired SNN with an LSTM, leveraging SNN’s energy efficiency with LSTM’s temporal modeling capabilities.




\subsection{Schizophrenia Identification}

\subsubsection{Task Description}
Schizophrenia (SZ) is a psychiatric disorder affecting 24 million people worldwide, characterized by cognitive deficits, delusions, and hallucinations~\cite{WHO_SZ}.  
SZ is associated with disruptions in structural and functional brain connectivity, marked by decreased global efficiency, weakened strength, and increased clustering~\cite{zalesky2011disrupted}. 
These abnormalities manifest in EEG signals, enabling reliable binary classification of SZ patients versus healthy controls. Publicly available datasets supporting this task are summarized in Table~\ref{tab:sz}.

\subsubsection{Approach overview}
EEG-based SZ diagnosis has been studied through three main strategies (Table~\ref{tab:sz-acc}).  
\textit{Time-series models} work directly on raw EEG, capturing temporal dynamics with relatively simple architectures but limited spectral–spatial representation~\cite{oh2019deep, supakar2022deep}.  
\textit{2D image-based approaches} transform EEG into spectrograms or scalograms, allowing CNNs to exploit richer spectral–spatial patterns~\cite{khare2023schizonet}.  
\textit{Transfer learning} builds on this idea by adapting pre-trained vision backbones (e.g., VGG, ResNet), whose hierarchical convolutional filters are well suited for capturing local and global patterns in spectrogram-like EEG representations, thereby achieving robust feature extraction even with limited data~\cite{SZ22}.
Reported results across representative studies show accuracies typically above 0.9 for all three strategies, suggesting that different modeling approaches can support reliable SZ classification under varied settings.


\subsection{Alzheimer's Disease Diagnosis}
\begin{table}[t]
 \renewcommand{\arraystretch}{1.2}
\caption{Public EEG Datasets for Alzheimer's Diagnosis, where \textbf{AD (n)} and \textbf{MCI (n)} represent the experimental groups, and \textbf{Ctrl (n)} represents the control group.}
\label{tab:ad}
\footnotesize
\centering
\begin{tabular}{p{50pt}p{20pt}p{20pt}p{20pt}p{30pt}p{30pt}}
\hline
\textbf{Dataset}            & \textbf{AD (n)} & \textbf{MCI (n)} & \textbf{Ctrl (n)} & \textbf{Frequency (Hz)} & \textbf{Channels} \\
\hline
\href{https://osf.io/2v5md/}{FSA}~\cite{ds_FSA_AD}        & 160            & -               & 24               & 128              & 21                \\
\href{https://openneuro.org/datasets/ds004504/versions/1.0.2}{AD-65}~\cite{ds004504:1.0.2} & 36             & -               & 29               & 250              & 19                \\
\href{https://github.com/tsyoshihara/Alzheimer-s-Classification-EEG}{Fiscon}~\cite{fiscon}         & 49             & 37              & 14               & 1024             & 19                \\
\href{https://figshare.com/articles/dataset/dataset_zip/5450293?file=9423433}{AD-59}~\cite{cejnek2021novelty} & 59            & 7               & 102              & 128-256          & 21                \\
\hline
\end{tabular}
\end{table}

\begin{table}[t]
\renewcommand{\arraystretch}{1.1}
\caption{Reported accuracies on private AD datasets with feature representations (chance level in parentheses).}
\label{tab:ad-private}
\centering
\begin{tabular}{cc}
\hline
 \textbf{Feature Representation} & \textbf{Accuracy} \\
\hline
Pearson correlation & \textbf{1.0000}~(0.5000)~\cite{AD1} \\
Wavelet Coherence &  0.9230~(0.5128)~\cite{shan2022spatial} \\
PSD image & 0.9295~(0.5000)~\cite{ieracitano2019convolutional} \\
\hline
\end{tabular}
\end{table}

\subsubsection{Task Description}
Alzheimer’s disease (AD) is a progressive neurodegenerative disorder that starts with mild memory loss and advances to severe cognitive impairment, affecting daily life. While medical interventions can improve quality of life, a definitive cure remains elusive~\cite{better2024alzheimer}. 
Alzheimer’s disease (AD) progresses through three stages: preclinical, mild cognitive impairment (MCI), and Alzheimer’s dementia.
Classification tasks typically distinguish MCI or Alzheimer’s dementia from healthy controls. 
EEG abnormalities, such as slowed brain rhythms and desynchronization, serve as biomarkers for AD-related neurodegeneration~\cite{labate2014eeg}. Table~\ref{tab:ad} summarizes publicly available datasets.

\subsubsection{Approach overview}
EEG abnormalities in Alzheimer’s disease, such as disrupted functional connectivity and altered brain rhythms, provide critical insights into the neurological changes. 
Representative strategies are summarized in Table~\ref{tab:ad-private}, noting that results on private datasets are not strictly comparable.
For instance, Alves et al.~\cite{AD1} employed Pearson correlation to construct connectivity matrices, achieving near-perfect discrimination between AD and healthy controls. 
Shan et al.~\cite{shan2022spatial} explored six functional connectivity measures for constructing adjacency matrices, reporting that wavelet coherence yielded the best performance for capturing spatial–temporal dependencies.  
Beyond connectivity, 2D spectral representations—such as PSD-based images—have been employed to enable feature learning for AD classification~\cite{ieracitano2019convolutional}.

\subsection{Parkinson's Disease Diagnosis}

\begin{table}[t]
\renewcommand{\arraystretch}{1.2}
\caption{Public EEG Datasets for Parkinson's Disease Diagnosis, where \textbf{Exp (n)} represents the number of patients and \textbf{Ctrl (n)} represents the healthy control group.}
\label{tab:pd}
\footnotesize
\centering
\begin{tabular}{lcccc}
\hline
\textbf{Dataset}            & \textbf{Exp (n)}   & \textbf{Ctrl (n)}   & \textbf{Frequency (Hz)} & \textbf{Channels} \\
\hline
\href{https://openneuro.org/datasets/ds002778/versions/1.0.5}{UCSD}~\cite{ds002778:1.0.5}  & 15                & 16                & 512              & 32                \\
\href{http://www.predictsite.com/}{UNM}~\cite{cavanagh2018diminished} & 27                & 27                & 500              & 64                \\
\href{https://narayanan.lab.uiowa.edu/}{UI}~\cite{singh2020frontal}   & 14                & 14                & 500              & 59                \\
\hline
\end{tabular}
\end{table}

\begin{table}[t]
\renewcommand{\arraystretch}{1.1}
\caption{Reported accuracies on the UCSD dataset with representative preprocessing and feature representations.}
\label{tab:pd-ucsd}
\centering
\begin{tabular}{ccc}
\hline
\textbf{Preprocessing} & \textbf{Feature Representation} & \textbf{Accuracy (UCSD)} \\
\hline
/ & Raw segments & 0.9800~\cite{shaban2021automated} \\
Gabor Transform & Spectrograms & 0.9946~\cite{PD2} \\
CWT & Scalograms & 0.9960~\cite{shaban2022resting} \\
SPWVD & TFR & \textbf{0.9997}~\cite{khare2021pdcnnet} \\
\hline
\textit{Chance Level} & -- & 0.6522 \\
\hline
\end{tabular}
\end{table}

\subsubsection{Task Description}
Parkinson’s disease (PD) is a progressive neurodegenerative disorder marked by motor symptoms (tremors, rigidity, bradykinesia) and non-motor symptoms (depression, sleep disturbances, cognitive decline). In 2019, over 8.5 million people worldwide were living with PD~\cite{who2023parkinson}.
EEG is widely used in PD research due to its noise resistance and sensitivity to neurological changes, such as slowing cortical oscillations and increased low-frequency power~\cite{morita2011relationship}. 
Most studies focus on supervised learning for binary classification, with some incorporating transfer learning. 
Table~\ref{tab:pd} summarizes publicly available datasets.

\subsubsection{Approach overview}
Transforming raw EEG signals into 2D representations is a well-established approach for PD classification ({Table~\ref{tab:pd-ucsd}).  
Time–frequency transformations such as Gabor Transform and CWT have been widely adopted: spectrograms~\cite{PD2} and scalograms~\cite{shaban2022resting} capture temporal–spectral dynamics more effectively than raw waveforms.  
More recently, advanced representations such as the Smoothed Pseudo Wigner–Ville Distribution (SPWVD)~\cite{khare2021pdcnnet} generate high-resolution time–frequency maps, allowing CNNs to exploit fine-grained signal structure.  
Collectively, these approaches illustrate a methodological shift from direct time-series analysis to progressively richer 2D representations, each achieving performance substantially above chance level.


\subsection{ADHD Identification}

\begin{table}[t]
\renewcommand{\arraystretch}{1}
\caption{Public EEG Datasets for ADHD Identification, where \textbf{Exp (n)} represents the number of ADHD patients and \textbf{Ctrl (n)} represents the healthy control group.}
\label{tab:adhd}
\footnotesize
\centering
\begin{tabular}{lcccc}
\hline
\textbf{Dataset}            & \textbf{Exp (n)}   & \textbf{Ctrl (n)}              & \textbf{Frequency (Hz)} & \textbf{Channels} \\
\hline
\href{https://data.mendeley.com/datasets/6k4g25fhzg/1}{ADHD-79}~\cite{sadeghibajestani2023dataset} & 37 & 42             & 256          & 2                \\
\href{https://ieee-dataport.org/open-access/eeg-data-adhd-control-children}{ADHD-121}~\cite{rzfh-zn36-20}      & 61 & 60    & 128          & 19                \\
\hline
\end{tabular}
\end{table}

\begin{table}[t]
\renewcommand{\arraystretch}{1.2}
\caption{Reported accuracies across frequency bands on the ADHD-121 dataset~\cite{bakhtyari2022adhd}.}
\label{tab:adhd-acc}
\centering
\begin{tabular}{cccccc}
\hline
Theta& Alpha & Beta & Gamma & Full & \textit{chance level} \\
\hline
0.9374 & 0.9724 & 0.9825 & 0.9725 & \textbf{0.9975} & 0.5825 \\
\hline
\end{tabular}
\end{table}

\subsubsection{Task Description}
Attention-deficit/hyperactivity disorder (ADHD) is a neurodevelopmental disorder affecting around 3.1\% of individuals aged 10–14 and 2.4\% of those aged 15–19~\cite{who_adolescent_mental_health}. 
It is categorized into three subtypes: Inattentive (ADHD-I), Hyperactive-Impulsive (ADHD-H), and Combined (ADHD-C)~\cite{nimh_adhd}.
EEG is widely used alongside neuroimaging and physiological measures for ADHD diagnosis. 
However, deep learning remains underexplored, with most existing approaches relying on supervised learning and feature-based classification. Research focuses on binary classification tasks, and Table~\ref{tab:adhd} lists two publicly available datasets.

\subsubsection{Approach overview}
Studies on ADHD diagnosis report elevated $\theta$ and reduced $\beta$ in children with ADHD, aligning with medical findings~\cite{chen2019use}.
More recent work compared models trained on individual bands with those using the full denoised range, showing that integrated inputs achieved the best results (Table~\ref{tab:adhd-acc}). This indicates that although $\theta$ and $\beta$ remain the most consistent group-level markers, combining multiband better captures individual variability and cross-band interactions, providing richer features for DL models~\cite{bakhtyari2022adhd}.

\begin{table*}[t]
\renewcommand{\arraystretch}{1.2}
\caption{Summary of pre-trained SSL frameworks for multi-task neurodiagnosis, focusing on relevant datasets and tasks, with paradigms such as Contrastive Learning (CL), Contrastive Predictive Coding (CPC), and Masked Autoencoding (MAE)}
\label{tab:pts}
\footnotesize
\centering
\begin{tabular}
{p{2.1cm}p{1.9cm}p{2.1cm}p{1.4cm}p{1.8cm}p{3.7cm}p{2.5cm}}
\hline
\textbf{Work}        & \textbf{SSL Paradigm}                                     & \textbf{Backbone}  & \textbf{Data Type} &\textbf{Partitioning}  & \textbf{pre-training Dataset}                          & \textbf{Downstream Tasks}                           \\
\hline
Banville et al.~\cite{banville2021uncovering} & CPC & CNN & EEG & dataset-specific & TUSZ, PC18  & Seizure, Sleep  \\
MBrain~\cite{cai2023mbrain} & CPC & CNN+LSTM+GNN & EEG, iEEG&dataset-specific & TUSZ, private & Seizure, etc. \\
TS-TCC~\cite{eldele2021time}               & CPC                       & CNN+Transformer   & EEG &cross-dataset     & Bonn, Sleep-EDF, etc.                                  & Seizure, Sleep, etc.                                    \\
SeqCLR~\cite{mohsenvand2020contrastive}               & CL                                            & CNN+GRU & EEG &mixed-dataset & TUSZ, Sleep-EDF, ISRUC, etc.      & Seizure, Sleep, etc. \\
TF-C~\cite{zhang2022self}                 & CL                              & CNN  & EEG                  &cross-dataset & Sleep-EDF, etc.         & Seizure, Sleep, etc.       \\
BIOT~\cite{yang2024biot}  & CL & Transformer & EEG, etc. & cross-dataset & SHHS, etc. & Seizure, etc \\
Jo et al.~\cite{jo2023channel} & Predictive & CNN  & EEG &mixed-dataset &CHB-MIT, Sleep-EDF & Seizure, Sleep  \\
neuro2vec~\cite{wu2022neuro2vec} & MAE & CNN+Transformer & EEG &cross-dataset& Bonn, Sleep-EDF, etc.         & Seizure, Sleep \\
CRT~\cite{zhang2023self} & MAE & Transformer & EEG &dataset-specific & Sleep-EDF, etc. & Sleep, etc.\\
NeuroBERT~\cite{wu2024neuro} & MAE & Transformer & EEG, etc. & dataset-specific & Bonn, SleepEDF, etc, & Seizure, Sleep,etc. \\
BENDR~\cite{Kostas2021BENDR} & CPC+MAE & CNN+Transformer & EEG &cross-dataset& TUEG & Sleep, etc. \\
CBRAMOD~\cite{wang2024cbramod} & MAE & Transformer & EEG & cross-dataset & TUEG & Seizure, Sleep, MDD \\
Brant~\cite{zhang2024brant} & MAE & Transformer & iEEG&cross-dataset & private & Seizure, etc.  \\  
Brainwave~\cite{yuan2024brainwavebrainsignalfoundation} & MAE & Transformer & EEG, iEEG&cross-dataset & TUEG, Siena, CCEP, Sleep-EDF, NIMH, FSA, private, etc.  & Seizure, Sleep,  MDD,\newline SZ, AD, ADHD \\ 
EEGFormer~\cite{chen2024eegformer} & VQ+MAE & Transformer & EEG & cross-dataset & TUEG & Seizure, etc. \\
LaBraM~\cite{jiang2024large} & VQ+MAE & Transformer & EEG & cross-dataset & TUEG, Siena, etc. & Seizure, etc. \\
NeuroLM~\cite{jiang2024neurolm} & VQ+MAE\newline+Predictive & Transformer & EEG & cross-dataset & TUEG, Siena, etc. & Seizure, Sleep, etc.\\
\hline
\end{tabular}
\end{table*}


\section{Universal Pre-trained Models}
\label{sec:bm}

In recent years, SSL has revolutionized EEG/iEEG analysis in neurological diagnosis. Emerging methods focus on generalizable SSL frameworks that integrate heterogeneous datasets during pre-training, overcoming the limitations of task- and dataset-specific models and enabling seamless adaptation to multiple downstream tasks.
These innovations bring us closer to the development of universal neurodiagnostic models capable of addressing challenges across diverse clinical settings.

Table~\ref{tab:pts} summarizes pre-trained SSL frameworks for multi-task neurodiagnosis, organized by the SSL paradigms to align with their technical evolution analyzed in this section. 
While some frameworks extend to broader time-series data, such as BCI signals and motion sensor data, we focus on datasets and tasks directly relevant to neurological applications.
Below, we explore these frameworks, examining their contributions to unified pre-training strategies, multitask adaptability, and their potential to impact real-world applications.

\subsection{Contrastive- and Predictive- Based Learning}
\paragraph{Contrastive Predictive Coding}
Early SSL approaches in EEG/iEEG analysis are largely based on the Contrastive Predictive Coding (CPC) paradigm~\cite{banville2021uncovering,cai2023mbrain}, which learns representations by predicting signal segments through contrastive learning. While these models employed generic architectures across neurophysiological tasks, they fail to achieve true cross-task generalization. As a result, they are trained separately on specific datasets, limiting their clinical applicability across diverse neurodiagnostic applications.
CPC variants like TS-TCC~\cite{eldele2021time} introduce a one-to-one feature transfer mechanism. This framework enables feature migration across tasks such as human activity recognition, sleep staging, and seizure detection, paving the way for broader multi-domain generalization.

Building on the foundational principles of CPC, two distinct approaches have emerged: contrastive learning (CL) and predictive-based variants. CL retains CPC’s contrastive framework but emphasizes explicit instance-level discrimination through hand-crafted augmentations for positive/negative pairs, instead of CPC’s autoregressive future state prediction.
Predictive variants inherit CPC’s structure but replace its auto-learned latent contexts with manually defined features.

\paragraph{Contrastive-Based learning}
SeqCLR~\cite{mohsenvand2020contrastive}, inspired by SimCLR, employs contrastive learning to EEG data, enhancing similarity between augmented views of the same channel through domain-specific transformations. Adopting a mixed-dataset training approach, it unifies diverse EEG datasets for robust representation learning.
TF-C~\cite{zhang2022self} incorporates dual time-frequency contrastive learning with a cross-domain consistency loss to align embeddings across temporal and spectral representations.
It further examines cross-dataset generalization, training on a source dataset and evaluating transferability to multiple targets, highlighting the potential of cross-task feature sharing for universal neural signal models.
BIOT~\cite{yang2024biot} integrates contrastive learning, unifying multimodal biosignals (e.g., EEG, ECG) via tokenization and linear attention to learn invariant physiological patterns for cross-task generalization.

\paragraph{Predictive-Based Learning}
Jo et al.~\cite{jo2023channel} proposes a channel-aware predictive-based framework, which leverages stopped band prediction for spectral feature learning and employs temporal trend identification to capture dynamic patterns. 
By integrating mix-dataset pretraining, it enhances generalization through cross-domain feature fusion. However, the pretraining scale remains limited.

\subsection{Reconstruction-Based Learning}
\paragraph{Masked Autoencoding}
The paradigm shift from CPC to masked reconstruction in SSL aims for higher data efficiency and scalability, inspired by cross-domain advances like masked language modeling in NLP (e.g., BERT~\cite{devlin2018bert}), with MAE's generative approach enhancing classification performance while avoiding complex negative sampling.

Neuro2vec~\cite{wu2022neuro2vec} extends masked reconstruction by integrating EEG-specific spatiotemporal recovery and spectral component prediction into a unified framework, utilizing a CNN-ViT hybrid architecture for patch embedding and reconstruction. 
CRT~\cite{zhang2023self} further introduces multi-domain reconstruction through cross-domain synchronization of temporal and spectral features, replacing conventional masking with adaptive input dropping to preserve data distribution integrity, thereby improving robustness in physiological signal modeling.
Neuro-BERT~\cite{wu2024neuro} introduces Fourier Inversion Prediction (FIP), reconstructing masked signals by predicting their Fourier amplitude and phase, then applying an inverse Fourier transform. The spectral-based prediction framework inherently matches the physiological nature of EEG signals.

\paragraph{Large-Scale Continuous-Reconstruction Models}
Transformer architectures are increasingly applied in neurodiagnostics, leveraging their scalability and attention mechanisms to capture global dependencies in irregular neural signals. BERT-style pretraining, particularly masked reconstruction, enhances neurodiagnostic classification by enforcing robust contextual learning of latent bioelectrical patterns, which is crucial for distinguishing subtle neurological signatures. Their parallelizable training and tokenized time-frequency representations pave the way for scalable foundation models, driving large-scale pretraining in neural signal analysis.

Inspired by Bert, BENDR~\cite{Kostas2021BENDR} integrates CPC with MAE-inspired reconstruction for temporal feature learning. 
Pretrained on the Temple University Hospital EEG Corpus—a diverse dataset containing 1.5 TB of raw clinical EEG from over 10,000 subjects—BENDR represents the emergence of large-scale pretraining for neurodiagnostics, showcasing the cross-subject scalability of transformers. 
It demonstrates how foundation models can unify heterogeneous neural signal paradigms, advancing generalized, scalable EEG analysis.
CBRAMOD~\cite{wang2024cbramod} introduces a criss-cross transformer framework to explicitly model EEG’s spatial-temporal heterogeneity.
Using patch-based masked reconstruction, it separately processes spatial and temporal patches through parallel attention, preserving the structural dependencies to EEG.

Brant~\cite{zhang2024brant} and Brainwave~\cite{yuan2024brainwavebrainsignalfoundation}
represent a unified effort to establish foundation models for neural signal analysis. 
Brant focuses on SEEG signals, employing a masked autoencoding framework with dual Transformer
encoders to capture temporal dependencies and spatial
correlations, enabling seizure detection and signal forecasting.
Brainwave pioneers large-scale pretraining with an unprecedented multimodal corpus of over 40,000 hours of EEG/iEEG data from 16,000 subjects, marking a significant milestone in neural signal foundation models. Its pre-training strategy follows a masked modeling paradigm that randomly masks time-frequency patches of neural signals, and the model is trained to reconstruct the missing regions. To enhance generalizability across neural data types, Brainwave employs a shared encoder for both EEG and iEEG, coupled with modality-specific reconstruction decoders. These innovations position Brainwave as the first comprehensive foundation model unifying EEG/iEEG analysis, with transformative implications for neuroscience research.
\paragraph{Large-Scale Discrete-Reconstruction Models}
Vector Quantized Variational Autoencoder (VQ-VAE) is a powerful framework for learning discrete representations of continuous data by mapping inputs to a predefined codebook, which has been widely adopted in domains like speech and image processing~\cite{van2017neural}. By tokenizing raw data into discrete codes, this approach enhances cross-subject generalization while preserving interpretable spatiotemporal patterns.

LaBraM~\cite{jiang2024large} trains its discrete codebook by reconstructing spectral magnitudes and phases of EEG segments, then pretrains with a symmetric masking task that predicts masked code indices bidirectionally.
NeuroLM~\cite{jiang2024neurolm} extends this approach by introducing VQ Temporal-Frequency Prediction, aligning EEG tokens with textual representations through adversarial training. After tokenization, it employs autoregressive modeling, enabling an LLM to predict the next EEG token analogous to language modeling.
EEGFormer~\cite{chen2024eegformer} focuses on reconstructing raw waveforms for codebook training, followed by BERT-style masked signal reconstruction pretraining.
These methods demonstrate how VQ-based tokenization adapts to EEG modeling—whether prioritizing spectral synchrony (LaBraM), fusing time-frequency features (NeuroLM), or preserving temporal fidelity (EEGFormer).

\subsection{BrainBenchmark}
The development of universal pre-trained frameworks represents a transformative advancement in healthcare, enabling the integration of heterogeneous datasets and generalization across diverse diagnostic tasks.
However, existing studies—whether supervised or self-supervised—often adopt inconsistent dataset usage, validation splits, and evaluation metrics.
These inconsistencies make it difficult to fairly compare different paradigms and accurately assess progress in the field.
To address this issue, we have established an open benchmark that provides a unified evaluation standard and toolset for the community.
It currently includes 8 representative models and 9 public datasets, with support for flexible model integration and dataset expansion.
Our goal is to encourage researchers to adopt this common framework for consistent, reproducible benchmarking and to lower the barrier for integrating new methods.
The implementation is publicly available at \href{https://github.com/ZJU-BrainNet/BrainBenchmark}{https://github.com/ZJU-BrainNet/BrainBenchmark}, and we hope it will serve as a foundation for advancing universal pre-trained frameworks in EEG/iEEG analysis.
\section{Conclusion}
This survey systematically reviews 448 studies and 46 public datasets to advance deep learning-driven analysis of EEG/iEEG signals across seven neurological diagnostic tasks: seizure detection, sleep staging and disorder, major depressive disorder, schizophrenia, Alzheimer’s disease, Parkinson’s disease, and ADHD.
Our work establishes three foundational contributions: First, we unify fragmented methodologies across neurological conditions by standardizing data processing, model architectures, and evaluation protocols.
Second, we identify self-supervised learning as the most promising paradigm for multi-task neurodiagnosis, providing a comprehensive overview of pre-trained SSL frameworks and their advancements. Third, we introduce \href{https://github.com/ZJU-BrainNet/BrainBenchmark}{BrainBenchmark}, a unified platform that standardizes evaluations and integrates neurological datasets with diverse models, aiming to improve comparability and reproducibility across studies.

Looking back, the pursuit of universal models capable of learning from diverse, multimodal data reflects the field's growing ambition, laying the groundwork for a new era of intelligent and adaptable healthcare systems. Over the past decades, significant progress has established a strong foundation for neurological diagnostics based on electrical brain signals. Key contributions include advances in signal preprocessing, curating large-scale, well-annotated datasets, and developing deep learning architectures for specific tasks.
The integration of self-supervised pretraining marks a paradigm shift, enabling models to extract rich and meaningful representations from vast amounts of unlabeled, heterogeneous data.

Looking forward, the ultimate goal is to develop genuinely universal and adaptable frameworks capable of transcending individual tasks and datasets to address a broader range of neurological disorders. These advancements will pave the way for intelligent diagnostic tools that deliver precise, efficient, and accessible healthcare solutions globally, driving transformative progress in biomedical research and clinical applications.
\section*{Acknowledgment}
This work is supported by NSFC (62322606) and Zhejiang NSF (LR22F020005).

\section*{Appendix}

In this section, we provide summaries of deep learning-based frameworks for the seven neurodiagnostic tasks mentioned earlier. These summaries include details on preprocessing methods, extracted features, deep learning backbones, training paradigms, downstream task datasets, classification tasks, data partitioning strategies, and reported performances. 
Before these task-specific summaries, Table~\ref{tab:xixdatasets} presents an overview of all publicly available EEG datasets referenced in this study.
The relevant tables are as follows: seizure detection in Table~\ref{tab:seizures}, sleep staging in Table~\ref{tab:sleeps}, depression identification in Table~\ref{tab:mdds}, schizophrenia identification in Table~\ref{tab:schis}, Alzheimer’s disease diagnosis in Table~\ref{tab:ads}, Parkinson’s disease diagnosis in Table~\ref{tab:pds}, and ADHD identification in Table~\ref{tab:adhds}.

\begin{table*}[ht]
\renewcommand{\arraystretch}{1.2}
\caption{Summary of publicly available EEG datasets}
\label{tab:xixdatasets}
\footnotesize
\begin{tabular}{p{1.2cm}p{1.8cm}p{1.2cm}p{1.2cm}p{4.8cm}p{5.0cm}}
\hline
\textbf{Ref.} & \textbf{Dataset} & \textbf{Disease} & \textbf{Year} & \textbf{Institution} & \textbf{URL} \\
\hline
~\cite{andrzejak2001indications} & Bonn & Epilepsy & 2001 & University Hospital Bonn & \url{https://www.ukbonn.de/epileptologie/arbeitsgruppen/ag-lehnertz-neurophysik/downloads/} \\
~\cite{ihle2012epilepsiae} & Freiburg & Epilepsy & 2003 & University of Freiburg & \url{https://epilepsy.uni-freiburg.de/freiburg-seizure-prediction-project/eeg-database} \\
~\cite{seizure-detection} & Mayo-Upenn & Epilepsy & 2014 & UPenn and Mayo Clinic & \url{https://www.kaggle.com/c/seizure-detection} \\
~\cite{guttag2010chb, shoeb2009application, goldberger2000physiobank} & CHB-MIT & Epilepsy & 2010 & The Massachusetts Institute of Technology & \url{https://physionet.org/content/chbmit/1.0.0/} \\
~\cite{andrzejak2012nonrandomness} & Bern Barcelona & Epilepsy & 2012 & Universitat Pompeu Fabra & \url{https://www.upf.edu/web/ntsa/downloads/-/asset_publisher/xvT6E4pczrBw/content/2012-nonrandomness-nonlinear-dependence-and-nonstationarity-of-electroencephalographic-recordings-from-epilepsy-patients} \\
~\cite{hauz} & Hauz Khas & Epilepsy & 2016 & Neurology and Sleep Centre, New Delhi & \url{https://www.researchgate.net/publication/308719109_EEG_Epilepsy_Datasets} \\
~\cite{melbourne} & Melbourne & Epilepsy & 2016 & The University of Melbourne & \url{https://www.epilepsyecosystem.org/} \\
\hline
\end{tabular}
\end{table*}

\begin{table*}[ht]
\renewcommand{\arraystretch}{1.2}
\caption*{(Continued) Summary of publicly available EEG datasets}
\footnotesize
\begin{tabular}{p{1.0cm}p{1.5cm}p{1.5cm}p{1.5cm}p{4.5cm}p{5.5cm}}
\hline
\textbf{Ref.} & \textbf{Dataset} & \textbf{Disease} & \textbf{Year} & \textbf{Institution} & \textbf{URL} \\
\hline
~\cite{kemp2000analysis, goldberger2000physiobank} & Sleep-EDF & Sleep & 2013\&2018 & MCH-Westeinde Hospital & \url{http://www.physionet.org/physiobank/
database/sleep-edfx/} \\
~\cite{oreilly2014montreal} & MASS & Sleep & 2014 & Center for Advanced Research in Sleep Medicine & \url{http://ceams-carsm.ca/en/MASS/} \\
~\cite{quan1997sleep, zhang2018national} & SHHS & Sleep & 2018 & National Heart Lung \& Blood Institute & \url{https://sleepdata.org/datasets/shhs} \\
~\cite{ucddb2007sleep, goldberger2000physiobank} & SVUH\_UCD & Sleep & 2007 & St. Vincent's University Hospital / University College Dublin & \url{https://physionet.org/content/ucddb/1.0.0/} \\
~\cite{Alvarez-Estevez2022, goldberger2000physiobank} & HMC & Sleep & 2022 & Haaglanden Medisch Centrum & \url{https://physionet.org/content/hmc-sleep-staging/1.1/} \\
~\cite{ghassemi2018you, goldberger2000physiobank} & PC18 & Sleep & 2018 & Massachusetts General Hospital & \url{https://physionet.org/content/challenge-2018/1.0.0/#files} \\
~\cite{ichimaru1999development, goldberger2000physiobank} & MIT-BIH & Sleep & 1999 & Boston's Beth Israel Hospital Sleep Laboratory & \url{https://physionet.org/content/slpdb/1.0.0/} \\
\hline
\end{tabular}
\end{table*}

\begin{table*}[h]
\renewcommand{\arraystretch}{1.2}
\caption{Summary of deep learning frameworks for seizure detection}
\label{tab:seizures}
\footnotesize
\begin{tabular}{p{0.4cm}p{2.8cm}p{2cm}p{1.5cm}p{1.9cm}p{1.9cm}p{0.8cm}p{1.8cm}p{1.5cm}}
\hline
\textbf{Ref.} & \textbf{Preprocessing} & \textbf{Feature} & \textbf{Backbone} & \textbf{Training} & \textbf{Dataset} & \textbf{Task} & \textbf{Partitioning} & \textbf{Accuracy} \\
\hline
~\cite{Seizure1} & Image generation & 2D Image & 2D-CNN & supervised & Bern-Barcelona,\newline private  & binary & mixed-subject & 100\% \\
~\cite{zhou2018epileptic} & FFT & Frequency-domain features & 2D-CNN & supervised & Freiburg,\newline CHB-MIT & binary \newline 
3-class & subject-specific & 98.2\%-99.4\% \newline
95.3\% \\
~\cite{Seizure3} & Filtering,Downsampling & Raw & 2D-CNN & supervised & private & binary & cross-subject & AUC=0.94 \\
~\cite{Seizure4} & FSST,WSST & Time-Frequency matrix & 2D-CNN & supervised & Bern-Barcelona & binary & mixed-subject & 99.94\% \\
~\cite{Seizure5} & Filtering,EMD,FWT,FT & Raw,IMFs,\newline Wavelet Coefficients,\newline Module Values & 2D-CNN & supervised & Bern-Barcelona,\newline private  & binary & mixed-subject & 98.9\% \\
~\cite{Seizure6} & Z-norm,STFT & 2D Spectrograms & 2D-CNN & supervised & Bern-Barcelona,\newline private  & binary & mixed-subject & 91.8\% \\
~\cite{Seizure7} & Filtering & 2D Images & 2D-CNN & supervised & Bonn & binary & mixed-subject & 99.6\% \\
\hline
\end{tabular}
\end{table*}

\begin{table*}[ht]
\renewcommand{\arraystretch}{1.2}
\caption*{(Continued) Summary of deep learning frameworks for seizure detection}
\footnotesize
\begin{tabular}{p{0.4cm}p{2.8cm}p{2cm}p{1.5cm}p{1.9cm}p{1.9cm}p{0.8cm}p{1.8cm}p{1.5cm}}
\hline
\textbf{Ref.} & \textbf{Preprocessing} & \textbf{Feature} & \textbf{Backbone} & \textbf{Training} & \textbf{Dataset} & \textbf{Task} & \textbf{Partitioning} & \textbf{Accuracy} \\
\hline
 ~\cite{Seizure34} & Segmentation,
FNSW & 2D Image & 2D-CNN & supervised & Bonn & binary
3-class
5-class & mixed-subject & 100\% \\
~\cite{Seizure35} & Segmentation,EMD,WOG & Graph\newline  representation & 2D-CNN & supervised & Bonn,\newline private & binary & mixed-subject & 100\% \newline
97.65\% \\
~\cite{Seizure36} & Z-norm,Windowing & RPS Image & 2D-CNN & supervised & Bonn & binary & mixed-subject & 92.3\% \\
~\cite{Seizure37} & Filtering,CWT & 2D Scalograms & 2D-CNN & supervised & Bonn & binary & mixed-subject,\newline cross-subject & 99.5\% \\
 ~\cite{yuan2018novel} & STFT & 2D Spectrograms & 2D-CNN\newline +Attention & supervised & CHB-MIT & binary & mixed-subject & 96.61\% \\
~\cite{Seizure39} & Filtering,Z-norm & Raw Segments & 2D-CNN\newline +Attention & supervised & SWEC-ETHZ,private & binary & subject-specific & AUC=0.92
AUC=0.96 \\
~\cite{choi2019novel} & STFT & Spectrograms & 3D-CNN & supervised & CHB-MIT,\newline private & binary & cross-subject & 99.4\% \\
\hline
\end{tabular}
\end{table*}

\begin{table*}[ht]
\renewcommand{\arraystretch}{1.2}
\caption*{(Continued) Summary of deep learning frameworks for seizure detection}
\footnotesize
\begin{tabular}{p{0.4cm}p{2.8cm}p{2cm}p{1.5cm}p{1.9cm}p{1.9cm}p{0.8cm}p{1.8cm}p{2cm}}
\hline
\textbf{Ref.} & \textbf{Preprocessing} & \textbf{Feature} & \textbf{Backbone} & \textbf{Training} & \textbf{Dataset} & \textbf{Task} & \textbf{Partitioning} & \textbf{Accuracy} \\
\hline
~\cite{Seizure67} & Filtering,DWT,\newline Power Spectrum Band Calculation,Frequency Band Calculation & 2D Image & CNN & supervised & Bonn & binary & cross-subject & 99.99\% \\
~\cite{Seizure68} & Segmentation,Filtering & ApEn and RQA vector & CNN & supervised & Bonn & binary & mixed-subject & 99.26\% \\
~\cite{Seizure69} & Normalization,CWT & 2D Scalograms & CNN & supervised & Bonn & binary & mixed-subject & 98.78\% \\
~\cite{Seizure70} & Filtering & Raw & CNN & supervised & Bonn & binary \newline
3-class & mixed-subject & 100\% \newline
99.8\% \\
~\cite{Seizure71} & - & Raw & CNN & supervised & Bonn & 3-class & mixed-subject & 98.67\% \\
 ~\cite{Seizure72} & Segmentation,\newline Data Augmentation & Raw Segments & CNN & supervised & Bonn & binary & mixed-subject & AUC=0.92\% \\
~\cite{Seizure73} & Z-norm & Raw & CNN & supervised & Bonn & binary \newline
5-class & mixed-subject & 99.93\% \newline
94.01\% \\
\hline
\end{tabular}
\end{table*}

\begin{table*}[ht]
\renewcommand{\arraystretch}{1.2}
\caption*{(Continued) Summary of deep learning frameworks for seizure detection}
\footnotesize
\begin{tabular}{p{0.4cm}p{2.8cm}p{2cm}p{1.5cm}p{1.9cm}p{1.9cm}p{0.8cm}p{1.8cm}p{1.5cm}}
\hline
\textbf{Ref.} & \textbf{Preprocessing} & \textbf{Feature} & \textbf{Backbone} & \textbf{Training} & \textbf{Dataset} & \textbf{Task} & \textbf{Partitioning} & \textbf{Accuracy} \\
\hline
~\cite{Seizure97} & Filtering,GPSO & Time- \& Freq-\newline domain featrues & CNN & supervised & Bonn & binary & mixed-subject & 99.65\% \\
~\cite{Seizure98} & Z-norm,FFT & Raw,features & CNN & supervised & Bonn & binary \newline
3-class & mixed-subject & 98.23\% \newline
96.33\% \\
~\cite{assali2023cnn} & Normalization,Filtering,\newline STFT & RPSD,SampEn,SI & CNN & supervised & CHB-MIT & binary & subject-specific & 94.5\% \\
~\cite{Seizure100} & Normalization,\newline Segmentation & Raw Segments & CNN & supervised & private & binary & mixed-subject & 99.61\% \\
~\cite{Seizure101} & Filtering,Segmentation & Freq-features,\newline Time-Freq Image & CNN \newline 3D-CNN & supervised & Helsinki & binary & cross-subject & 90.06\% \\
~\cite{Seizure102} & Filtering,Segmentation,\newline Artifact rejection
 & Raw Segments & CNN\newline 2D-CNN & supervised & private & binary & cross-subject & 96.39\% \\
~\cite{Seizure103} & Normalization & Raw & CNN\newline CNN-LSTM & supervised & CHB-MIT & binary & subject-specific & 91.50\% \newline
92.11\% \\
\hline
\end{tabular}
\end{table*}

\begin{table*}[ht]
\renewcommand{\arraystretch}{1.2}
\caption*{(Continued) Summary of deep learning frameworks for seizure detection}
\footnotesize
\begin{tabular}{p{0.4cm}p{2.8cm}p{2cm}p{1.5cm}p{1.9cm}p{1.9cm}p{0.8cm}p{1.8cm}p{1.5cm}}
\hline
\textbf{Ref.} & \textbf{Preprocessing} & \textbf{Feature} & \textbf{Backbone} & \textbf{Training} & \textbf{Dataset} & \textbf{Task} & \textbf{Partitioning} & \textbf{Accuracy} \\
\hline
~\cite{Seizure128} & Filtering,Segmentation,\newline CWT,STFT & 2D Spectrogram,\newline Scalogram & CNN-LSTM & supervised & Bonn \newline CHB-MIT \newline Bern-Barcelona & binary & mixed-subject & 99.94\% \newline
93.77\% \newline
95.08\% \\
~\cite{Seizure129} & Filtering,STFT & 2D Spectrograms & CNN-LSTM & supervised & CHB-MIT & binary & subject-specific & 94.5\% \\
~\cite{Seizure130} & Filtering,Difference & Raw,Differential signal & CNN-LSTM\newline +Attention & supervised & Bonn & binary \newline
5-class & mixed-subject & 98.87\% \newline
90.17\% \\
~\cite{Seizure131} & Filtering,Resampling,TCP & Segments & CNN-RNN & supervised & TUSZ & binary & cross-subject & 82.27\% \\
~\cite{Seizure132} & Segmentation & Raw Segments & CNN-RNN \newline +Attention & supervised & CHB-MIT & binary & mixed-subject & SEN=92.88\% \\
~\cite{Seizure133} & Normalization & Raw Segments & CNN-Transformer & supervised & TUSZ & various & cross-subject & AUC=0.72 \\
~\cite{Seizure134} & Channel selection, \newline Windowing & Raw Segments & CNN-Transformer & supervised & CHB-MIT & binary & cross-subject, \newline subject-specific & SEN=65.5\% \\
\hline
\end{tabular}
\end{table*}

\begin{table*}[ht]
\renewcommand{\arraystretch}{1.2}
\caption*{(Continued) Summary of deep learning frameworks for seizure detection}
\footnotesize
\begin{tabular}{p{0.4cm}p{2.8cm}p{2cm}p{1.5cm}p{1.9cm}p{1.9cm}p{0.8cm}p{1.8cm}p{2cm}}
\hline
\textbf{Ref.} & \textbf{Preprocessing} & \textbf{Feature} & \textbf{Backbone} & \textbf{Training} & \textbf{Dataset} & \textbf{Task} & \textbf{Partitioning} & \textbf{Accuracy} \\
\hline
~\cite{Seizure156} & Segmentation,Network construction & Adjacency matrix & GCN & supervised & CHB-MIT & binary & subject-specific & 98.38\% \\
~\cite{Seizure157} & Filtering,Z-norm & Node Feature \newline matrix,Adjacency matrix & GCN+Attention & supervised & CHB-MIT & binary & subject-specific & 98.7\% \\
~\cite{Seizure158} & Filtering,Windowing & Raw Segments & GCN-Transformer & supervised & CHB-MIT & binary & subject-specific & AUC=0.935 \\
~\cite{rahmani2023meta} & Filtering,Segmentation,FFT & Node Feature \newline matrix,Adjacency matrix & GNN & supervised & TUSZ & binary & cross-subject & 81.77\% \\
 ~\cite{Seizure160} & Filtering,Segmentation,\newline Network construction & NaN & GNN+ Transformer & supervised & CHB-MIT & binary & subject-specific & 98.43\% \\
~\cite{Seizure161} & Segmentation & Raw Segments & GRU & supervised & Bonn & 3-class & mixed-subject & 98\% \\
~\cite{Seizure162} & DWT & Wavelet Coefficients & GRU & supervised & Bonn & binary & subject-specific & 98.5\% \\
\hline
\end{tabular}
\end{table*}

\begin{table*}[ht]
\renewcommand{\arraystretch}{1.2}
\caption*{(Continued) Summary of deep learning frameworks for seizure detection}
\footnotesize
\begin{tabular}{p{0.4cm}p{2.8cm}p{2cm}p{1.5cm}p{1.9cm}p{1.9cm}p{0.8cm}p{1.8cm}p{1.5cm}}
\hline
\textbf{Ref.} & \textbf{Preprocessing} & \textbf{Feature} & \textbf{Backbone} & \textbf{Training} & \textbf{Dataset} & \textbf{Task} & \textbf{Partitioning} & \textbf{Accuracy} \\
\hline
~\cite{yuan2024ppi} & Downsampling,\newline Segmentation & Time- \& Freq- domain featrues, Raw & CNN-LSTM & self-supervised & Mayo-UPenn,\newline FNUSA, \newline private & binary & cross-subject & F1=85.6\% \newline F1=82.3\% \newline F1=87.1\% \\
~\cite{tang2021self} & Z-norm,FFT & Adjacency matrix,Frequency-domain features & GNN & self-supervised & TUSZ & binary & cross-subject & AUC=0.875 \\
~\cite{ho2023self} & FFT,graph construction & EEG Network & GNN & self-supervised & TUSZ & binary & cross-subject & F1=0.534\% \\
~\cite{lih2023epilepsynet} & Segmentation,PCC & PCC matrix & Transformer & self-supervised & Turkish & binary & cross-subject & 85\% \\
~\cite{XIAO2024105464} & Filtering,Z-norm,\newline Windowing & Raw Segments & Transformer & self-supervised & CHB-MIT & binary & cross-subject & 97.07\% \\
~\cite{li2022spp} & Filtering,Segmentation & Raw Segments &	CNN	& self-supervised & TUSZ & binary & cross-subject &	- \\
~\cite{peng2023wavelet2vec} & DWPT	& Wavelets & Transformer &	self-supervised & TUSZ & 4-class &	cross-subject & 73\%\\
\hline
\end{tabular}
\end{table*}

\begin{table*}[ht]
\renewcommand{\arraystretch}{1.2}
\caption{Summary of deep learning frameworks for sleep staging}
\label{tab:sleeps}
\footnotesize
\begin{tabular}{p{0.4cm}p{2.8cm}p{2cm}p{1.5cm}p{1.9cm}p{1.9cm}p{0.8cm}p{1.8cm}p{2cm}}
\hline
\textbf{Ref.} & \textbf{Preprocessing} & \textbf{Feature} & \textbf{Backbone} & \textbf{Training} & \textbf{Dataset} & \textbf{Task} & \textbf{Partitioning} & \textbf{Accuracy} \\
\hline
~\cite{Sleep1} & Filtering,Feature \newline Extraction & Time- \& Freq- features & 3D-CNN & supervised & ISRUC & 5-class & cross-subject & 82\%-83.2\% \\
~\cite{eldele2021attention} & AFR & Raw Segments & CNN & supervised & Sleep-EDF, \newline SHHS & 5-class & cross-subject & 82.9\%-86.6\% \\
~\cite{Sleep3} & Filtering,Resampling & Raw Segments & CNN-Transformer & supervised & Sleep-EDF, \newline ISRUC, \newline private & 5-class & cross-subject & 84.76\%-86.32\% \\
~\cite{Sleep4} & Resampling,Segmentation & Raw Segments & CNN & supervised & Sleep-EDF, \newline SHHS  & 5-class  & cross-subject & 85.3\% \newline 88.1\% \\
~\cite{Sleep5} & DCT & DCT Coefficients & CNN-LSTM & supervised & SleepEDF, \newline DRM-SUB, \newline ISRUC & 5-class & cross-subject & 85.47\%-87.11\% \\
~\cite{Sleep6} & Segmentation & Raw Segments & CNN-BiLSTM & supervised & Sleep-EDF, \newline MASS & 5-class  & cross-subject & 82.0\% \newline 86.2\% \\
~\cite{Sleep7} & Filtering,Spectrogram Generation & 2D Spectrogram & 2D-CNN & supervised & Sleep-EDF, \newline SHHS & 5-class & mixed-subject & 83.02\%-94.17\% \\
\hline
\end{tabular}
\end{table*}

\begin{table*}[ht]
\renewcommand{\arraystretch}{1.2}
\caption*{(Continued) Summary of deep learning frameworks for sleep staging}
\footnotesize
\begin{tabular}{p{0.4cm}p{2.8cm}p{2cm}p{1.5cm}p{1.9cm}p{1.9cm}p{0.8cm}p{1.8cm}p{2cm}}
\hline
\textbf{Ref.} & \textbf{Preprocessing} & \textbf{Feature} & \textbf{Backbone} & \textbf{Training} & \textbf{Dataset} & \textbf{Task} & \textbf{Partitioning} & \textbf{Accuracy} \\
\hline
\cite{alvarez2021inter} & Downsampling,\newline Normalization,\newline Segmentation & Raw segments & CNN-LSTM & supervised & SHHS, ISRUC, DRM-SUB, SVUH\_UCD, HMC, Sleep-EDF& 5-class &cross-subject & $\kappa$=0.8 \\
~\cite{Sleep35} & Downsampling,STFT & Time-Frequency Image & CNN & supervised & Sleep-EDF, \newline MASS & 5-class  & cross-subject & 82.3\% \newline 83.6\% \\
~\cite{Sleep36} & Segmentation & Raw Segments & CNN & supervised & Sleep-EDF & 5-class  & cross-subject & 92.67\% \\
~\cite{Sleep37} & Z-norm & Raw Segments & CNN+Attention & supervised & Sleep-EDF & 5-class  & mixed-subject & 82.8\%-93.7\% \\
~\cite{Sleep38} & DE Calculation & DE matrix & CNN-GCN & supervised & Sleep-EDF, \newline ISRUC & 5-class  & cross-subject & 91.0\% \newline
87.4\% \\
~\cite{Sleep39} & Windowing,STFT,PSD Calculation & Spectral \& Temporal features & LSTM & supervised & MASS & 5-class  & cross-subject & 89.4\% \\
~\cite{Sleep44} & Filtering,Segmentation,\newline Normalization & Raw Segments & CNN & supervised & ISRUC & (2-5)-class & mixed-subject & 98.93\%-99.24\% \\
\end{tabular}

~\cite{Sleep67} & Filtering,Normalization,\newline Segmentation & Raw Segments & CNN-BiGRU & semi-supervised & Sleep-EDF, \newline DRM-SUB & 5-class & mixed-subject & 82.3\% \newline
81.6\% \\
~\cite{Sleep68} & Multi-tapered Spectrogram Generation & Time-Frequency Image & GMM & semi-supervised & Sleep-EDF & 4-class & subject-specific & 73\% \\
~\cite{Sleep69} & Filtering & Raw Segments & CNN & semi-supervised & Sleep-EDF & 5-class & mixed-subject & 80\% \\
~\cite{Sleep70} & Filtering,Downsampling & Raw Segments & CNN & unsupervised & Sleep-EDF, \newline UCD & 5-class  & cross-subject & 83.4\% \newline 77.2\% \\
~\cite{Sleep71} & Segmentation,Filtering & Complex Values & CNN & unsupervised & UCD, \newline MIT-BIH & 5-class  & cross-subject & 87\% \\
~\cite{Sleep72} & Filtering,Segmentation,\newline feature extraction & Time-Frequency domain features & AE & unsupervised & Piryatinska & 3-class & mixed-subject & 80.4\% \\
~\cite{Sleep73} & Filtering,Downsampling,\newline Segmentation & Raw Segments & DBN & unsupervised & UCD & 5-class  & cross-subject & 91.31\% \\
~\cite{Sleep74} & Morlet Calculation,\newline Normalization & SSAE-based & DNN & unsupervised & Sleep-EDF & 5-class  & cross-subject & 78\% \\
~\cite{Sleep75} & Filtering,feature extraction & Time- \& Freq- features,Raw & DBN & unsupervised & UCD & 5-class  & cross-subject & 67.4\%-72.2\% \\
\hline
\end{tabular}
\end{table*}

\begin{table*}[ht]
\renewcommand{\arraystretch}{1.2}
\caption{Summary of deep learning frameworks for depression identification}
\label{tab:mdds}
\footnotesize
\begin{tabular}{p{0.4cm}p{2.8cm}p{2cm}p{1.5cm}p{1.9cm}p{1.9cm}p{0.8cm}p{1.8cm}p{2cm}}
\hline
\textbf{Ref.} & \textbf{Preprocessing} & \textbf{Feature} & \textbf{Backbone} & \textbf{Training} & \textbf{Dataset} & \textbf{Task} & \textbf{Partitioning} & \textbf{Accuracy} \\
\hline
~\cite{MDD2} & Filtering,ICA,STFT & Connectivity Graph & GCN-LSTM & supervised & PRED+CT, \newline MODMA & binary & cross-subject & 90.38\% \newline 90.57\% \\
~\cite{MDD3} & ICA,DWT,Segmentation & Frequency-domain matrix & CNN-LSTM & supervised & HUSM & binary & mixed-subject & 99.15\% \\
~\cite{MDD4} & Filtering,ICA,Z-norm,\newline STFT & 2D Spectrogram & CNN-LSTM & supervised & HUSM & binary & mixed-subject & 99.9\% \\
~\cite{sharma2021dephnn} & ICA,FFT,Windowing & Time-Frequency features & CNN-LSTM & supervised & private & binary & mixed-subject & 99.1\% \\
~\cite{seal2021deprnet} & Filtering,ICA,Segmentation & Raw Segments & CNN & supervised & private & binary & mixed-subject & 99.37\% \\
~\cite{MDD7} & Filtering,Segmentation & Raw Segments & CNN-Transformer & supervised & HUSM,\newline
private & binary & cross-subject & 93.7\% \newline 96.2\% \\
~\cite{MDD8} & Filtering,ICA,Band \newline Filter,CSP & Raw Segments & Transformer & supervised & private & binary & mixed-subject & 92.25\% \\
\hline
\end{tabular}
\end{table*}

\begin{table*}[ht]
\renewcommand{\arraystretch}{1.2}
\caption*{(Continued) Summary of deep learning frameworks for depression identification}
\footnotesize
\begin{tabular}{p{0.4cm}p{2.8cm}p{2cm}p{1.5cm}p{1.9cm}p{1.9cm}p{0.8cm}p{1.8cm}p{1.5cm}}
\hline
\textbf{Ref.} & \textbf{Preprocessing} & \textbf{Feature} & \textbf{Backbone} & \textbf{Training} & \textbf{Dataset} & \textbf{Task} & \textbf{Partitioning} & \textbf{Accuracy} \\
\hline
~\cite{MDD34} & Filtering,STFT & 2D Spectrogram & 2D-CNN & supervised & private & binary & mixed-subject & 96.43\% \\
~\cite{MDD35} & Filtering,ICA,Segmentation & Mixed Feature matrix & CNN & supervised & private & binary & mixed-subject & 94.13\% \\
~\cite{li2019depression} & ICA,LMS,AR,Hjorth & 2D Image & CNN & supervised & private & binary & cross-subject & 84.75\% \\
~\cite{MDD37} & Filtering,Segmentation & Raw Segments & CNN & supervised & private & binary & mixed-subject & 75.29\% \\
~\cite{MDD38} & Filtering,Segmentation,\newline PLV,PLI & Connectivity \newline matrix & 2D-CNN & supervised & private & binary & cross-subject & 80.74\% \\
~\cite{MDD39} & Filtering,PLI & Connectivity \newline matrix & 2D-CNN & supervised & private & binary & mixed-subject & 67.67\% \\
~\cite{khan2021automated} & Denoising,Segmentation,\newline PDC matrix Calculation & 3D CPC & 3D-CNN & supervised & private & binary & cross-subject & 100\% \\
\hline
\end{tabular}
\end{table*}

\begin{table*}[ht]
\renewcommand{\arraystretch}{1.2}
\caption{Summary of deep learning frameworks for schizophrenia identifiaction}
\label{tab:schis}
\footnotesize
\begin{tabular}{p{0.4cm}p{2.8cm}p{2cm}p{1.5cm}p{1.9cm}p{1.9cm}p{0.8cm}p{1.8cm}p{2cm}}
\hline
\textbf{Ref.} & \textbf{Preprocessing} & \textbf{Feature} & \textbf{Backbone} & \textbf{Training} & \textbf{Dataset} & \textbf{Task} & \textbf{Partitioning} & \textbf{Accuracy} \\
\hline
~\cite{phang2019multi} & Connectivity Measures,Complex Network construction & VAR,PDC,CN & CNN & supervised & MHRC & binary & cross-subject & 91.69\% \\
~\cite{SZ2} & Z-norm,Segmentation & Raw Segments & CNN & supervised & CeonRepod & binary & mixed-subject & 98.07\% \\
~\cite{SZ3} & Segmentation,\newline Margenau–Hill & Time-Frequency Image & CNN & supervised & MHRC, \newline CeonRepod, \newline NIMH & binary & mixed-subject & 96.35\%-99.75\% \\
~\cite{SZ4} & Connectivity Networks Construction & WOC-Based \newline features & CNN & supervised & MHRC & binary & cross-subject & 90\% \\
~\cite{SZ5} & Filtering,Segmentation,\newline Welch Method & Spectrum matrix & CNN & supervised & private & binary & cross-subject & 91.12\% \\
~\cite{SZ6} & Filtering & Raw Segments & CNN & supervised & CeonRepod & binary & mixed-subject & 98.05\% \\
~\cite{10023506} & Filtering,Segmentation,\newline ASR,ICA & Connectivity \newline features & CNN & supervised & CeonRepod & binary & mixed-subject &  99.84\% \\
\hline
\end{tabular}
\end{table*}

\begin{table*}[ht]
\renewcommand{\arraystretch}{1.2}
\caption*{(Continued) Summary of deep learning frameworks for schizophrenia identification}
\footnotesize
\begin{tabular}{p{0.4cm}p{2.8cm}p{2cm}p{1.5cm}p{1.9cm}p{1.9cm}p{0.8cm}p{1.8cm}p{2cm}}
\hline
\textbf{Ref.} & \textbf{Preprocessing} & \textbf{Feature} & \textbf{Backbone} & \textbf{Training} & \textbf{Dataset} & \textbf{Task} & \textbf{Partitioning} & \textbf{Accuracy} \\
\hline
~\cite{sun2021hybrid} & Filtering,Segmentation,\newline Baseline Correction,\newline Ocular Correction & FuzzyEn RGB Image & CNN-LSTM & supervised & private & binary & mixed-subject & 99.22\% \\
~\cite{SZ36} & Filtering,Segmentation & Raw Segments & CNN-LSTM & supervised & MHRC, \newline CeonRepod & binary & cross-subject & 91\% \newline 96.1\% \\
~\cite{SZ37} & MSST & Time-Frequency Feature Image & CNN-LSTM & supervised & CeonRepod & binary & mixed-subject & 84.42\% \\
~\cite{SZ38} & Filtering,TE & 2D Image & CNN-LSTM & supervised & CeonRepod & binary & mixed-subject & 99.9\% \\
~\cite{SZ39} & Artifact Removal,\newline Filtering & Raw & CNN-LSTM & supervised & NIMH & binary & cross-subject & 98.2\% \\
~\cite{SZ40} & Segmentation,Z-norm & Raw Segments & CNN-LSTM & supervised & CeonRepod & binary & mixed-subject & 99.25\% \\
~\cite{SZ41} & Filtering,PCA,ICA & Raw,features & CNN-TCN & supervised & CeonRepod & binary & mixed-subject & 99.57\% \\
\hline
\end{tabular}
\end{table*}

\begin{table*}[ht]
\renewcommand{\arraystretch}{1.2}
\caption{Summary of deep learning frameworks for Alzheimer's Disease Diagnosis}
\label{tab:ads}
\footnotesize
\begin{tabular}{p{0.4cm}p{2.8cm}p{2cm}p{1.5cm}p{1.9cm}p{1.9cm}p{0.8cm}p{1.8cm}p{2cm}}
\hline
\textbf{Ref.} & \textbf{Preprocessing} & \textbf{Feature} & \textbf{Backbone} & \textbf{Training} & \textbf{Dataset} & \textbf{Task} & \textbf{Partitioning} & \textbf{Accuracy} \\
\hline
~\cite{AD1} & Filtering,Segmentation,\newline Connections Calculation & Connection \newline matrix & 2D-CNN & Supervised & private & binary & mixed-subject & 100\% \\
~\cite{AD2} & Filtering,Segmentation & Raw Segments & 2D-CNN & Supervised & FSA\_Alzheimer’s & binary & mixed-subject & 97.9\% \\
~\cite{shan2022spatial} & Filtering,Segmentation,\newline Network construction & Adjacency matrix,Segments & GCN & Supervised & private & binary & mixed-subject & 92.3\% \\
~\cite{AD4} & Filtering,Segmentation & Raw Segments & 2D-CNN & Supervised & private & binary & mixed-subject & 69.03\%-85.78\% \\
~\cite{AD5} & Filtering,Segmentation,\newline CWT & Time-Frequency features & 2D-CNN & Supervised & private & binary\newline 3-class & cross-subject & 85\%\newline82\% \\
~\cite{AD6} & Filtering,FFT & 2D Spectrograms & 2D-CNN & Supervised & private & binary\newline 3-class & mixed-subject & 97.11\%\newline95.04\% \\
~\cite{AD7} & Filtering,FFT & Frequency-domain features & 2D-CNN & Supervised & private & binary& - & 93.7\% \\
\hline
\end{tabular}
\end{table*}

\begin{table*}[ht]
\renewcommand{\arraystretch}{1.2}
\caption{Summary of deep learning frameworks for Parkinson's Disease Diagnosis}
\label{tab:pds}
\footnotesize
\begin{tabular}{p{0.4cm}p{2.8cm}p{2cm}p{1.5cm}p{1.9cm}p{1.9cm}p{0.8cm}p{1.8cm}p{2cm}}
\hline
\textbf{Ref.} & \textbf{Preprocessing} & \textbf{Feature} & \textbf{Backbone} & \textbf{Training} & \textbf{Dataset} & \textbf{Task} & \textbf{Partitioning} & \textbf{Accuracy} \\
\hline
~\cite{PD1} & Segmentation,Embedding Reconstruction & Reconstructed Segments & CNN-LSTM & supervised & UNM & binary & mixed-subject & 99.22\% \\
~\cite{PD2} & Gabor Transform & 2D Spectrograms & 2D-CNN & supervised & UCSD & 3-class & mixed-subject & 92.6\%-99.46\% \\
~\cite{PD3} & SPWVD,Artifact Removal,Segmentation & TFR & 2D-CNN & supervised & UCSD, \newline private & binary & mixed-subject & 99.84\%-100\% \\
~\cite{PD5} & Denoising,TQWT,WPT & Time-Frequency features & CNN & supervised & private & 3-class & mixed-subject & 92.59\%-99.92\% \\
~\cite{PD6} & Artifact rejection,\newline Filtering,Segmentation & Raw Segments & CNN-RNN & supervised & private & binary & cross-subject & 82.89\% \\
~\cite{shaban2022resting} & CWT,Segmentation & 2D Image & CNN & supervised & UCSD & 3-class & mixed-subject & 99.6\%-99.9\% \\
~\cite{PD8} & ICA,Filtering,P-Welch & PSD Image & 2D-CNN & supervised & private & binary & mixed-subject & 99.87\% \\
\hline
\end{tabular}
\end{table*}

\begin{table*}[ht]
\renewcommand{\arraystretch}{1.2}
\caption{Summary of deep learning frameworks for ADHD identification}
\label{tab:adhds}
\footnotesize
\begin{tabular}{p{0.4cm}p{2.8cm}p{2cm}p{1.5cm}p{1.9cm}p{1.9cm}p{0.8cm}p{1.8cm}p{1.5cm}}
\hline
\textbf{Ref.} & \textbf{Preprocessing} & \textbf{Feature} & \textbf{Backbone} & \textbf{Training} & \textbf{Dataset} & \textbf{Task} & \textbf{Partitioning} &
\textbf{Accuracy} \\
\hline
~\cite{chen2019use} & segment screening & PSD & CNN & Supervised & private & Binary & mixed-subject & 90.29\% \\
~\cite{dubreuil2020deep} & Filtering,Segmentation,\newline wavelet transform & Spectrogram & CNN & Supervised & private & Binary & cross-subject & 88\% \\
~\cite{MOGHADDARI2020105738} & Resampling,filtering,\newline ASR,windowing,\newline Freq. bands separation & Frequency bands, RGB Images & CNN & Supervised & ADHD-Child & Binary & cross-subject & 98.48\% \\
~\cite{tosun2021effects} & PSD & PSD,SE & LSTM & Supervised & private & Binary & mixed-subject & 92.15\% \\
~\cite{ADHD5} & Filtering,Segmentation,\newline ICA,segment screening & End-to-end & CNN & Supervised & private & 3-class & mixed-subject & 99.46\% \\
~\cite{ADHD6} & FIR,filtering,ICA,\newline Segmentation & Dynamic connectivity tensor (DCT) & ConvLSTM +Attention & Supervised & ADHD-Child & Binary & mixed-subject & 99.75\% \\
~\cite{nouri2024detection} & Re-referencing,filtering,\newline Baseline rejection,\newline downsampling,Segmentation & PSD & CNN & Supervised & ADHD-Child & Binary & mixed-subject & 94.52\% \\
~\cite{ADHD8} & Filtering,Segmentation,\newline CWT & Time-Frequency Image & ConvMixer,\newline ResNet50,\newline ResNet18 & Supervised & ADHD-Child & Binary & mixed-subject & 72.58\% \\
\hline
\end{tabular}
\end{table*}



\ifCLASSOPTIONcaptionsoff
  \newpage
\fi



%
\clearpage  
\bibliography{reference}

\begin{thebibliography}{100}
\providecommand{\url}[1]{#1}
\csname url@samestyle\endcsname
\providecommand{\newblock}{\relax}
\providecommand{\bibinfo}[2]{#2}
\providecommand{\BIBentrySTDinterwordspacing}{\spaceskip=0pt\relax}
\providecommand{\BIBentryALTinterwordstretchfactor}{4}
\providecommand{\BIBentryALTinterwordspacing}{\spaceskip=\fontdimen2\font plus
\BIBentryALTinterwordstretchfactor\fontdimen3\font minus
  \fontdimen4\font\relax}
\providecommand{\BIBforeignlanguage}[2]{{%
\expandafter\ifx\csname l@#1\endcsname\relax
\typeout{** WARNING: IEEEtran.bst: No hyphenation pattern has been}%
\typeout{** loaded for the language `#1'. Using the pattern for}%
\typeout{** the default language instead.}%
\else
\language=\csname l@#1\endcsname
\fi
#2}}
\providecommand{\BIBdecl}{\relax}
\BIBdecl

\bibitem{WHO2024}
{World Health Organization}, ``Over 1 in 3 people affected by neurological
  conditions, the leading cause of illness and disability worldwide,'' 2024.

\bibitem{better2024alzheimer}
M.~A. BETTER, ``Alzheimer’s disease facts and figures,'' \emph{Alzheimer’s
  Dement}, vol.~20, pp. 3708--3821, 2024.

\bibitem{WHO_epilepsy}
{World Health Organization}, ``Epilepsy,'' 2024.

\bibitem{recoveryvillage_sleep_statistics_2023}
{The Recovery Village}, ``Sleep disorders statistics,'' 2023.

\bibitem{who2023parkinson}
{World Health Organization}, ``Parkinson disease,'' 2023.

\bibitem{WHO_SZ}
------. (2024) Schizophrenia.

\bibitem{WHO_depression}
------. (2024) Depression.

\bibitem{who_adolescent_mental_health}
------. (2021) Adolescent mental health.

\bibitem{WHO_MentalHealth}
------, ``Mental health: Neurological disorders.''

\bibitem{berger1929elektroenkephalogramm}
H.~Berger, ``{\"U}ber das elektroenkephalogramm des menschen,'' \emph{Archiv
  f{\"u}r psychiatrie und nervenkrankheiten}, vol.~87, no.~1, pp. 527--570,
  1929.

\bibitem{ramantani2016correlation}
G.~Ramantani and et~al., ``Correlation of invasive eeg and scalp eeg,''
  \emph{Seizure}, vol.~41, pp. 196--200, 2016.

\bibitem{ACHARYA2018270}
U.~R. Acharya and et~al., ``Deep convolutional neural network for the automated
  detection and diagnosis of seizure using eeg signals,'' \emph{Computers in
  Biology and Medicine}, vol. 100, pp. 270--278, 2018.

\bibitem{acharya2018automated}
------, ``Automated eeg-based screening of depression using deep convolutional
  neural network,'' \emph{Computer methods and programs in biomedicine}, vol.
  161, pp. 103--113, 2018.

\bibitem{lecun1995convolutional}
Y.~LeCun and et~al., ``Convolutional networks for images, speech, and time
  series,'' \emph{The handbook of brain theory and neural networks}, vol. 3361,
  no.~10, p. 1995, 1995.

\bibitem{vaswani2017attention}
A.~Vaswani and et~al., ``Attention is all you need,'' \emph{Advances in neural
  information processing systems}, vol.~30, 2017.

\bibitem{turk2019epilepsy}
{\"O}.~T{\"u}rk and M.~S. {\"O}zerdem, ``Epilepsy detection by using scalogram
  based convolutional neural network from eeg signals,'' \emph{Brain sciences},
  vol.~9, no.~5, p. 115, 2019.

\bibitem{Kostas2021BENDR}
D.~Kostas and et~al., ``Bendr: Using transformers and a contrastive
  self-supervised learning task to learn from massive amounts of eeg data,''
  \emph{Frontiers in Human Neuroscience}, vol.~15, p. 653659, 2021.

\bibitem{roy2019deep}
Y.~Roy and et~al., ``Deep learning-based electroencephalography analysis: a
  systematic review,'' \emph{Journal of neural engineering}, vol.~16, no.~5, p.
  051001, 2019.

\bibitem{amer2023eeg}
N.~S. Amer and S.~B. Belhaouari, ``Eeg signal processing for medical diagnosis,
  healthcare, and monitoring: A comprehensive review,'' \emph{IEEE Access},
  2023.

\bibitem{zhang2021survey}
X.~Zhang and et~al., ``A survey on deep learning-based non-invasive brain
  signals: recent advances and new frontiers,'' \emph{Journal of neural
  engineering}, vol.~18, no.~3, p. 031002, 2021.

\bibitem{khan2021machine}
P.~Khan and et~al., ``Machine learning and deep learning approaches for brain
  disease diagnosis: principles and recent advances,'' \emph{Ieee Access},
  vol.~9, pp. 37\,622--37\,655, 2021.

\bibitem{shoeibi2021epileptic}
A.~Shoeibi and et~al., ``Epileptic seizures detection using deep learning
  techniques: a review,'' \emph{International journal of environmental research
  and public health}, vol.~18, no.~11, p. 5780, 2021.

\bibitem{rahul2024systematic}
J.~Rahul and et~al., ``A systematic review of eeg based automated schizophrenia
  classification through machine learning and deep learning,'' \emph{Frontiers
  in Human Neuroscience}, vol.~18, p. 1347082, 2024.

\bibitem{hossain2023status}
K.~M. Hossain and et~al., ``Status of deep learning for eeg-based
  brain--computer interface applications,'' \emph{Frontiers in computational
  neuroscience}, vol.~16, p. 1006763, 2023.

\bibitem{weng2024self}
W.~Weng and et~al., ``Self-supervised learning for electroencephalogram: A
  systematic survey,'' \emph{arXiv preprint arXiv:2401.05446}, 2024.

\bibitem{chen2024con4m}
J.~Chen and et~al., ``Con4m: Context-aware consistency learning framework for
  segmented time series classification,'' \emph{The Thirty-Eighth Annual
  Conference on Neural Information Processing Systems}, 2024.

\bibitem{banville2021uncovering}
H.~Banville and et~al., ``Uncovering the structure of clinical eeg signals with
  self-supervised learning,'' \emph{Journal of Neural Engineering}, vol.~18,
  no.~4, p. 046020, 2021.

\bibitem{oh2020deep}
S.~L. Oh and et~al., ``A deep learning approach for parkinson’s disease
  diagnosis from eeg signals,'' \emph{Neural Computing and Applications},
  vol.~32, pp. 10\,927--10\,933, 2020.

\bibitem{9353630}
C.~Chatzichristos and et~al., ``Epileptic seizure detection in eeg via fusion
  of multi-view attention-gated u-net deep neural networks,'' in \emph{2020
  IEEE Signal Processing in Medicine and Biology Symposium (SPMB)}, 2020, pp.
  1--7.

\bibitem{ay2019automated}
B.~Ay and et~al., ``Automated depression detection using deep representation
  and sequence learning with eeg signals,'' \emph{Journal of medical systems},
  vol.~43, pp. 1--12, 2019.

\bibitem{10023506}
N.~Grover and et~al., ``Schizo-net: A novel schizophrenia diagnosis framework
  using late fusion multimodal deep learning on electroencephalogram-based
  brain connectivity indices,'' \emph{IEEE Transactions on Neural Systems and
  Rehabilitation Engineering}, vol.~31, pp. 464--473, 2023.

\bibitem{MOGHADDARI2020105738}
M.~Moghaddari and et~al., ``Diagnose adhd disorder in children using
  convolutional neural network based on continuous mental task eeg,''
  \emph{Computer Methods and Programs in Biomedicine}, vol. 197, p. 105738,
  2020.

\bibitem{sun2021hybrid}
J.~Sun and et~al., ``A hybrid deep neural network for classification of
  schizophrenia using eeg data,'' \emph{Scientific Reports}, vol.~11, no.~1, p.
  4706, 2021.

\bibitem{nouri2024detection}
A.~Nouri and Z.~Tabanfar, ``Detection of adhd disorder in children using
  layer-wise relevance propagation and convolutional neural network: An eeg
  analysis,'' \emph{Frontiers in Biomedical Technologies}, vol.~11, no.~1, pp.
  14--21, 2024.

\bibitem{Seizure49}
A.~Antoniades and et~al., ``Deep learning for epileptic intracranial eeg
  data,'' in \emph{2016 IEEE 26th International Workshop on Machine Learning
  for Signal Processing (MLSP)}.\hskip 1em plus 0.5em minus 0.4em\relax IEEE,
  2016, pp. 1--6.

\bibitem{wen2018deep}
T.~Wen and Z.~Zhang, ``Deep convolution neural network and autoencoders-based
  unsupervised feature learning of eeg signals,'' \emph{IEEE Access}, vol.~6,
  pp. 25\,399--25\,410, 2018.

\bibitem{ko2022eeg}
W.~Ko and H.-I. Suk, ``Eeg-oriented self-supervised learning and cluster-aware
  adaptation,'' in \emph{Proceedings of the 31st ACM International Conference
  on Information \& Knowledge Management}, 2022, pp. 4143--4147.

\bibitem{mousavi2019deep}
Z.~Mousavi and et~al., ``Deep convolutional neural network for classification
  of sleep stages from single-channel eeg signals,'' \emph{Journal of
  neuroscience methods}, vol. 324, p. 108312, 2019.

\bibitem{ZHANG2020105089}
J.~Zhang and et~al., ``Orthogonal convolutional neural networks for automatic
  sleep stage classification based on single-channel eeg,'' \emph{Computer
  Methods and Programs in Biomedicine}, vol. 183, p. 105089, 2020.

\bibitem{SZ27}
S.~Siuly and et~al., ``Exploring deep residual network based features for
  automatic schizophrenia detection from eeg,'' \emph{Physics in Engineering
  and Science Medicine}, vol.~46, pp. 561--574, 2023.

\bibitem{sharma2021dephnn}
G.~Sharma and et~al., ``Dephnn: a novel hybrid neural network for
  electroencephalogram (eeg)-based screening of depression,'' \emph{Biomedical
  signal processing and control}, vol.~66, p. 102393, 2021.

\bibitem{seal2021deprnet}
A.~Seal and et~al., ``Deprnet: A deep convolution neural network framework for
  detecting depression using eeg,'' \emph{IEEE Transactions on Instrumentation
  and Measurement}, vol.~70, pp. 1--13, 2021.

\bibitem{iwama2023two}
S.~Iwama and et~al., ``Two common issues in synchronized multimodal recordings
  with eeg: Jitter and latency,'' \emph{Neuroscience Research}, 2023.

\bibitem{jasper1958ten}
H.~H. Jasper, ``Ten-twenty electrode system of the international federation,''
  \emph{Electroencephalogr Clin Neurophysiol}, vol.~10, pp. 371--375, 1958.

\bibitem{buzsaki2004neuronal}
G.~Buzsaki and A.~Draguhn, ``Neuronal oscillations in cortical networks,''
  \emph{science}, vol. 304, no. 5679, pp. 1926--1929, 2004.

\bibitem{jeong2004eeg}
J.~Jeong, ``Eeg dynamics in patients with alzheimer's disease,'' \emph{Clinical
  neurophysiology}, vol. 115, no.~7, pp. 1490--1505, 2004.

\bibitem{9713847}
N.~Sobahi and et~al., ``A new signal to image mapping procedure and
  convolutional neural networks for efficient schizophrenia detection in eeg
  recordings,'' \emph{IEEE Sensors Journal}, vol.~22, no.~8, pp. 7913--7919,
  2022.

\bibitem{PD2}
H.~W. Loh and et~al., ``Gaborpdnet: Gabor transformation and deep neural
  network for parkinson’s disease detection using eeg signals,''
  \emph{Electronics}, vol.~10, no.~14, p. 1740, 2021.

\bibitem{zulfikar2022empirical}
A.~Z{\"u}lfikar and A.~Mehmet, ``Empirical mode decomposition and convolutional
  neural network-based approach for diagnosing psychotic disorders from eeg
  signals,'' \emph{Applied Intelligence}, vol.~52, no.~11, pp.
  12\,103--12\,115, 2022.

\bibitem{li2019eeg}
X.~Li and et~al., ``Eeg-based mild depression recognition using convolutional
  neural network,'' \emph{Medical \& biological engineering \& computing},
  vol.~57, pp. 1341--1352, 2019.

\bibitem{Seizure214}
A.~M. Karim and et~al., ``A new framework using deep auto-encoder and energy
  spectral density for medical waveform data classification and processing,''
  \emph{Biocybernetics and Biomedical Engineering}, vol.~39, no.~1, pp.
  148--159, 2019.

\bibitem{khan2021automated}
D.~M. Khan and et~al., ``Automated diagnosis of major depressive disorder using
  brain effective connectivity and 3d convolutional neural network,''
  \emph{Ieee Access}, vol.~9, pp. 8835--8846, 2021.

\bibitem{zhu2019multimodal}
J.~Zhu and et~al., ``Multimodal mild depression recognition based on eeg-em
  synchronization acquisition network,'' \emph{IEEE Access}, vol.~7, pp.
  28\,196--28\,210, 2019.

\bibitem{Seizure7}
J.~Liu and B.~Woodson, ``Deep learning classification for epilepsy detection
  using a single channel electroencephalography (eeg),'' in \emph{Proceedings
  of the 2019 3rd International Conference on Deep Learning Technologies},
  2019, pp. 23--26.

\bibitem{li2019depression}
X.~Li and et~al., ``Depression recognition using machine learning methods with
  different feature generation strategies,'' \emph{Artificial intelligence in
  medicine}, vol.~99, p. 101696, 2019.

\bibitem{Seizure67}
I.~Bhattacherjee, ``Epileptic seizure detection using multicolumn convolutional
  neural network,'' in \emph{2020 7th International Conference on Computing for
  Sustainable Global Development (INDIACom)}.\hskip 1em plus 0.5em minus
  0.4em\relax IEEE, 2020, pp. 58--63.

\bibitem{tosun2021effects}
M.~Tosun, ``Effects of spectral features of eeg signals recorded with different
  channels and recording statuses on adhd classification with deep learning,''
  \emph{Physical and Engineering Sciences in Medicine}, vol.~44, no.~3, pp.
  693--702, 2021.

\bibitem{aslan2022deep}
Z.~Aslan and M.~Akin, ``A deep learning approach in automated detection of
  schizophrenia using scalogram images of eeg signals,'' \emph{Physical and
  Engineering Sciences in Medicine}, vol.~45, no.~1, pp. 83--96, 2022.

\bibitem{choi2019novel}
G.~Choi and et~al., ``A novel multi-scale 3d cnn with deep neural network for
  epileptic seizure detection,'' in \emph{2019 IEEE International Conference on
  Consumer Electronics (ICCE)}, 2019, pp. 1--2.

\bibitem{vilamala2017deep}
A.~Vilamala and et~al., ``Deep convolutional neural networks for interpretable
  analysis of eeg sleep stage scoring,'' in \emph{2017 IEEE 27th international
  workshop on machine learning for signal processing (MLSP)}, 2017, pp. 1--6.

\bibitem{Seizure109}
J.~Xu and et~al., ``Epileptic seizure detection based on feature extraction and
  cnn- bigru network with attention mechanism,'' in \emph{International
  Conference on Intelligent Computing}.\hskip 1em plus 0.5em minus 0.4em\relax
  Springer, 2023, pp. 308--319.

\bibitem{phang2019multi}
C.-R. Phang and et~al., ``A multi-domain connectome convolutional neural
  network for identifying schizophrenia from eeg connectivity patterns,''
  \emph{IEEE journal of biomedical and health informatics}, vol.~24, no.~5, pp.
  1333--1343, 2019.

\bibitem{electronics11142265}
D.-W. Ko and J.-J. Yang, ``Eeg-based schizophrenia diagnosis through time
  series image conversion and deep learning,'' \emph{Electronics}, vol.~11,
  no.~14, 2022.

\bibitem{sohrabpour2020noninvasive}
A.~Sohrabpour and et~al., ``Noninvasive electromagnetic source imaging of
  spatiotemporally distributed epileptogenic brain sources,'' \emph{Nature
  communications}, vol.~11, no.~1, p. 1946, 2020.

\bibitem{sun2022deep}
R.~Sun and et~al., ``Deep neural networks constrained by neural mass models
  improve electrophysiological source imaging of spatiotemporal brain
  dynamics,'' \emph{Proceedings of the National Academy of Sciences}, vol. 119,
  no.~31, p. e2201128119, 2022.

\bibitem{Zhan2020EpilepsyDetection}
Q.~Zhan and W.~Hu, ``An epilepsy detection method using multiview clustering
  algorithm and deep features,'' \emph{Computational and Mathematical Methods
  in Medicine}, vol. 2020, p. 5128729, 2020.

\bibitem{ho2023self}
T.~K.~K. Ho and N.~Armanfard, ``Self-supervised learning for anomalous channel
  detection in eeg graphs: Application to seizure analysis,'' in
  \emph{Proceedings of the AAAI conference on artificial intelligence},
  vol.~37, no.~7, 2023, pp. 7866--7874.

\bibitem{9047940}
K.~A. Robbins and et~al., ``How sensitive are eeg results to preprocessing
  methods: A benchmarking study,'' \emph{IEEE Transactions on Neural Systems
  and Rehabilitation Engineering}, vol.~28, no.~5, pp. 1081--1090, 2020.

\bibitem{akut2019wavelet}
R.~Akut, ``Wavelet based deep learning approach for epilepsy detection,''
  \emph{Health information science and systems}, vol.~7, no.~1, p.~8, 2019.

\bibitem{oh2014novel}
S.-H. Oh and et~al., ``A novel eeg feature extraction method using hjorth
  parameter,'' \emph{International Journal of Electronics and Electrical
  Engineering}, vol.~2, no.~2, pp. 106--110, 2014.

\bibitem{zancanaro2021cnn}
A.~Zancanaro and et~al., ``Cnn-based approaches for cross-subject
  classification in motor imagery: From the state-of-the-art to dynamicnet,''
  in \emph{2021 IEEE conference on computational intelligence in bioinformatics
  and computational biology (CIBCB)}.\hskip 1em plus 0.5em minus 0.4em\relax
  IEEE, 2021, pp. 1--7.

\bibitem{elman1990finding}
J.~L. Elman, ``Finding structure in time,'' \emph{Cognitive science}, vol.~14,
  no.~2, pp. 179--211, 1990.

\bibitem{4700287}
F.~Scarselli and et~al., ``The graph neural network model,'' \emph{IEEE
  Transactions on Neural Networks}, vol.~20, no.~1, pp. 61--80, 2009.

\bibitem{hinton1993autoencoders}
G.~E. Hinton and R.~Zemel, ``Autoencoders, minimum description length and
  helmholtz free energy,'' \emph{Advances in neural information processing
  systems}, vol.~6, 1993.

\bibitem{goodfellow2014generative}
I.~Goodfellow and et~al., ``Generative adversarial nets,'' \emph{Advances in
  neural information processing systems}, vol.~27, 2014.

\bibitem{maass1997networks}
W.~Maass, ``Networks of spiking neurons: the third generation of neural network
  models,'' \emph{Neural networks}, vol.~10, no.~9, pp. 1659--1671, 1997.

\bibitem{andrzejak2001indications}
R.~G. Andrzejak and et~al., ``{Indications of nonlinear deterministic and
  finite-dimensional structures in time series of brain electrical activity:
  Dependence on recording region and brain state},'' \emph{Physical Review E},
  vol.~64, no.~6, p. 061907, 2001.

\bibitem{ihle2012epilepsiae}
M.~Ihle and et~al., ``{Epilepsiae -- A European epilepsy database},''
  \emph{Computer Methods and Programs in Biomedicine}, vol. 106, no.~3, pp.
  127--138, 2012.

\bibitem{seizure-detection}
bbrinkm and et~al., ``Upenn and mayo clinic's seizure detection challenge,''
  2014.

\bibitem{guttag2010chb}
J.~Guttag, ``{CHB-MIT Scalp EEG Database (version 1.0.0)},'' 2010.

\bibitem{shoeb2009application}
A.~Shoeb, ``{Application of Machine Learning to Epileptic Seizure Onset
  Detection and Treatment},'' Ph.D. dissertation, Massachusetts Institute of
  Technology, September 2009.

\bibitem{goldberger2000physiobank}
A.~Goldberger and et~al., ``{PhysioBank, PhysioToolkit, and PhysioNet:
  Components of a new research resource for complex physiologic signals},''
  \emph{Circulation}, vol. 101, no.~23, pp. e215--e220, 2000.

\bibitem{andrzejak2012nonrandomness}
R.~G. Andrzejak and et~al., ``{Nonrandomness, nonlinear dependence, and
  nonstationarity of electroencephalographic recordings from epilepsy
  patients},'' \emph{Physical Review E}, vol.~86, p. 046206, 2012.

\bibitem{hauz}
P.~Swami and et~al., ``Eeg epilepsy datasets,'' 09 2016.

\bibitem{melbourne}
L.~Kuhlmann and et~al., ``Melbourne university aes/mathworks/nih seizure
  prediction,'' 2016.

\bibitem{shah2018temple}
V.~Shah and et~al., ``{The Temple University Hospital Seizure Detection
  Corpus},'' \emph{Frontiers in Neuroinformatics}, vol.~12, p.~83, 2018.

\bibitem{burrello2018oneshot}
A.~Burrello and et~al., ``One-shot learning for ieeg seizure detection using
  end-to-end binary operations: Local binary patterns with hyperdimensional
  computing,'' in \emph{Proceedings of the IEEE Biomedical Circuits and Systems
  Conference (BioCAS)}, October 17-19 2018.

\bibitem{burrello2019hdc}
------, ``Hyperdimensional computing with local binary patterns: One-shot
  learning of seizure onset and identification of ictogenic brain regions using
  short-time ieeg recordings,'' \emph{IEEE Transactions on Biomedical
  Engineering (TBME)}, 2019.

\bibitem{stevenson2019dataset}
N.~Stevenson and et~al., ``{A dataset of neonatal EEG recordings with seizure
  annotations},'' \emph{Scientific Data}, vol.~6, p. 190039, 2019.

\bibitem{Nejedly2020}
P.~Nejedly and et~al., ``Multicenter intracranial eeg dataset for
  classification of graphoelements and artifactual signals,'' \emph{Scientific
  Data}, vol.~7, no.~1, p. 179, June 2020.

\bibitem{detti2020eeg}
P.~Detti and et~al., ``Eeg synchronization analysis for seizure prediction: A
  study on data of noninvasive recordings,'' \emph{Processes}, vol.~8, no.~7,
  p. 846, 2020.

\bibitem{nasreddine2021epileptic}
W.~Nasreddine, ``Epileptic eeg dataset,'' 2021.

\bibitem{HUP}
J.~M. Bernabei and et~al., ``"hup ieeg epilepsy dataset",'' 2022.

\bibitem{ds004080:1.2.4}
D.~van Blooijs, M.~van~den Boom, J.~van~der Aar, G.~Huiskamp, G.~Castegnaro,
  M.~Demuru, W.~Zweiphenning, P.~van Eijsden, K.~J. Miller, F.~Leijten, and
  D.~Hermes, ``"ccep ecog dataset across age 4-51",'' 2023.

\bibitem{mohsenvand2020contrastive}
M.~N. Mohsenvand and et~al., ``Contrastive representation learning for
  electroencephalogram classification,'' in \emph{Machine Learning for Health},
  2020, pp. 238--253.

\bibitem{banville2019self}
H.~Banville and et~al., ``Self-supervised representation learning from
  electroencephalography signals,'' in \emph{2019 IEEE 29th International
  Workshop on Machine Learning for Signal Processing (MLSP)}, 2019, pp. 1--6.

\bibitem{oord2018representation}
A.~v.~d. Oord and et~al., ``Representation learning with contrastive predictive
  coding,'' \emph{arXiv preprint arXiv:1807.03748}, 2018.

\bibitem{wu2022neuro2vec}
D.~Wu and et~al., ``neuro2vec: Masked fourier spectrum prediction for
  neurophysiological representation learning,'' \emph{arXiv preprint
  arXiv:2204.12440}, 2022.

\bibitem{cai2023mbrain}
D.~Cai and et~al., ``Mbrain: A multi-channel self-supervised learning framework
  for brain signals,'' in \emph{Proceedings of the 29th ACM SIGKDD Conference
  on Knowledge Discovery and Data Mining}, 2023, pp. 130--141.

\bibitem{ULLAH201861}
I.~Ullah and et~al., ``An automated system for epilepsy detection using eeg
  brain signals based on deep learning approach,'' \emph{Expert Systems with
  Applications}, vol. 107, pp. 61--71, 2018.

\bibitem{abdelhameed2018epileptic}
A.~M. Abdelhameed and et~al., ``Epileptic seizure detection using deep
  convolutional autoencoder,'' in \emph{2018 IEEE international workshop on
  signal processing systems (SiPS)}, 2018, pp. 223--228.

\bibitem{zhou2018epileptic}
M.~Zhou and et~al., ``Epileptic seizure detection based on eeg signals and
  cnn,'' \emph{Frontiers in neuroinformatics}, vol.~12, p.~95, 2018.

\bibitem{dutta2024deep}
A.~K. Dutta and et~al., ``Deep learning-based multi-head self-attention model
  for human epilepsy identification from eeg signal for biomedical traits,''
  \emph{Multimedia Tools and Applications}, pp. 1--23, 2024.

\bibitem{zhao2023hybrid}
Y.~Zhao, , and et~al., ``Hybrid attention network for epileptic eeg
  classification,'' \emph{International Journal of Neural Systems}, vol.~33,
  no.~06, p. 2350031, 2023.

\bibitem{XIAO2024105464}
T.~Xiao and et~al., ``Self-supervised learning with attention mechanism for
  eeg-based seizure detection,'' \emph{Biomedical Signal Processing and
  Control}, vol.~87, p. 105464, 2024.

\bibitem{zheng2022task}
Y.~Zheng and et~al., ``Task-oriented self-supervised learning for anomaly
  detection in electroencephalography,'' in \emph{International Conference on
  Medical Image Computing and Computer-Assisted Intervention}.\hskip 1em plus
  0.5em minus 0.4em\relax Springer, 2022, pp. 193--203.

\bibitem{raghu2020eeg}
S.~Raghu and et~al., ``Eeg based multi-class seizure type classification using
  convolutional neural network and transfer learning,'' \emph{Neural Networks},
  vol. 124, pp. 202--212, 2020.

\bibitem{tang2021self}
S.~Tang and et~al., ``Self-supervised graph neural networks for improved
  electroencephalographic seizure analysis,'' \emph{10th International
  Conference on Learning Representations (ICLR’22)}, 2022.

\bibitem{peng2023wavelet2vec}
R.~Peng and et~al., ``Wavelet2vec: a filter bank masked autoencoder for
  eeg-based seizure subtype classification,'' in \emph{ICASSP 2023-2023 IEEE
  International Conference on Acoustics, Speech and Signal Processing
  (ICASSP)}, 2023, pp. 1--5.

\bibitem{xu2023patient}
X.~Xu and et~al., ``Patient-specific method for predicting epileptic seizures
  based on drsn-gru,'' \emph{Biomedical Signal Processing and Control},
  vol.~81, p. 104449, 2023.

\bibitem{sui2019localization}
L.~Sui and et~al., ``Localization of epileptic foci by using convolutional
  neural network based on ieeg,'' in \emph{IFIP International Conference on
  Artificial Intelligence Applications and Innovations}.\hskip 1em plus 0.5em
  minus 0.4em\relax Springer, 2019, pp. 331--339.

\bibitem{worrell2000localization}
G.~A. Worrell and et~al., ``Localization of the epileptic focus by
  low-resolution electromagnetic tomography in patients with a lesion
  demonstrated by mri,'' \emph{Brain topography}, vol.~12, no.~4, pp. 273--282,
  2000.

\bibitem{lih2023epilepsynet}
O.~S. Lih and et~al., ``Epilepsynet: Novel automated detection of epilepsy
  using transformer model with eeg signals from 121 patient population,''
  \emph{Computers in Biology and Medicine}, vol. 164, p. 107312, 2023.

\bibitem{zhang2024brant}
D.~Zhang and et~al., ``Brant: Foundation model for intracranial neural
  signal,'' \emph{Advances in Neural Information Processing Systems}, vol.~36,
  2024.

\bibitem{9345750}
J.~Guo and et~al., ``Detecting high frequency oscillations for
  stereoelectroencephalography in epilepsy via hypergraph learning,''
  \emph{IEEE Transactions on Neural Systems and Rehabilitation Engineering},
  vol.~29, pp. 587--596, 2021.

\bibitem{rahmani2023meta}
A.~Rahmani and et~al., ``A meta-gnn approach to personalized seizure detection
  and classification,'' in \emph{ICASSP 2023-2023 IEEE International Conference
  on Acoustics, Speech and Signal Processing (ICASSP)}.\hskip 1em plus 0.5em
  minus 0.4em\relax IEEE, 2023, pp. 1--5.

\bibitem{sun2022continuous}
Y.~Sun and et~al., ``Continuous seizure detection based on transformer and
  long-term ieeg,'' \emph{IEEE Journal of Biomedical and Health Informatics},
  vol.~26, no.~11, pp. 5418--5427, 2022.

\bibitem{tudmnet}
S.~Tu and et~al., ``Dmnet: Self-comparison driven model for subject-independent
  seizure detection,'' in \emph{The Thirty-eighth Annual Conference on Neural
  Information Processing Systems}.

\bibitem{you2020unsupervised}
S.~You and et~al., ``Unsupervised automatic seizure detection for focal-onset
  seizures recorded with behind-the-ear eeg using an anomaly-detecting
  generative adversarial network,'' \emph{Computer Methods and Programs in
  Biomedicine}, vol. 193, p. 105472, 2020.

\bibitem{li2022spp}
X.~Li and V.~Metsis, ``Spp-eegnet: An input-agnostic self-supervised eeg
  representation model for inter-dataset transfer learning,'' in
  \emph{International Conference on Computing and Information Technology},
  2022, pp. 173--182.

\bibitem{chen2022brainnet}
J.~Chen and et~al., ``Brainnet: Epileptic wave detection from seeg with
  hierarchical graph diffusion learning,'' pp. 2741--2751, 2022.

\bibitem{yuan2024ppi}
Z.~Yuan and et~al., ``Ppi: Pretraining brain signal model for
  patient-independent seizure detection,'' \emph{Advances in Neural Information
  Processing Systems}, vol.~36, 2024.

\bibitem{kemp2000analysis}
B.~Kemp and et~al., ``{Analysis of a sleep-dependent neuronal feedback loop:
  The slow-wave microcontinuity of the EEG},'' \emph{IEEE Transactions on
  Biomedical Engineering}, vol.~47, no.~9, pp. 1185--1194, 2000.

\bibitem{oreilly2014montreal}
C.~O'Reilly and et~al., ``{Montreal Archive of Sleep Studies: An open-access
  resource for instrument benchmarking and exploratory research},''
  \emph{Journal of Sleep Research}, vol.~23, no.~6, pp. 628--635, 2014.

\bibitem{quan1997sleep}
S.~F. Quan and et~al., ``{The Sleep Heart Health Study: Design, rationale, and
  methods},'' \emph{Sleep}, vol.~20, no.~12, pp. 1077--1085, 1997.

\bibitem{zhang2018national}
G.~Q. Zhang and et~al., ``{The National Sleep Research Resource: Towards a
  sleep data commons},'' \emph{Journal of the American Medical Informatics
  Association}, vol.~25, no.~10, pp. 1351--1358, October 2018.

\bibitem{ucddb2007sleep}
``{St. Vincent's University Hospital / University College Dublin Sleep Apnea
  Database},'' 2007.

\bibitem{Alvarez-Estevez2022}
D.~Alvarez-Estevez and R.~Rijsman, ``Haaglanden medisch centrum sleep staging
  database (version 1.1),'' 2022.

\bibitem{ghassemi2018you}
M.~M. Ghassemi and et~al., ``{You Snooze, You Win: The PhysioNet/Computing in
  Cardiology Challenge 2018},'' in \emph{2018 Computing in Cardiology
  Conference (CinC)}, 2018, pp. 1--4.

\bibitem{ichimaru1999development}
Y.~Ichimaru and G.~B. Moody, ``{Development of the polysomnographic database on
  CD-ROM},'' \emph{Psychiatry and Clinical Neurosciences}, vol.~53, pp.
  175--177, April 1999.

\bibitem{dod_dataset}
A.~Guillot and et~al., ``Dreem open datasets: Multi-scored sleep datasets to
  compare human and automated sleep staging,'' \emph{IEEE Transactions on
  Neural Systems and Rehabilitation Engineering}, vol.~PP, pp. 1--1, 07 2020.

\bibitem{khalighi2016isruc}
S.~Khalighi and et~al., ``{ISRUC-Sleep: A comprehensive public dataset for
  sleep researchers},'' \emph{Computer Methods and Programs in Biomedicine},
  vol. 124, pp. 180--192, 2016.

\bibitem{biswal2018expert}
S.~Biswal and et~al., ``{Expert-level sleep scoring with deep neural
  networks},'' \emph{Journal of the American Medical Informatics Association},
  vol.~25, no.~12, pp. 1643--1650, 2018.

\bibitem{piryatinska2009automated}
A.~Piryatinska and et~al., ``Automated detection of neonate eeg sleep stages,''
  \emph{Computer methods and programs in biomedicine}, vol.~95, no.~1, pp.
  31--46, 2009.

\bibitem{devuyst2005dreams}
S.~Devuyst, ``The dreams databases and assessment algorithm,'' 2005.

\bibitem{xiang2023resting}
C.~Xiang and et~al., ``A resting-state eeg dataset for sleep deprivation,''
  \url{https://doi.org/10.18112/openneuro.ds004902.v1.0.5},
  https://doi.org/10.18112/openneuro.ds004902.v1.0.5.

\bibitem{eldele2021attention}
E.~Eldele and et~al., ``An attention-based deep learning approach for sleep
  stage classification with single-channel eeg,'' \emph{IEEE Transactions on
  Neural Systems and Rehabilitation Engineering}, vol.~29, pp. 809--818, 2021.

\bibitem{Sleep28}
K.~H and et~al., ``Accurate deep learning-based sleep staging in a clinical
  population with suspected obstructive sleep apnea,'' \emph{IEEE journal of
  biomedical and health informatics}, vol. 24(7), pp. 2073--2081, 2019.

\bibitem{yang2023self}
C.~Yang and et~al., ``Self-supervised electroencephalogram representation
  learning for automatic sleep staging: model development and evaluation
  study,'' \emph{JMIR AI}, vol.~2, no.~1, p. e46769, 2023.

\bibitem{kumar2022muleeg}
V.~Kumar and et~al., ``muleeg: a multi-view representation learning on eeg
  signals,'' in \emph{International Conference on Medical Image Computing and
  Computer-Assisted Intervention}, 2022, pp. 398--407.

\bibitem{ye2021cosleep}
J.~Ye and et~al., ``Cosleep: A multi-view representation learning framework for
  self-supervised learning of sleep stage classification,'' \emph{IEEE Signal
  Processing Letters}, vol.~29, pp. 189--193, 2021.

\bibitem{zhang2024brantx}
D.~Zhang and et~al., ``Brant-x: A unified physiological signal alignment
  framework,'' in \emph{Proceedings of the 30th ACM SIGKDD Conference on
  Knowledge Discovery and Data Mining}, 2024, pp. 4155--4166.

\bibitem{tsinalis2016automatic}
O.~Tsinalis and et~al., ``Automatic sleep stage scoring with single-channel eeg
  using convolutional neural networks,'' \emph{arXiv preprint
  arXiv:1610.01683}, 2016.

\bibitem{Mumtaz2016}
W.~Mumtaz, ``{MDD Patients and Healthy Controls EEG Data (New)},'' 11 2016.

\bibitem{cavanagh2017patient}
J.~Cavanagh and et~al., ``The patient repository for eeg data + computational
  tools (pred+ct),'' \emph{Frontiers in Neuroinformatics}, vol.~11, p.~67, Nov
  2017.

\bibitem{yang2023automatic}
L.~Yang and et~al., ``Automatic feature learning model combining functional
  connectivity network and graph regularization for depression detection,''
  \emph{Biomedical Signal Processing and Control}, vol.~82, p. 104520, 2023.

\bibitem{cai2022multi}
H.~Cai and et~al., ``A multi-modal open dataset for mental-disorder analysis,''
  \emph{Scientific Data}, vol.~9, no.~1, p. 178, 2022.

\bibitem{MUMTAZ2019103983}
W.~Mumtaz and A.~Qayyum, ``A deep learning framework for automatic diagnosis of
  unipolar depression,'' \emph{International Journal of Medical Informatics},
  vol. 132, p. 103983, 2019.

\bibitem{li2024eeg}
L.~Li and et~al., ``An eeg-based marker of functional connectivity: detection
  of major depressive disorder,'' \emph{Cognitive Neurodynamics}, vol.~18,
  no.~4, pp. 1671--1687, 2024.

\bibitem{wang2023depression}
B.~Wang, Y.~Kang, D.~Huo, D.~Chen, W.~Song, and F.~Zhang, ``Depression signal
  correlation identification from different eeg channels based on cnn feature
  extraction,'' \emph{Psychiatry Research: Neuroimaging}, vol. 328, p. 111582,
  2023.

\bibitem{thoduparambil2020eeg}
P.~P. Thoduparambil and et~al., ``Eeg-based deep learning model for the
  automatic detection of clinical depression,'' \emph{Physical and Engineering
  Sciences in Medicine}, vol.~43, no.~4, pp. 1349--1360, 2020.

\bibitem{yang2023gated}
L.~Yang and et~al., ``A gated temporal-separable attention network for
  eeg-based depression recognition,'' \emph{Computers in Biology and Medicine},
  vol. 157, p. 106782, 2023.

\bibitem{sun2023multi}
X.~Sun and et~al., ``Multi-granularity graph convolution network for major
  depressive disorder recognition,'' \emph{IEEE Transactions on Neural Systems
  and Rehabilitation Engineering}, vol.~32, pp. 559--569, 2023.

\bibitem{zhang2024novel}
Z.~Zhang and et~al., ``A novel eeg-based graph convolution network for
  depression detection: incorporating secondary subject partitioning and
  attention mechanism,'' \emph{Expert Systems with Applications}, vol. 239, p.
  122356, 2024.

\bibitem{sam2023depression}
A.~Sam and et~al., ``Depression identification using eeg signals via a hybrid
  of lstm and spiking neural networks,'' \emph{IEEE Transactions on Neural
  Systems and Rehabilitation Engineering}, vol.~31, pp. 4725--4737, 2023.

\bibitem{olejarczyk2017graph}
E.~Olejarczyk and W.~Jernajczyk, ``Graph-based analysis of brain connectivity
  in schizophrenia,'' \emph{PloS one}, vol.~12, no.~11, p. e0188629, 2017.

\bibitem{ford2014did}
J.~M. Ford and et~al., ``Did i do that? abnormal predictive processes in
  schizophrenia when button pressing to deliver a tone,'' \emph{Schizophrenia
  bulletin}, vol.~40, no.~4, pp. 804--812, 2014.

\bibitem{borisov2005analysis}
S.~Borisov and et~al., ``Analysis of eeg structural synchrony in adolescents
  with schizophrenic disorders,'' \emph{Human Physiology}, vol.~31, pp.
  255--261, 2005.

\bibitem{oh2019deep}
S.~L. Oh and et~al., ``Deep convolutional neural network model for automated
  diagnosis of schizophrenia using eeg signals,'' \emph{Applied Sciences},
  vol.~9, no.~14, p. 2870, 2019.

\bibitem{guo2021deep}
Z.~Guo and et~al., ``Deep neural network classification of eeg data in
  schizophrenia,'' in \emph{2021 IEEE 10th Data Driven Control and Learning
  Systems Conference (DDCLS)}.\hskip 1em plus 0.5em minus 0.4em\relax IEEE,
  2021, pp. 1322--1327.

\bibitem{supakar2022deep}
R.~Supakar and et~al., ``A deep learning based model using rnn-lstm for the
  detection of schizophrenia from eeg data,'' \emph{Computers in Biology and
  Medicine}, vol. 151, p. 106225, 2022.

\bibitem{khare2023schizonet}
S.~K. Khare and et~al., ``Schizonet: a robust and accurate margenau--hill
  time-frequency distribution based deep neural network model for schizophrenia
  detection using eeg signals,'' \emph{Physiological Measurement}, vol.~44,
  no.~3, p. 035005, 2023.

\bibitem{SZ22}
A.~Shalbaf and et~al., ``Transfer learning with deep convolutional neural
  network for automated detection of schizophrenia from eeg signals,''
  \emph{Physical and Engineering Sciences in Medicine}, vol.~43, pp.
  1229--1239, 2020.

\bibitem{khare2021spwvd}
S.~K. Khare and et~al., ``Spwvd-cnn for automated detection of schizophrenia
  patients using eeg signals,'' \emph{IEEE Transactions on Instrumentation and
  Measurement}, vol.~70, pp. 1--9, 2021.

\bibitem{shah2019deep}
D.~Shah and et~al., ``Deep learning of eeg data in the neucube brain-inspired
  spiking neural network architecture for a better understanding of
  depression,'' in \emph{Neural Information Processing: 26th International
  Conference, ICONIP 2019, Sydney, NSW, Australia, December 12--15, 2019,
  Proceedings, Part III 26}, 2019, pp. 195--206.

\bibitem{zalesky2011disrupted}
A.~Zalesky and et~al., ``Disrupted axonal fiber connectivity in
  schizophrenia,'' \emph{Biological psychiatry}, vol.~69, no.~1, pp. 80--89,
  2011.

\bibitem{ds_FSA_AD}
M.~L. Vicchietti and et~al., ``Data from: Computational methods of eeg signals
  analysis for alzheimer’s disease classification,'' Feb 2023.

\bibitem{ds004504:1.0.2}
A.~Miltiadous and et~al., ``"a dataset of 88 eeg recordings from: Alzheimer's
  disease, frontotemporal dementia and healthy subjects",'' 2023.

\bibitem{fiscon}
G.~Fiscon and et~al., ``In alzheimer’s disease patients classification
  through eeg signals processing,'' in \emph{Computational Intelligence \& Data
  Mining}, 2014, pp. 105--112.

\bibitem{cejnek2021novelty}
M.~Cejnek and et~al., ``Novelty detection-based approach for alzheimer’s
  disease and mild cognitive impairment diagnosis from eeg,'' \emph{Medical \&
  Biological Engineering \& Computing}, vol.~59, no.~11, pp. 2287--2296, 2021.

\bibitem{AD1}
A.~C. L and et~al., ``Eeg functional connectivity and deep learning for
  automatic diagnosis of brain disorders: Alzheimer’s disease and
  schizophrenia,'' \emph{Journal of Physics: complexity}, vol. 3(2), p. 025001,
  2022.

\bibitem{shan2022spatial}
X.~Shan and et~al., ``Spatial--temporal graph convolutional network for
  alzheimer classification based on brain functional connectivity imaging of
  electroencephalogram,'' \emph{Human Brain Mapping}, vol.~43, no.~17, pp.
  5194--5209, 2022.

\bibitem{ieracitano2019convolutional}
C.~Ieracitano and et~al., ``A convolutional neural network approach for
  classification of dementia stages based on 2d-spectral representation of eeg
  recordings,'' \emph{Neurocomputing}, vol. 323, pp. 96--107, 2019.

\bibitem{labate2014eeg}
D.~Labate and et~al., ``Eeg complexity modifications and altered
  compressibility in mild cognitive impairment and alzheimer’s disease,'' in
  \emph{Recent Advances of Neural Network Models and Applications: Proceedings
  of the 23rd Workshop of the Italian Neural Networks Society (SIREN)}, 2014,
  pp. 163--173.

\bibitem{ds002778:1.0.5}
A.~P. Rockhill and et~al., ``"uc san diego resting state eeg data from patients
  with parkinson's disease",'' 2021.

\bibitem{cavanagh2018diminished}
J.~F. Cavanagh and et~al., ``Diminished eeg habituation to novel events
  effectively classifies parkinson's patients,'' \emph{Clinical
  Neurophysiology}, vol. 129, no.~2, pp. 409--418, Feb 2018.

\bibitem{singh2020frontal}
A.~Singh and et~al., ``Frontal theta and beta oscillations during lower-limb
  movement in parkinson’s disease,'' \emph{Clinical Neurophysiology}, vol.
  131, pp. 694--702, 2020.

\bibitem{shaban2021automated}
M.~Shaban, ``Automated screening of parkinson's disease using deep learning
  based electroencephalography,'' in \emph{2021 10th international IEEE/EMBS
  conference on neural engineering (NER)}.\hskip 1em plus 0.5em minus
  0.4em\relax IEEE, 2021, pp. 158--161.

\bibitem{shaban2022resting}
M.~Shaban and A.~W. Amara, ``Resting-state electroencephalography based
  deep-learning for the detection of parkinson’s disease,'' \emph{Plos one},
  vol.~17, no.~2, p. e0263159, 2022.

\bibitem{khare2021pdcnnet}
S.~K. Khare and et~al., ``Pdcnnet: An automatic framework for the detection of
  parkinson’s disease using eeg signals,'' \emph{IEEE Sensors Journal},
  vol.~21, no.~15, pp. 17\,017--17\,024, 2021.

\bibitem{morita2011relationship}
A.~Morita and et~al., ``Relationship between slowing of the eeg and cognitive
  impairment in parkinson disease,'' \emph{Journal of Clinical
  Neurophysiology}, vol.~28, no.~4, pp. 384--387, 2011.

\bibitem{sadeghibajestani2023dataset}
G.~S. Bajestani and et~al., ``A dataset of eeg signals from adults with adhd
  and healthy controls: Resting state, cognitive function, and sound listening
  paradigm,'' 2023.

\bibitem{rzfh-zn36-20}
M.~Nasrabadi and et~al., ``Eeg data for adhd / control children,'' 2020.

\bibitem{bakhtyari2022adhd}
M.~Bakhtyari and S.~Mirzaei, ``Adhd detection using dynamic connectivity
  patterns of eeg data and convlstm with attention framework,''
  \emph{Biomedical Signal Processing and Control}, vol.~76, p. 103708, 2022.

\bibitem{nimh_adhd}
{National Institute of Mental Health}. (2023) Attention-deficit/hyperactivity
  disorder (adhd). U.S. Department of Health and Human Services.

\bibitem{chen2019use}
H.~Chen and et~al., ``Use of deep learning to detect personalized
  spatial-frequency abnormalities in eegs of children with adhd,''
  \emph{Journal of neural engineering}, vol.~16, no.~6, p. 066046, 2019.

\bibitem{eldele2021time}
E.~Eldele and et~al., ``Time-series representation learning via temporal and
  contextual contrasting,'' \emph{Proceedings of the Thirtieth International
  Joint Conference on Artificial Intelligence, IJCAI-21}, 2021.

\bibitem{zhang2022self}
X.~Zhang and et~al., ``Self-supervised contrastive pre-training for time series
  via time-frequency consistency,'' \emph{Advances in Neural Information
  Processing Systems}, vol.~35, pp. 3988--4003, 2022.

\bibitem{yang2024biot}
C.~Yang and et~al., ``Biot: Biosignal transformer for cross-data learning in
  the wild.''

\bibitem{jo2023channel}
S.~Jo and et~al., ``Channel-aware self-supervised learning for eeg-based bci,''
  in \emph{2023 11th International Winter Conference on Brain-Computer
  Interface (BCI)}.\hskip 1em plus 0.5em minus 0.4em\relax IEEE, 2023, pp.
  1--4.

\bibitem{zhang2023self}
W.~Zhang and et~al., ``Self-supervised time series representation learning via
  cross reconstruction transformer,'' \emph{IEEE Transactions on Neural
  Networks and Learning Systems}, 2023.

\bibitem{wu2024neuro}
D.~Wu and et~al., ``Neuro-bert: Rethinking masked autoencoding for
  self-supervised neurological pretraining,'' \emph{IEEE Journal of Biomedical
  and Health Informatics}, 2024.

\bibitem{wang2024cbramod}
J.~Wang and et~al., ``Cbramod: A criss-cross brain foundation model for eeg
  decoding,'' \emph{The Thirteenth International Conference on Learning
  Representations}, 2025.

\bibitem{yuan2024brainwavebrainsignalfoundation}
Z.~Yuan and et~al., ``Brainwave: A brain signal foundation model for clinical
  applications,'' 2024.

\bibitem{chen2024eegformer}
Y.~Chen and et~al., ``Eegformer: Towards transferable and interpretable
  large-scale eeg foundation model,'' in \emph{AAAI 2024 Spring Symposium on
  Clinical Foundation Models}, 2024.

\bibitem{jiang2024large}
W.-B. Jiang and et~al., ``Large brain model for learning generic
  representations with tremendous eeg data in bci,'' \emph{The Twelfth
  International Conference on Learning Representations}, 2024.

\bibitem{jiang2024neurolm}
W.~Jiang and et~al., ``Neurolm: A universal multi-task foundation model for
  bridging the gap between language and eeg signals,'' \emph{The Thirteenth
  International Conference on Learning Representations}, 2025.

\bibitem{devlin2018bert}
J.~Devlin, ``Bert: Pre-training of deep bidirectional transformers for language
  understanding,'' \emph{arXiv preprint arXiv:1810.04805}, 2018.

\bibitem{van2017neural}
A.~Van Den~Oord and et~al., ``Neural discrete representation learning,''
  \emph{Advances in neural information processing systems}, vol.~30, 2017.

\bibitem{Seizure1}
A.~M. Taqi and et~al., ``Classification and discrimination of focal and
  non-focal eeg signals based on deep neural network,'' in \emph{2017
  International Conference on Current Research in Computer Science and
  Information Technology (ICCIT)}.\hskip 1em plus 0.5em minus 0.4em\relax IEEE,
  2017, pp. 86--92.

\bibitem{Seizure3}
M.~C. Tjepkema-Cloostermans and et~al., ``Deep learning for detection of focal
  epileptiform discharges from scalp eeg recordings,'' \emph{Clinical
  Neurophysiology}, vol. 129, no.~10, pp. 2191--2196, 2018.

\bibitem{Seizure4}
S.~Madhavan and et~al., ``Time-frequency domain deep convolutional neural
  network for the classification of focal and non-focal eeg signals,''
  \emph{IEEE Sensors Journal}, vol.~20, no.~6, pp. 3078--3086, 2019.

\bibitem{Seizure5}
R.~San-Segundo and et~al., ``Classification of epileptic eeg recordings using
  signal transforms and convolutional neural networks,'' \emph{Computers in
  Biology and Medicine}, vol. 109, pp. 148--158, 2019.

\bibitem{Seizure6}
L.~Sui and et~al., ``Localization of epileptic foci by using convolutional
  neural network based on ieeg,'' in \emph{IFIP International Conference on
  Artificial Intelligence Applications and Innovations}.\hskip 1em plus 0.5em
  minus 0.4em\relax Springer, 2019, pp. 331--339.

\bibitem{Seizure9}
P.~Boonyakitanont and et~al., ``A comparison of deep neural networks for
  seizure detection in eeg signals,'' \emph{bioRxiv}, p. 702654, 2019.

\bibitem{Seizure10}
B.~Bouaziz and et~al., ``Epileptic seizure detection using a convolutional
  neural network,'' in \emph{Digital Health Approach for Predictive,
  Preventive, Personalised and Participatory Medicine}.\hskip 1em plus 0.5em
  minus 0.4em\relax Springer, 2019, pp. 79--86.

\bibitem{hossain2019applying}
M.~S. Hossain and et~al., ``Applying deep learning for epilepsy seizure
  detection and brain mapping visualization,'' \emph{ACM Transactions on
  Multimedia Computing, Communications, and Applications (TOMM)}, vol.~15,
  no.~1s, pp. 1--17, 2019.

\bibitem{Seizure12}
X.~Tian and et~al., ``Deep multi-view feature learning for eeg-based epileptic
  seizure detection,'' \emph{IEEE Transactions on Neural Systems and
  Rehabilitation Engineering}, vol.~27, no.~10, pp. 1962--1972, 2019.

\bibitem{Seizure13}
J.~Cao and et~al., ``Epileptic signal classification with deep eeg features by
  stacked cnns,'' \emph{IEEE Transactions on Cognitive and Developmental
  Systems}, vol.~12, no.~4, pp. 709--722, 2019.

\bibitem{Seizure14}
P.~Z. Yan and et~al., ``Automated spectrographic seizure detection using
  convolutional neural networks,'' \emph{Seizure}, vol.~71, pp. 124--131, 2019.

\bibitem{Seizure15}
A.~Emami and et~al., ``Seizure detection by convolutional neural network-based
  analysis of scalp electroencephalography plot images,'' \emph{NeuroImage:
  Clinical}, vol.~22, p. 101684, 2019.

\bibitem{Seizure16}
R.~Zuo and et~al., ``Automated detection of high-frequency oscillations in
  epilepsy based on a convolutional neural network,'' \emph{Frontiers in
  Computational Neuroscience}, vol.~13, p.~6, 2019.

\bibitem{Seizure17}
U.~Asif and et~al., ``Seizurenet: A deep convolutional neural network for
  accurate seizure type classification and seizure detection,'' arXiv preprint
  arXiv:1903.03232, 2019.

\bibitem{Seizure18}
N.~Ilakiyaselvan and et~al., ``Deep learning approach to detect seizure using
  reconstructed phase space images,'' \emph{Journal of Biomedical Research},
  vol.~34, no.~3, p. 240, 2020.

\bibitem{Seizure19}
W.~Mao and et~al., ``Eeg dataset classification using cnn method,'' in
  \emph{Journal of Physics: Conference Series}, vol. 1456.\hskip 1em plus 0.5em
  minus 0.4em\relax IOP Publishing, 2020, p. 012017.

\bibitem{Seizure20}
A.~Shankar and et~al., ``Epileptic seizure classification based on gramian
  angular field transformation and deep learning,'' in \emph{2020 IEEE Applied
  Signal Processing Conference (ASPCON)}.\hskip 1em plus 0.5em minus
  0.4em\relax IEEE, 2020, pp. 147--151.

\bibitem{Seizure21}
A.~Singh and et~al., ``Cnn-based epilepsy detection using image like features
  of eeg signals,'' in \emph{2020 International Conference on Electrical and
  Electronics Engineering (ICE3)}.\hskip 1em plus 0.5em minus 0.4em\relax IEEE,
  2020, pp. 280--284.

\bibitem{Seizure22}
P.~Boonyakitanont and et~al., ``Automatic epileptic seizure onset-offset
  detection based on cnn in scalp eeg,'' in \emph{ICASSP 2020-2020 IEEE
  International Conference on Acoustics, Speech and Signal Processing
  (ICASSP)}.\hskip 1em plus 0.5em minus 0.4em\relax IEEE, 2020, pp. 1225--1229.

\bibitem{Seizure23}
Y.~Gao and et~al., ``Deep convolutional neural network-based epileptic
  electroencephalogram (eeg) signal classification,'' \emph{Frontiers in
  Neurology}, vol.~11, 2020.

\bibitem{Seizure24}
F.~George and et~al., ``Epileptic seizure prediction using eeg images,'' in
  \emph{2020 International Conference on Communication and Signal Processing
  (ICCSP)}.\hskip 1em plus 0.5em minus 0.4em\relax IEEE, 2020, pp. 1595--1598.

\bibitem{Seizure25}
Y.~Qin and et~al., ``Patient-specific seizure prediction with scalp eeg using
  convolutional neural network and extreme learning machine,'' in \emph{2020
  39th Chinese Control Conference (CCC)}.\hskip 1em plus 0.5em minus
  0.4em\relax IEEE, 2020, pp. 7622--7625.

\bibitem{Seizure26}
S.~M. Usman and et~al., ``Epileptic seizures prediction using deep learning
  techniques,'' \emph{IEEE Access}, vol.~8, pp. 39\,998--40\,007, 2020.

\bibitem{Seizure27}
B.~Zhang and et~al., ``Cross-subject seizure detection in eegs using deep
  transfer learning,'' \emph{Computational and Mathematical Methods in
  Medicine}, vol. 2020, 2020.

\bibitem{Seizure28}
D.~Hu and et~al., ``Epileptic state classification by fusing hand-crafted and
  deep learning eeg features,'' \emph{IEEE Transactions on Circuits and Systems
  II: Express Briefs}, 2020.

\bibitem{Seizure29}
G.~Liu and et~al., ``Automatic seizure detection based on s-transform and deep
  convolutional neural network,'' \emph{International Journal of Neural
  Systems}, vol.~30, no.~04, p. 1950024, 2020.

\bibitem{Seizure30}
R.~Hussein and et~al., ``Epileptic seizure prediction: A semi-dilated
  convolutional neural network architecture,'' arXiv preprint arXiv:2007.11716,
  2020.

\bibitem{Seizure31}
S.~Raghu and et~al., ``Eeg based multi-class seizure type classification using
  convolutional neural network and transfer learning,'' in \emph{Neural
  Networks}, vol. 124, 2020, pp. 202--212.

\bibitem{Seizure32}
T.~Uyttenhove and et~al., ``Interpretable epilepsy detection in routine,
  interictal eeg data using deep learning,'' in \emph{Machine Learning for
  Health}.\hskip 1em plus 0.5em minus 0.4em\relax PMLR, 2020, pp. 355--366.

\bibitem{Seizure33}
M.~Rashed-Al-Mahfuz and et~al., ``A deep convolutional neural network method to
  detect seizures and characteristic frequencies using epileptic
  electroencephalogram (eeg) data,'' \emph{IEEE Journal of Translational
  Engineering in Health and Medicine}, vol.~9, pp. 1--12, 2021.

\bibitem{Seizure34}
M.~Zeng and et~al., ``Grp-dnet: A gray recurrence plot-based densely connected
  convolutional network for classification of epileptiform eeg,'' \emph{Journal
  of Neuroscience Methods}, vol. 347, p. 108953, 2021.

\bibitem{Seizure35}
T.~Luo and et~al., ``Emd-wog-2dcnn based eeg signal processing for rolandic
  seizure classification,'' \emph{Comput. Methods Biomech. Biomed. Eng.},
  vol.~25, no.~14, pp. 1565--1575, 2022.

\bibitem{Seizure36}
T.~Jagadesh and et~al., ``Early prediction of epileptic seizure using deep
  learning algorithm,'' in \emph{Brain‐Computer Interface}.\hskip 1em plus
  0.5em minus 0.4em\relax Wiley, 2023, pp. 157--177.

\bibitem{Seizure37}
T.~Kaur and T.~K. Gandhi, ``Automated diagnosis of epileptic seizures using eeg
  image representations and deep learning,'' \emph{Neuroscience Informatics},
  vol.~3, no.~3, p. 100139, 2023.

\bibitem{yuan2018novel}
Y.~Yuan and et~al., ``A novel channel-aware attention framework for
  multi-channel eeg seizure detection via multi-view deep learning,'' in
  \emph{2018 IEEE EMBS international conference on biomedical \& health
  informatics (BHI)}, 2018, pp. 206--209.

\bibitem{Seizure39}
X.~Si and et~al., ``Patient-independent seizure detection based on long-term
  ieeg and a novel lightweight cnn,'' \emph{J. Neural Eng.}, vol.~20, no.~1, p.
  016037, 2023.

\bibitem{zhang2022epileptic}
Y.~Zhang and et~al., ``Epileptic seizure detection based on bidirectional gated
  recurrent unit network,'' \emph{IEEE Transactions on Neural Systems and
  Rehabilitation Engineering}, vol.~30, pp. 135--145, 2022.

\bibitem{Seizure42}
B.~Ganti and et~al., ``Time-series generative adversarial network approach of
  deep learning improves seizure detection from the human thalamic seeg,''
  \emph{Frontiers in Neurology}, vol.~13, p. 755094, 2022.

\bibitem{Seizure43}
X.~Hu and Q.~Yuan, ``Epileptic eeg identification based on deep bi-lstm
  network,'' in \emph{2019 IEEE 11th International Conference on Advanced
  Infocomm Technology (ICAIT)}.\hskip 1em plus 0.5em minus 0.4em\relax IEEE,
  2019, pp. 63--66.

\bibitem{Seizure44}
L.~Fraiwan and M.~Alkhodari, ``Classification of focal and non-focal epileptic
  patients using single channel eeg and long short-term memory learning
  system,'' \emph{IEEE Access}, vol.~8, pp. 77\,255--77\,262, 2020.

\bibitem{Seizure45}
X.~Hu and et~al., ``Scalp eeg classification using deep bi-lstm network for
  seizure detection,'' \emph{Computers in Biology and Medicine}, vol. 124, p.
  103919, 2020.

\bibitem{Seizure46}
M.~Geng and et~al., ``Epileptic seizure detection based on stockwell transform
  and bidirectional long short-term memory,'' \emph{IEEE Transactions on Neural
  Systems and Rehabilitation Engineering}, vol.~28, no.~3, pp. 573--580, 2020.

\bibitem{Seizure47}
X.~Yao and et~al., ``A robust deep learning approach for automatic
  classification of seizures against non-seizures,'' \emph{Biomedical Signal
  Processing and Control}, vol.~64, p. 102215, 2021.

\bibitem{Seizure48}
E.~Tuncer and E.~Bolat, ``Classification of epileptic seizures from
  electroencephalogram (eeg) data using bidirectional short-term memory
  (bi-lstm) network architecture,'' \emph{Biomed. Signal Process. Control},
  vol.~73, p. 103462, 2022.

\bibitem{Seizure50}
A.~O'Shea and et~al., ``Neonatal seizure detection using convolutional neural
  networks,'' in \emph{2017 IEEE 27th International Workshop on Machine
  Learning for Signal Processing (MLSP)}.\hskip 1em plus 0.5em minus
  0.4em\relax IEEE, 2017, pp. 1--6.

\bibitem{Seizure52}
H.~G. Daoud and et~al., ``Automatic epileptic seizure detection based on
  empirical mode decomposition and deep neural network,'' in \emph{2018 IEEE
  14th international colloquium on signal processing \& its applications
  (CSPA)}.\hskip 1em plus 0.5em minus 0.4em\relax IEEE, 2018, pp. 182--186.

\bibitem{Seizure53}
I.~Ullah and et~al., ``An automated system for epilepsy detection using eeg
  brain signals based on deep learning approach,'' \emph{Expert Systems with
  Applications}, vol. 107, pp. 61--71, 2018.

\bibitem{Seizure54}
J.~Zhang and et~al., ``A new approach for classification of epilepsy eeg
  signals based on temporal convolutional neural networks,'' in \emph{2018 11th
  International Symposium on Computational Intelligence and Design (ISCID)},
  vol.~2.\hskip 1em plus 0.5em minus 0.4em\relax IEEE, 2018, pp. 80--84.

\bibitem{Seizure55}
R.~Yuvaraj and et~al., ``A deep learning scheme for automatic seizure detection
  from long-term scalp eeg,'' in \emph{2018 52nd Asilomar Conference on
  Signals, Systems, and Computers}.\hskip 1em plus 0.5em minus 0.4em\relax
  IEEE, 2018, pp. 368--372.

\bibitem{Seizure56}
J.~Thomas and et~al., ``Eeg classification via convolutional neural
  network-based interictal epileptiform event detection,'' in \emph{2018 40th
  Annual International Conference of the IEEE Engineering in Medicine and
  Biology Society (EMBC)}.\hskip 1em plus 0.5em minus 0.4em\relax IEEE, 2018,
  pp. 3148--3151.

\bibitem{Seizure57}
O.~Yıldırım and et~al., ``A deep convolutional neural network model for
  automated identification of abnormal eeg signals,'' \emph{Neural Computing
  and Applications}, pp. 1--12, 2018.

\bibitem{Seizure58}
R.~Akut, ``Wavelet based deep learning approach for epilepsy detection,''
  \emph{Health Information Science and Systems}, vol.~7, no.~1, pp. 1--9, 2019.

\bibitem{Seizure59}
D.~Lu and J.~Triesch, ``Residual deep convolutional neural network for eeg
  signal classification in epilepsy,'' \emph{arXiv preprint arXiv:1903.08100},
  2019.

\bibitem{Seizure60}
Z.~Wei and et~al., ``Automatic epileptic eeg detection using convolutional
  neural network with improvements in time-domain,'' \emph{Biomedical Signal
  Processing and Control}, vol.~53, p. 101551, 2019.

\bibitem{Seizure61}
J.~Craley and et~al., ``Integrating convolutional neural networks and
  probabilistic graphical modeling for epileptic seizure detection in
  multichannel eeg,'' in \emph{International Conference on Information
  Processing in Medical Imaging}.\hskip 1em plus 0.5em minus 0.4em\relax
  Springer, 2019, pp. 291--303.

\bibitem{Seizure62}
A.~H. Ansari and et~al., ``Neonatal seizure detection using deep convolutional
  neural networks,'' \emph{International Journal of Neural Systems}, vol.~29,
  no.~04, p. 1850011, 2019.

\bibitem{Seizure63}
M.~T. Avcu and et~al., ``Seizure detection using least eeg channels by deep
  convolutional neural network,'' in \emph{2019 IEEE International Conference
  on Acoustics, Speech and Signal Processing (ICASSP)}.\hskip 1em plus 0.5em
  minus 0.4em\relax IEEE, 2019, pp. 1120--1124.

\bibitem{Seizure64}
K.~Fukumori and et~al., ``Fully data-driven convolutional filters with deep
  learning models for epileptic spike detection,'' in \emph{2019 IEEE
  International Conference on Acoustics, Speech and Signal Processing
  (ICASSP)}.\hskip 1em plus 0.5em minus 0.4em\relax IEEE, 2019, pp. 2772--2776.

\bibitem{Seizure65}
S.~Wang and et~al., ``Time-resnext for epilepsy recognition based on eeg
  signals in wireless networks,'' \emph{EURASIP Journal on Wireless
  Communications and Networking}, vol. 2020, no.~1, pp. 1--12, 2020.

\bibitem{Seizure66}
X.~Zhao and et~al., ``Classification of epileptic ieeg signals by cnn and data
  augmentation,'' in \emph{2020 IEEE International Conference on Acoustics,
  Speech and Signal Processing (ICASSP)}.\hskip 1em plus 0.5em minus
  0.4em\relax IEEE, 2020, pp. 926--930.

\bibitem{Seizure68}
X.~Gao and et~al., ``Automatic detection of epileptic seizure based on
  approximate entropy, recurrence quantification analysis and convolutional
  neural networks,'' \emph{Artificial Intelligence in Medicine}, vol. 102, p.
  101711, 2020.

\bibitem{Seizure69}
D.~Kaya, ``The mrmr-cnn based influential support decision system approach to
  classify eeg signals,'' \emph{Measurement}, vol. 156, p. 107602, 2020.

\bibitem{Seizure70}
J.~Lian and et~al., ``Pair-wise matching of eeg signals for epileptic
  identification via convolutional neural network,'' \emph{IEEE Access},
  vol.~8, pp. 40\,008--40\,017, 2020.

\bibitem{Seizure71}
H.~Qin and et~al., ``Deep multi-scale feature fusion convolutional neural
  network for automatic epilepsy detection using eeg signals,'' in \emph{2020
  39th Chinese Control Conference (CCC)}.\hskip 1em plus 0.5em minus
  0.4em\relax IEEE, 2020, pp. 7061--7066.

\bibitem{Seizure72}
A.~Torfi and E.~A. Fox, ``Corgan: Correlation-capturing convolutional
  generative adversarial networks for generating synthetic healthcare
  records,'' \emph{arXiv e-prints}, pp. arXiv--2001, 2020.

\bibitem{Seizure73}
G.~Zhang and et~al., ``Mnl-network: A multi-scale non-local network for
  epilepsy detection from eeg signals,'' \emph{Frontiers in Neuroscience},
  vol.~14, 2020.

\bibitem{Seizure74}
W.~Zhao and et~al., ``Seizurenet: A model for robust detection of epileptic
  seizures based on convolutional neural network,'' \emph{Cognitive Computation
  and Systems}, vol.~2, no.~3, pp. 119--124, 2020.

\bibitem{Seizure75}
------, ``A novel deep neural network for robust detection of seizures using
  eeg signals,'' \emph{Computational and Mathematical Methods in Medicine},
  vol. 2020, p. Article 2020, 2020.

\bibitem{Seizure76}
R.~Abiyev and et~al., ``Identification of epileptic eeg signals using
  convolutional neural networks,'' \emph{Applied Sciences}, vol.~10, no.~12, p.
  4089, 2020.

\bibitem{Seizure77}
Y.~Li and et~al., ``Epileptic seizure detection in eeg signals using a unified
  temporal-spectral squeeze-and-excitation network,'' \emph{IEEE Transactions
  on Neural Systems and Rehabilitation Engineering}, vol.~28, no.~4, pp.
  782--794, 2020.

\bibitem{Seizure78}
H.~Daoud and et~al., ``Iot based efficient epileptic seizure prediction system
  using deep learning,'' in \emph{2020 IEEE 6th World Forum on Internet of
  Things (WF-IoT)}.\hskip 1em plus 0.5em minus 0.4em\relax IEEE, 2020, pp.
  1--6.

\bibitem{Seizure79}
G.~C. Jana and et~al., ``A cnn-spectrogram based approach for seizure detection
  from eeg signal,'' \emph{Procedia Computer Science}, vol. 167, pp. 403--412,
  2020.

\bibitem{Seizure80}
O.~Kaziha and T.~Bonny, ``A convolutional neural network for seizure
  detection,'' in \emph{2020 Advances in Science and Engineering Technology
  International Conferences (ASET)}.\hskip 1em plus 0.5em minus 0.4em\relax
  IEEE, 2020, pp. 1--5.

\bibitem{Seizure81}
S.~Khalilpour and et~al., ``Application of 1-d cnn to predict epileptic
  seizures using eeg records,'' in \emph{2020 6th International Conference on
  Web Research (ICWR)}.\hskip 1em plus 0.5em minus 0.4em\relax IEEE, 2020, pp.
  314--318.

\bibitem{Seizure82}
R.~V. Sharan and S.~Berkovsky, ``Epileptic seizure detection using multichannel
  eeg wavelet power spectra and 1-d convolutional neural networks,'' in
  \emph{2020 42nd Annual International Conference of the IEEE Engineering in
  Medicine \& Biology Society (EMBC)}.\hskip 1em plus 0.5em minus 0.4em\relax
  IEEE, 2020, pp. 545--548.

\bibitem{Seizure83}
A.~H. Thomas and et~al., ``Noise-resilient and interpretable epileptic seizure
  detection,'' in \emph{2020 IEEE International Symposium on Circuits and
  Systems (ISCAS)}.\hskip 1em plus 0.5em minus 0.4em\relax IEEE, 2020, pp.
  1--5.

\bibitem{Seizure84}
S.~Zhao and et~al., ``Binary single-dimensional convolutional neural network
  for seizure prediction,'' in \emph{2020 IEEE International Symposium on
  Circuits and Systems (ISCAS)}.\hskip 1em plus 0.5em minus 0.4em\relax IEEE,
  2020, pp. 1--5.

\bibitem{Seizure85}
M.~Abou~Jaoude and et~al., ``Detection of mesial temporal lobe epileptiform
  discharges on intracranial electrodes using deep learning,'' \emph{Clinical
  Neurophysiology}, vol. 131, no.~1, pp. 133--141, 2020.

\bibitem{Seizure86}
L.-C. Lin and et~al., ``Alternative diagnosis of epilepsy in children without
  epileptiform discharges using deep convolutional neural networks,''
  \emph{International Journal of Neural Systems}, vol.~30, no.~05, p. 1850060,
  2020.

\bibitem{Seizure87}
F.~Pisano and et~al., ``Convolutional neural network for seizure detection of
  nocturnal frontal lobe epilepsy,'' \emph{Complexity}, 2020.

\bibitem{Seizure88}
T.~Sakai and et~al., ``Scalpnet: Detection of spatiotemporal abnormal intervals
  in epileptic eeg using convolutional neural networks,'' in \emph{ICASSP
  2020-2020 IEEE International Conference on Acoustics, Speech and Signal
  Processing (ICASSP)}.\hskip 1em plus 0.5em minus 0.4em\relax IEEE, 2020, pp.
  1244--1248.

\bibitem{Seizure89}
J.~Thomas and et~al., ``Automated detection of interictal epileptiform
  discharges from scalp electroencephalograms by convolutional neural
  networks,'' \emph{International Journal of Neural Systems}, vol.~30, no.~11,
  p. 2050030, 2020.

\bibitem{Seizure90}
C.~Vance and et~al., ``Learning to detect the onset of slow activity after a
  generalized tonic–clonic seizure,'' \emph{BMC Medical Informatics and
  Decision Making}, vol.~20, no.~12, pp. 1--8, 2020.

\bibitem{Seizure91}
K.-O. Cho and H.-J. Jang, ``Comparison of different input modalities and
  network structures for deep learning-based seizure detection,''
  \emph{Scientific Reports}, vol.~10, no.~1, pp. 1--11, 2020.

\bibitem{Seizure92}
Y.~Xu and et~al., ``An end-to-end deep learning approach for epileptic seizure
  prediction,'' in \emph{2020 2nd IEEE International Conference on Artificial
  Intelligence Circuits and Systems (AICAS)}.\hskip 1em plus 0.5em minus
  0.4em\relax IEEE, 2020, pp. 266--270.

\bibitem{Seizure93}
T.~Ieˇsmantas and R.~Alzbutas, ``Convolutional neural network for detection
  and classification of seizures in clinical data,'' \emph{Medical \&
  Biological Engineering \& Computing}, vol.~58, no.~9, pp. 1919--1932, 2020.

\bibitem{Seizure94}
X.~Zhang and et~al., ``Adversarial representation learning for robust
  patient-independent epileptic seizure detection,'' \emph{IEEE Journal of
  Biomedical and Health Informatics}, vol.~24, no.~10, pp. 2852--2859, 2020.

\bibitem{Seizure95}
S.~Ramakrishnan and et~al., ``Seizure detection with local binary pattern and
  cnn classifier,'' in \emph{Journal of Physics: Conference Series}, vol.
  1767.\hskip 1em plus 0.5em minus 0.4em\relax IOP Publishing, 2021, p. 012029.

\bibitem{Seizure96}
K.~Singh and J.~Malhotra, ``Prediction of epileptic seizures from spectral
  features of intracranial eeg recordings using deep learning approach,''
  \emph{Multimed. Tools Appl.}, vol.~81, no.~20, pp. 28\,875--28\,898, Aug
  2022.

\bibitem{Seizure97}
D.~K. Atal and M.~Singh, ``Effectual seizure detection using mbbf-gpso with cnn
  network,'' \emph{Cogn. Neurodyn}, pp. 1--12, 2023.

\bibitem{Seizure98}
Y.~Zaid, M.~Sah, and C.~Direkoglu, ``Pre-processed and combined eeg data for
  epileptic seizure classification using deep learning,'' \emph{Biomedical
  Signal Processing and Control}, vol.~84, p. 104738, 2023.

\bibitem{assali2023cnn}
I.~Assali and et~al., ``Cnn-based classification of epileptic states for
  seizure prediction using combined temporal and spectral features,''
  \emph{Biomedical Signal Processing and Control}, vol.~82, p. 104519, 2023.

\bibitem{Seizure100}
S.~Mekruksavanich and A.~Jitpattanakul, ``Deep learning approaches for
  epileptic seizures recognition based on eeg signal,'' in \emph{46th
  International Conference on Telecommunications and Signal Processing
  (TSP)}.\hskip 1em plus 0.5em minus 0.4em\relax IEEE, 2023, pp. 33--36.

\bibitem{Seizure101}
D.~Raab, A.~Theissler, and M.~Spiliopoulou, ``Xai4eeg: Spectral and
  spatio-temporal explanation of deep learning-based seizure detection in eeg
  time series,'' \emph{Neural Comput. Appl.}, vol.~35, no.~14, pp.
  10\,051--10\,068, May 2023.

\bibitem{Seizure102}
T.~Prasanth and et~al., ``Deep learning for interictal epileptiform spike
  detection from scalp eeg frequency sub bands,'' in \emph{Annual International
  Conference of the IEEE Engineering in Medicine \& Biology Society
  (EMBC)}.\hskip 1em plus 0.5em minus 0.4em\relax IEEE, 2020, pp. 3703--3706.

\bibitem{Seizure103}
S.~Poorani and P.~Balasubramanie, ``Deep learning based epileptic seizure
  detection with eeg data,'' \emph{Int. J. Syst. Assur. Eng. Manag.}, pp.
  1--10, 2023.

\bibitem{Seizure104}
X.~Liu and A.~G. Richardson, ``Embedded deep learning for neural implants,''
  \emph{arXiv preprint arXiv:2012.00307}, 2020.

\bibitem{Seizure105}
X.~Chen, J.~Ji, T.~Ji, and P.~Li, ``Cost-sensitive deep active learning for
  epileptic seizure detection,'' in \emph{ACM International Conference on
  Bioinformatics, Computational Biology, and Health Informatics}, 2018, pp.
  226--235.

\bibitem{Seizure106}
Y.~Yuan and K.~Jia, ``Fusionatt: Deep fusional attention networks for
  multi-channel biomedical signals,'' \emph{Sensors}, vol.~19, no.~11, p. 2429,
  2019.

\bibitem{Seizure107}
D.~Y. Isaev and et~al., ``Attention-based network for weak labels in neonatal
  seizure detection,'' in \emph{Proc. Mach. Learn. Res.}, vol. 126, 2020, p.
  479.

\bibitem{Seizure108}
M.~Natu and et~al., ``Hcla\_cbigru: Hybrid convolutional bidirectional gru
  based model for epileptic seizure detection,'' \emph{Neurosci. Inform.}, p.
  100135, 2023.

\bibitem{Seizure110}
J.~Craley, E.~Johnson, C.~Jouny, and A.~Venkataraman, ``Automated inter-patient
  seizure detection using multichannel convolutional and recurrent neural
  networks,'' \emph{Biomed. Signal Process. Control}, vol.~64, p. 102360, 2021.

\bibitem{Seizure111}
I.~Ahmad and et~al., ``A hybrid deep learning approach for epileptic seizure
  detection in eeg signals,'' \emph{IEEE Journal of Biomedical and Health
  Informatics}, 2023.

\bibitem{Seizure112}
Y.~Wang and et~al., ``Seeg-net: An explainable and deep learning-based
  cross-subject pathological activity detection method for drug-resistant
  epilepsy,'' \emph{Computers in Biology and Medicine}, vol. 148, p. 105703,
  2022.

\bibitem{Seizure114}
F.~A. Jibon and et~al., ``Epileptic seizure detection from electroencephalogram
  (eeg) signals using linear graph convolutional network and densenet based
  hybrid framework,'' \emph{J. Radiat. Res. Appl. Sci.}, vol.~16, no.~3, p.
  100607, 2023.

\bibitem{Seizure115}
S.~Roy and et~al., ``Chrononet: A deep recurrent neural network for abnormal
  eeg identification,'' in \emph{Conference on Artificial Intelligence in
  Medicine in Europe}.\hskip 1em plus 0.5em minus 0.4em\relax Springer, 2019,
  pp. 47--56.

\bibitem{Seizure116}
L.~Tang and et~al., ``Seizure prediction using multiview features and improved
  convolutional gated recurrent network,'' \emph{IEEE Access}, vol.~8, pp.
  172\,352--172\,361, 2020.

\bibitem{Seizure117}
S.~P. Kumar and et~al., ``Automatic detection of epilepsy using cnn-gru hybrid
  model,'' in \emph{Biomed. Signals Based Comput.-Aided Diagn. Neurol.
  Disord.}\hskip 1em plus 0.5em minus 0.4em\relax Springer, 2022, pp. 165--186.

\bibitem{Seizure119}
P.~Thodoroff and et~al., ``Learning robust features using deep learning for
  automatic seizure detection,'' in \emph{Machine Learning for Healthcare
  Conference}.\hskip 1em plus 0.5em minus 0.4em\relax PMLR, 2016, pp. 178--190.

\bibitem{Seizure120}
M.~Golmohammadi and et~al., ``Deep architectures for automated seizure
  detection in scalp eegs,'' \emph{arXiv preprint arXiv:1712.09776}, 2017.

\bibitem{Seizure121}
U.~B. Baloglu and et~al., ``Convolutional long-short term memory networks model
  for long duration eeg signal classification,'' \emph{J. Mech. Med. Biol.},
  vol.~19, no.~01, p. 1940005, 2019.

\bibitem{Seizure122}
Y.~Liu and et~al., ``Deep c-lstm neural network for epileptic seizure and tumor
  detection using high-dimension eeg signals,'' \emph{IEEE Access}, vol.~8, pp.
  37\,495--37\,504, 2020.

\bibitem{Seizure123}
G.~Xu and et~al., ``A one-dimensional cnn-lstm model for epileptic seizure
  recognition using eeg signal analysis,'' \emph{Frontiers in Neuroscience},
  vol.~14, p. 1253, 2020.

\bibitem{Seizure124}
Y.~Li and et~al., ``Automatic seizure detection using fully convolutional
  nested lstm,'' \emph{International Journal of Neural Systems}, vol.~30,
  no.~04, p. 2050019, 2020.

\bibitem{Seizure125}
W.~Liang and et~al., ``Scalp eeg epileptogenic zone recognition and
  localization based on long-term recurrent convolutional network,''
  \emph{Neurocomputing}, vol. 396, pp. 569--576, 2020.

\bibitem{Seizure126}
W.~Hussain and et~al., ``Epileptic seizure detection using 1d-convolutional
  long short-term memory neural networks,'' \emph{Appl. Acoust.}, vol. 177, p.
  107941, 2021.

\bibitem{Seizure127}
Y.~Yang and et~al., ``Video-based detection of generalized tonic-clonic
  seizures using deep learning,'' \emph{IEEE Journal of Biomedical and Health
  Informatics}, 2021.

\bibitem{Seizure128}
M.~Varlı and et~al., ``Multiple classification of eeg signals and epileptic
  seizure diagnosis with combined deep learning,'' \emph{J. Comput. Sci.},
  vol.~67, p. 101943, 2023.

\bibitem{Seizure129}
Y.~P. Singh and et~al., ``Automatic prediction of epileptic seizure using
  hybrid deep resnet-lstm model,'' \emph{AI Commun.}, vol.~36, no.~1, pp.
  57--72, Feb 2023.

\bibitem{Seizure130}
X.~Qiu and et~al., ``A difference attention resnet-lstm network for epileptic
  seizure detection using eeg signal,'' \emph{Biomed. Signal Process. Control},
  vol.~83, p. 104652, 2023.

\bibitem{Seizure131}
S.~Roy and et~al., ``Deep learning enabled automatic abnormal eeg
  identification,'' in \emph{2018 40th Annual International Conference of the
  IEEE Engineering in Medicine and Biology Society (EMBC)}.\hskip 1em plus
  0.5em minus 0.4em\relax IEEE, 2018, pp. 2756--2759.

\bibitem{Seizure132}
C.~Huang and et~al., ``Automatic epileptic seizure detection via
  attention-based cnn-birnn,'' in \emph{2019 IEEE International Conference on
  Bioinformatics and Biomedicine (BIBM)}.\hskip 1em plus 0.5em minus
  0.4em\relax IEEE, 2019, pp. 660--663.

\bibitem{Seizure133}
D.~Kostas and et~al., ``Bendr: Using transformers and a contrastive
  self-supervised learning task to learn from massive amounts of eeg data,''
  \emph{Frontiers in Human Neuroscience}, vol.~15, p. 653659, 2021.

\bibitem{Seizure134}
P.~Busia and et~al., ``Eegformer: Transformer-based epilepsy detection on raw
  eeg traces for low-channel-count wearable continuous monitoring devices,'' in
  \emph{2022 IEEE Biomedical Circuits and Systems Conference (BioCAS)}.\hskip
  1em plus 0.5em minus 0.4em\relax IEEE, 2022, pp. 640--644.

\bibitem{Seizure136}
S.~Hu and et~al., ``Exploring the applicability of transfer learning and
  feature engineering in epilepsy prediction using hybrid transformer model,''
  \emph{IEEE Transactions on Neural Systems and Rehabilitation Engineering},
  vol.~31, pp. 1321--1332, 2023.

\bibitem{Seizure137}
Z.~Deng and et~al., ``Eeg-based seizure prediction via hybrid vision
  transformer and data uncertainty learning,'' \emph{Engineering Applications
  of Artificial Intelligence}, vol. 123, p. 106401, 2023.

\bibitem{Seizure138}
W.~Y. Peh and et~al., ``Six-center assessment of cnn-transformer with belief
  matching loss for patient-independent seizure detection in eeg,''
  \emph{International Journal of Neural Systems}, vol.~33, no.~03, p. 2350012,
  2023.

\bibitem{Seizure139}
C.~Dong and et~al., ``Eeg-based patient-specific seizure prediction based on
  spatial–temporal hypergraph attention transformer,'' \emph{Biomedical
  Signal Processing and Control}, vol. 100, p. 107075, 2025.

\bibitem{Seizure140}
Y.~Sun and et~al., ``Multi-task transformer network for subject-independent
  ieeg seizure detection,'' \emph{Expert Systems with Applications}, p. 126282,
  2024.

\bibitem{Seizure141}
T.~X. Le and et~al., ``Deep learning for epileptic spike detection,'' \emph{VNU
  Journal of Science: Computer Science and Communication Engineering}, vol.~33,
  no.~2, pp. 1--13, 2018.

\bibitem{Seizure142}
K.~P. Thanaraj and et~al., ``Implementation of deep neural networks to classify
  eeg signals using gramian angular summation field for epilepsy diagnosis,''
  \emph{arXiv preprint arXiv:2003.04534}, 2020.

\bibitem{Seizure143}
K.~Akyol, ``Stacking ensemble based deep neural networks modeling for effective
  epileptic seizure detection,'' \emph{Expert Systems with Applications}, vol.
  148, p. 113239, 2020.

\bibitem{Seizure144}
A.~Guha and et~al., ``Epileptic seizure recognition using deep neural
  network,'' in \emph{Emerging Technology in Modelling and Graphics}.\hskip 1em
  plus 0.5em minus 0.4em\relax Springer, 2020, pp. 21--28.

\bibitem{Seizure145}
R.~Sharma and et~al., ``Seizures classification based on higher order
  statistics and deep neural network,'' \emph{Biomedical Signal Processing and
  Control}, vol.~59, p. 101921, 2020.

\bibitem{Seizure146}
H.~A. Glory and et~al., ``Ahw-bgoa-dnn: A novel deep learning model for
  epileptic seizure detection,'' \emph{Neural Computing and Applications}, pp.
  1--29, 2020.

\bibitem{Seizure147}
Z.~Zhang and et~al., ``Dwt-net: Seizure detection system with structured eeg
  montage and multiple feature extractor in convolution neural network,''
  \emph{Journal of Sensors}, 2020.

\bibitem{Seizure148}
Y.~Zhao and et~al., ``Graph attention network with focal loss for seizure
  detection on electroencephalography signals,'' \emph{International Journal of
  Neural Systems}, vol.~31, no.~07, p. 2150027, 2021.

\bibitem{Seizure149}
Y.~Wang and et~al., ``A spatiotemporal graph attention network based on
  synchronization for epileptic seizure prediction,'' \emph{IEEE Journal of
  Biomedical and Health Informatics}, vol.~27, no.~2, pp. 900--911, 2023.

\bibitem{Seizure150}
Y.~Zhao and et~al., ``Hybrid attention network for epileptic eeg
  classification,'' \emph{International Journal of Neural Systems}, vol.~33,
  no.~06, p. 2350031, 2023.

\bibitem{Seizure151}
J.~He and et~al., ``Spatial–temporal seizure detection with graph attention
  network and bi-directional lstm architecture,'' \emph{Biomedical Signal
  Processing and Control}, vol.~78, p. 103908, 2022.

\bibitem{Seizure152}
X.~Chen and et~al., ``Epilepsy classification for mining deeper relationships
  between eeg channels based on gcn,'' in \emph{2020 International Conference
  on Computer Vision, Image and Deep Learning (CVIDL)}.\hskip 1em plus 0.5em
  minus 0.4em\relax IEEE, 2020, pp. 701--706.

\bibitem{Seizure153}
J.~Wang and et~al., ``A sequential graph convolutional network with
  frequency-domain complex network of eeg signals for epilepsy detection,'' in
  \emph{2020 IEEE International Conference on Bioinformatics and Biomedicine
  (BIBM)}.\hskip 1em plus 0.5em minus 0.4em\relax IEEE, 2020, pp. 785--792.

\bibitem{zhao2021eeg}
Y.~Zhao and et~al., ``Eeg-based seizure detection using linear graph
  convolution network with focal loss,'' \emph{Computer methods and programs in
  biomedicine}, vol. 208, p. 106277, 2021.

\bibitem{Seizure155}
D.~Nhu and et~al., ``Graph convolutional network for generalized epileptiform
  abnormality detection on eeg,'' in \emph{2021 IEEE Signal Processing in
  Medicine and Biology Symposium (SPMB)}.\hskip 1em plus 0.5em minus
  0.4em\relax IEEE, 2021, pp. 1--6.

\bibitem{Seizure156}
J.~Lian and F.~Xu, ``Spatial enhanced pattern through graph convolutional
  neural network for epileptic eeg identification,'' \emph{International
  Journal of Neural Systems}, vol.~32, no.~09, p. 2250033, 2022.

\bibitem{Seizure157}
C.~Dong and et~al., ``Attention-based graph resnet with focal loss for
  epileptic seizure detection,'' \emph{Journal of Ambient Intelligence and
  Smart Environments}, vol.~14, no.~1, pp. 61--73, 2022.

\bibitem{Seizure158}
Y.~Wang and et~al., ``Dynamic multi-graph convolution based channel-weighted
  transformer feature fusion network for epileptic seizure prediction,''
  \emph{IEEE Transactions on Neural Systems and Rehabilitation Engineering},
  2023.

\bibitem{Seizure160}
J.~Lian and F.~Xu, ``Epileptic eeg classification via graph transformer
  network,'' \emph{International Journal of Neural Systems}, vol.~33, no.~8, p.
  2350042, 2023.

\bibitem{Seizure161}
S.~S. Talathi, ``Deep recurrent neural networks for seizure detection and early
  seizure detection systems,'' arXiv preprint arXiv:1706.03283, 2017.

\bibitem{Seizure162}
A.~Verma and R.~R. Janghel, ``Epileptic seizure detection using deep recurrent
  neural networks in eeg signals,'' in \emph{Advances in Biomedical Engineering
  and Technology}.\hskip 1em plus 0.5em minus 0.4em\relax Springer, 2021, pp.
  189--198.

\bibitem{Seizure163}
O.~Ramwala and et~al., ``Gru-based parameter-efficient epileptic seizure
  detection,'' \emph{Biomedical Signal Processing and Artificial Intelligence},
  pp. 73--86, 2023.

\bibitem{Seizure164}
R.~Hussein and et~al., ``Robust detection of epileptic seizures using deep
  neural networks,'' in \emph{2018 IEEE International Conference on Acoustics,
  Speech and Signal Processing (ICASSP)}.\hskip 1em plus 0.5em minus
  0.4em\relax IEEE, 2018, pp. 2546--2550.

\bibitem{Seizure165}
D.~Ahmedt-Aristizabal and et~al., ``Deep classification of epileptic signals,''
  in \emph{2018 40th Annual International Conference of the IEEE}, 2018.

\bibitem{Seizure166}
R.~Hussein and et~al., ``Optimized deep neural network architecture for robust
  detection of epileptic seizures using eeg signals,'' \emph{Clinical
  Neurophysiology}, vol. 130, no.~1, pp. 25--37, 2019.

\bibitem{Seizure167}
M.~U. Abbasi and et~al., ``Detection of epilepsy seizures in neo-natal eeg
  using lstm architecture,'' \emph{IEEE Access}, vol.~7, pp.
  179\,074--179\,085, 2019.

\bibitem{Seizure169}
I.~Aliyu and C.~Lim, ``Selection of optimal wavelet features for epileptic eeg
  signal classification with lstm,'' \emph{Neural Computing and Applications},
  pp. 1--21, 2021.

\bibitem{Seizure170}
K.~Singh and J.~Malhotra, ``Two-layer lstm network-based prediction of
  epileptic seizures using eeg spectral features,'' \emph{Complex Intelligence
  Systems}, vol.~8, no.~3, pp. 2405--2418, Jun 2022.

\bibitem{Seizure171}
E.~Tuncer and E.~Bolat, ``Channel based epilepsy seizure type detection from
  electroencephalography (eeg) signals with machine learning techniques,''
  \emph{Biocybernetics and Biomedical Engineering}, vol.~42, no.~2, pp.
  575--595, 2022.

\bibitem{Seizure172}
A.~Pandey and et~al., ``An intelligent optimized deep learning model to achieve
  early prediction of epileptic seizures,'' \emph{Biomedical Signal Processing
  and Control}, vol.~84, p. 104798, Jul 2023.

\bibitem{Seizure173}
L.~Vidyaratne and et~al., ``Deep recurrent neural network for seizure
  detection,'' in \emph{2016 International Joint Conference on Neural Networks
  (IJCNN)}.\hskip 1em plus 0.5em minus 0.4em\relax IEEE, 2016, pp. 1202--1207.

\bibitem{Seizure174}
X.~Yao, Q.~Cheng, and G.-Q. Zhang, ``Automated classification of seizures
  against nonseizures: A deep learning approach,'' arXiv preprint
  arXiv:1906.02745, 2019.

\bibitem{Seizure175}
X.~Yao and et~al., ``A novel independent rnn approach to classification of
  seizures against non-seizures,'' arXiv preprint arXiv:1903.09326, 2019.

\bibitem{Seizure176}
Y.~Singh and D.~Lobiyal, ``A comparative study of deep learning algorithms for
  epileptic seizure classification,'' in \emph{2022 International Conference on
  Computing, Communication Security and Intelligent Systems (IC3SIS)}.\hskip
  1em plus 0.5em minus 0.4em\relax IEEE, 2022, pp. 1--6.

\bibitem{Seizure177}
R.~Chiranjeevi and et~al., ``Identification of epileptic seizures using
  recurrent neural networks and time series transformer,'' in \emph{2024 7th
  International Conference on Circuit Power and Computing Technologies
  (ICCPCT)}.\hskip 1em plus 0.5em minus 0.4em\relax IEEE, 2024, pp. 1546--1553.

\bibitem{Seizure178}
R.~Zhu and et~al., ``Epileptic seizure prediction via multidimensional
  transformer and recurrent neural network fusion,'' \emph{Journal of
  Translational Medicine}, vol.~22, no.~1, p. 895, 2024.

\bibitem{Seizure180}
I.~C. Covert and et~al., ``Temporal graph convolutional networks for automatic
  seizure detection,'' in \emph{Machine Learning for Healthcare
  Conference}.\hskip 1em plus 0.5em minus 0.4em\relax PMLR, 2019, pp. 160--180.

\bibitem{Seizure181}
J.~Pedoeem and et~al., \emph{Tabs: Transformer based Seizure Detection}.\hskip
  1em plus 0.5em minus 0.4em\relax Cham: Springer International Publishing,
  2022.

\bibitem{Seizure182}
Y.~Ma and et~al., ``Tsd: Transformers for seizure detection,'' \emph{bioRxiv},
  p. 2023.01.24.525308, 2023.

\bibitem{meisel2019identifying}
C.~Meisel and K.~A. Bailey, ``Identifying signal-dependent information about
  the preictal state: A comparison across ecog, eeg and ekg using deep
  learning,'' \emph{EBioMedicine}, vol.~45, pp. 422--431, 2019.

\bibitem{guo2021detecting}
J.~Guo and et~al., ``Detecting high frequency oscillations for
  stereoelectroencephalography in epilepsy via hypergraph learning,''
  \emph{IEEE Transactions on Neural Systems and Rehabilitation Engineering},
  vol.~29, pp. 587--596, 2021.

\bibitem{wagh2021domain}
N.~Wagh and et~al., ``Domain-guided self-supervision of eeg data improves
  downstream classification performance and generalizability,'' in
  \emph{Machine Learning for Health}, 2021, pp. 130--142.

\bibitem{Seizure193}
A.~Gogna and et~al., ``Semi-supervised stacked label consistent autoencoder for
  reconstruction and analysis of biomedical signals,'' \emph{IEEE Transactions
  on Biomedical Engineering}, vol.~64, no.~9, pp. 2196--2205, 2016.

\bibitem{Seizure194}
C.~Park and et~al., ``Epileptic seizure detection for multi-channel eeg with
  deep convolutional neural network,'' in \emph{2018 International Conference
  on Electronics, Information, and Communication (ICEIC)}.\hskip 1em plus 0.5em
  minus 0.4em\relax IEEE, 2018, pp. 1--5.

\bibitem{barry2021high}
W.~Barry, S.~Arcot~Desai, T.~K. Tcheng, and M.~J. Morrell, ``A high accuracy
  electrographic seizure classifier trained using semi-supervised labeling
  applied to a large spectrogram dataset,'' \emph{Frontiers in neuroscience},
  vol.~15, p. 667373, 2021.

\bibitem{Seizure195}
Y.~Yuan and et~al., ``A multi-view deep learning framework for eeg seizure
  detection,'' \emph{IEEE Journal of Biomedical and Health Informatics},
  vol.~23, no.~1, pp. 83--94, 2018.

\bibitem{Seizure196}
P.~N. Bhagat and et~al., ``Robust prior stage epileptic seizure diagnosis
  system using resnet and backpropagation techniques,'' \emph{International
  Journal}, vol.~8, no.~5, 2020.

\bibitem{Seizure199}
N.~D. Truong and et~al., ``Epileptic seizure forecasting with generative
  adversarial networks,'' \emph{IEEE Access}, vol.~7, pp. 143\,999--144\,009,
  2019.

\bibitem{Seizure201}
H.~Takahashi and et~al., ``Convolutional neural network with
  autoencoder-assisted multiclass labelling for seizure detection based on
  scalp electroencephalography,'' \emph{Computers in Biology and Medicine},
  vol. 125, p. 104016, 2020.

\bibitem{Seizure202}
A.~Shoeibi and et~al., ``A comprehensive comparison of handcrafted features and
  convolutional autoencoders for epileptic seizures detection in eeg signals,''
  \emph{Expert Systems with Applications}, vol. 163, p. 113788, 2021.

\bibitem{Seizure203}
D.~Wulsin and et~al., ``Modeling electroencephalography waveforms with
  semi-supervised deep belief nets: Fast classification and anomaly
  measurement,'' \emph{J. Neural Eng.}, vol.~8, no.~3, p. 036015, 2011.

\bibitem{turner2014deep}
J.~Turner and et~al., ``Deep belief networks used on high resolution
  multichannel electroencephalography data for seizure detection,'' in
  \emph{2014 aaai spring symposium series}, 2014.

\bibitem{Seizure205}
H.~Daoud and M.~Bayoumi, ``Deep learning approach for epileptic focus
  localization,'' \emph{IEEE Transactions on Biomedical Circuits and Systems},
  vol.~14, no.~2, pp. 209--220, 2019.

\bibitem{Seizure206}
Q.~Lin and et~al., ``Classification of epileptic eeg signals with stacked
  sparse autoencoder based on deep learning,'' in \emph{International
  Conference on Intelligent Computing}.\hskip 1em plus 0.5em minus 0.4em\relax
  Springer, 2016, pp. 802--810.

\bibitem{Seizure207}
B.~Yan and et~al., ``An eeg signal classification method based on sparse
  auto-encoders and support vector machine,'' in \emph{2016 IEEE/CIC
  International Conference on Communications in China (ICCC)}.\hskip 1em plus
  0.5em minus 0.4em\relax IEEE, 2016, pp. 1--6.

\bibitem{Seizure208}
Y.~Yuan and et~al., ``A multi-view deep learning method for epileptic seizure
  detection using short-time fourier transform,'' in \emph{Proceedings of the
  8th ACM International Conference on Bioinformatics, Computational Biology,
  and Health Informatics}, 2017, pp. 213--222.

\bibitem{Seizure209}
A.~M. Karim and et~al., ``A new generalized deep learning framework combining
  sparse autoencoder and taguchi method for novel data classification and
  processing,'' \emph{Mathematical Problems in Engineering}, vol. 2018, 2018.

\bibitem{Seizure210}
Y.~Qiu and et~al., ``Denoising sparse autoencoder-based ictal eeg
  classification,'' \emph{IEEE Transactions on Neural Systems and
  Rehabilitation Engineering}, vol.~26, no.~9, pp. 1717--1726, 2018.

\bibitem{Seizure211}
V.~Sharathappriyaa and et~al., ``Auto-encoder based automated epilepsy
  diagnosis,'' in \emph{2018 International Conference on Advances in Computing,
  Communications and Informatics (ICACCI)}.\hskip 1em plus 0.5em minus
  0.4em\relax IEEE, 2018, pp. 976--982.

\bibitem{Seizure212}
Y.~Yuan and et~al., ``Wave2vec: Deep representation learning for clinical
  temporal data,'' \emph{Neurocomputing}, vol. 324, pp. 31--42, 2019.

\bibitem{Seizure213}
A.~Emami and et~al., ``Autoencoding of long-term scalp electroencephalogram to
  detect epileptic seizure for diagnosis support system,'' \emph{Computers in
  Biology and Medicine}, vol. 110, pp. 227--233, 2019.

\bibitem{9096344}
T.~Siddharth and et~al., ``Eeg-based detection of focal seizure area using
  fbse-ewt rhythm and sae-svm network,'' \emph{IEEE Sensors Journal}, vol.~20,
  no.~19, pp. 11\,421--11\,428, 2020.

\bibitem{Sleep1}
J.~X and et~al., ``3dsleepnet: A multi-channel bio-signal based sleep stages
  classification method using deep learning,'' \emph{IEEE Transactions on
  Neural Systems and Rehabilitation Engineering}, 2023.

\bibitem{Sleep3}
W.~J and et~al., ``Caresleepnet: a hybrid deep learning network for automatic
  sleep staging,'' \emph{IEEE Journal of Biomedical and Health Informatics},
  2024.

\bibitem{Sleep4}
L.~F and et~al., ``End-to-end sleep staging using convolutional neural network
  in raw single-channel eeg,'' \emph{Biomedical Signal Processing and Control},
  vol.~63, p. 102203, 2021.

\bibitem{Sleep5}
E.~E and O.~S, ``Cosleepnet: Automated sleep staging using a hybrid cnn-lstm
  network on imbalanced eeg-eog datasets,'' \emph{Biomedical Signal Processing
  and Control}, vol.~80, p. 104299, 2023.

\bibitem{Sleep6}
S.~A and et~al., ``Deepsleepnet: A model for automatic sleep stage scoring
  based on raw single-channel eeg,'' \emph{IEEE transactions on neural systems
  and rehabilitation engineering}, vol. 25(11), pp. 1998--2008, 2017.

\bibitem{Sleep7}
L.~C and et~al., ``A deep learning method approach for sleep stage
  classification with eeg spectrogram,'' \emph{International Journal of
  Environmental Research and Public Health}, vol. 19(10), p. 6322, 2022.

\bibitem{Sleep8}
Z.~J and W.~Y, ``A new method for automatic sleep stage classification,''
  \emph{IEEE transactions on biomedical circuits and systems}, vol. 11(5), pp.
  1097--1110, 2017.

\bibitem{Sleep9}
Z.~Jia and et~al., ``Graphsleepnet: Adaptive spatial-temporal graph
  convolutional networks for sleep stage classification.'' in \emph{Ijcai},
  vol. 2021, 2020, pp. 1324--1330.

\bibitem{Sleep10}
F.~Y and et~al., ``A dual-stream deep neural network integrated with adaptive
  boosting for sleep staging,'' \emph{Biomedical Signal Processing and
  Control}, vol.~79, p. 104150, 2023.

\bibitem{Sleep11}
S.~H and et~al., ``Intra-and inter-epoch temporal context network (iitnet)
  using sub-epoch features for automatic sleep scoring on raw single-channel
  eeg,'' \emph{Biomedical signal processing and control}, vol.~61, p. 102037,
  2020.

\bibitem{Sleep12}
Y.~C and et~al., ``Lwsleepnet: A lightweight attention-based deep learning
  model for sleep staging with singlechannel eeg,'' \emph{Digital Health},
  vol.~9, p. 20552076231188206, 2023.

\bibitem{Sleep13}
L.~G and et~al., ``Micro sleepnet: efficient deep learning model for mobile
  terminal real-time sleep staging,'' \emph{Frontiers in Neuroscience},
  vol.~17, p. 1218072, 2023.

\bibitem{Sleep14}
J.~Z and et~al., ``Multi-view spatial-temporal graph convolutional networks
  with domain generalization for sleep stage classification,'' \emph{IEEE
  Transactions on Neural Systems and Rehabilitation Engineering}, vol.~29, pp.
  1977--1986, 2021.

\bibitem{Sleep16}
J.~Lu and et~al., ``Pearnet: A pearson correlation-based graph attention
  network for sleep stage recognition,'' in \emph{2022 IEEE 9th International
  Conference on Data Science and Advanced Analytics (DSAA)}.\hskip 1em plus
  0.5em minus 0.4em\relax IEEE, 2022, pp. 1--8.

\bibitem{phan2019seqsleepnet}
H.~Phan and et~al., ``Seqsleepnet: end-to-end hierarchical recurrent neural
  network for sequence-to-sequence automatic sleep staging,'' \emph{IEEE
  Transactions on Neural Systems and Rehabilitation Engineering}, vol.~27,
  no.~3, pp. 400--410, 2019.

\bibitem{Sleep18}
M.~S and et~al., ``Sleepeegnet: Automated sleep stage scoring with sequence to
  sequence deep learning approach,'' \emph{PloS one}, vol. 14(5), p. e0216456,
  2019.

\bibitem{Sleep19}
B.~S and et~al., ``Sleepnet: automated sleep staging system via deep
  learning,'' \emph{arXiv preprint arXiv:1707.08262}, 2017.

\bibitem{Sleep20}
D.~M and et~al., ``Sleepxai: An explainable deep learning approach for
  multi-class sleep stage identification,'' \emph{Applied Intelligence}, vol.
  53(13), pp. 16\,830--16\,843, 2023.

\bibitem{Sleep21}
A.~Supratak and Y.~Guo, ``Tinysleepnet: An efficient deep learning model for
  sleep stage scoring based on raw single-channel eeg,'' in \emph{2020 42nd
  Annual International Conference of the IEEE Engineering in Medicine \&
  Biology Society (EMBC)}.\hskip 1em plus 0.5em minus 0.4em\relax IEEE, 2020,
  pp. 641--644.

\bibitem{Sleep23}
J.~N and et~al., ``Zleepanlystnet: a novel deep learning model for automatic
  sleep stage scoring based on single-channel raw eeg data using separating
  training,'' \emph{Scientific Reports}, vol. 14(1), p. 9859, 2024.

\bibitem{Sleep24}
F.~M and et~al., ``Deep learning in automatic sleep staging with a single
  channel electroencephalography,'' \emph{Frontiers in Physiology}, vol.~12, p.
  628502, 2021.

\bibitem{Sleep25}
H.~M. N and K.~I, ``Mixed-input deep learning approach to sleep/wake state
  classification by using eeg signals,'' \emph{Diagnostics}, vol. 13(14), p.
  2358, 2023.

\bibitem{Sleep27}
S.~C and et~al., ``A hierarchical neural network for sleep stage classification
  based on comprehensive feature learning and multi-flow sequence learning,''
  \emph{IEEE journal of biomedical and health informatics}, vol. 24(5), pp.
  1351--1366, 2019.

\bibitem{yao2023cnntransformer}
Z.~Yao and X.~Liu, ``A cnn-transformer deep learning model for real-time sleep
  stage classification in an energy-constrained wireless device,'' in
  \emph{2023 11th International IEEE/EMBS Conference on Neural Engineering
  (NER)}.\hskip 1em plus 0.5em minus 0.4em\relax IEEE, 2023, pp. 1--4.

\bibitem{phan2018automatic}
H.~Phan and et~al., ``Automatic sleep stage classification using single-channel
  eeg: Learning sequential features with attention-based recurrent neural
  networks,'' in \emph{2018 40th annual international conference of the IEEE
  engineering in medicine and biology society (EMBC)}, 2018, pp. 1452--1455.

\bibitem{chambon2018deep}
S.~Chambon and et~al., ``A deep learning architecture for temporal sleep stage
  classification using multivariate and multimodal time series,'' \emph{IEEE
  Transactions on Neural Systems and Rehabilitation Engineering}, vol.~26,
  no.~4, pp. 758--769, 2018.

\bibitem{Sleep33}
S.~A and et~al., ``A convolutional neural network for sleep stage scoring from
  raw single-channel eeg,'' \emph{Biomedical Signal Processing and Control},
  vol.~42, pp. 107--114, 2018.

\bibitem{Sleep34}
S.~M and et~al., ``Deep learning for automated feature discovery and
  classification of sleep stages,'' \emph{IEEE/ACM transactions on
  computational biology and bioinformatics}, vol. 17(6), pp. 1835--1845, 2019.

\bibitem{alvarez2021inter}
D.~Alvarez-Estevez and R.~M. Rijsman, ``Inter-database validation of a deep
  learning approach for automatic sleep scoring,'' \emph{PloS one}, vol.~16,
  no.~8, p. e0256111, 2021.

\bibitem{Sleep35}
P.~H and et~al., ``Joint classification and prediction cnn framework for
  automatic sleep stage classification,'' \emph{IEEE Transactions on Biomedical
  Engineering}, vol. 66(5), pp. 1285--1296, 2018.

\bibitem{Sleep36}
F.-B. E and et~al., ``Convolutional neural networks for sleep stage scoring on
  a two-channel eeg signal,'' \emph{Soft Computing}, vol.~24, pp. 4067--4079,
  2020.

\bibitem{Sleep37}
Z.~T and et~al., ``Convolution-and attention-based neural network for automated
  sleep stage classification,'' \emph{International Journal of Environmental
  Research and Public Health}, vol. 17(11), p. 4152, 2020.

\bibitem{Sleep38}
L.~M and et~al., ``An attention-guided spatiotemporal graph convolutional
  network for sleep stage classification,'' \emph{Life}, vol. 12(5), p. 622,
  2022.

\bibitem{Sleep39}
D.~H and et~al., ``Mixed neural network approach for temporal sleep stage
  classification,'' \emph{IEEE Transactions on Neural Systems and
  Rehabilitation Engineering}, vol. 26(2), pp. 324--333, 2017.

\bibitem{Sleep44}
S.~S. K and L.~D, ``Automated classification of multi-class sleep stages
  classification using polysomnography signals: a nine-layer 1d-convolution
  neural network approach,'' \emph{Multimedia Tools and Applications}, vol.
  82(6), pp. 8049--8091, 2023.

\bibitem{Sleep47}
C.~S and et~al., ``Dssnet: a deep sequential sleep network for self-supervised
  representation learning based on single-channel eeg,'' \emph{IEEE Signal
  Processing Letters}, vol.~29, pp. 2143--2147, 2022.

\bibitem{Sleep48}
C.~H.~Y. S and et~al., ``Maeeg: Masked auto-encoder for eeg representation
  learning,'' \emph{arXiv preprint arXiv:2211.02625}, 2022.

\bibitem{Sleep51}
Y.~Y and et~al., ``Psnsleep: a self-supervised learning method for sleep
  staging based on siamese networks with only positive sample pairs,''
  \emph{Frontiers in Neuroscience}, vol.~17, p. 1167723, 2023.

\bibitem{Sleep53}
Q.~Xiao and et~al., ``Self-supervised learning for sleep stage classification
  with predictive and discriminative contrastive coding,'' in \emph{ICASSP
  2021-2021 IEEE International Conference on Acoustics, Speech and Signal
  Processing (ICASSP)}.\hskip 1em plus 0.5em minus 0.4em\relax IEEE, 2021, pp.
  1290--1294.

\bibitem{Sleep54}
H.~Zhang and et~al., ``Expert knowledge inspired contrastive learning for sleep
  staging,'' in \emph{2022 International Joint Conference on Neural Networks
  (IJCNN)}.\hskip 1em plus 0.5em minus 0.4em\relax IEEE, 2022, pp. 1--6.

\bibitem{Sleep55}
H.~Lee, E.~Seong, and D.~K. Chae, ``Self-supervised learning with
  attention-based latent signal augmentation for sleep staging with limited
  labeled data,'' in \emph{IJCAI}, 2022, pp. 3868--3876.

\bibitem{jiang2021self}
X.~Jiang and et~al., ``Self-supervised contrastive learning for eeg-based sleep
  staging,'' in \emph{2021 International Joint Conference on Neural Networks
  (IJCNN)}, 2021, pp. 1--8.

\bibitem{Sleep63}
T.~Brüsch and et~al., ``Multi-view self-supervised learning for multivariate
  variable-channel time series,'' in \emph{2023 IEEE 33rd International
  Workshop on Machine Learning for Signal Processing (MLSP)}.\hskip 1em plus
  0.5em minus 0.4em\relax IEEE, 2023, pp. 1--6.

\bibitem{Sleep64}
L.~Y and et~al., ``Adversarial learning for semi-supervised pediatric sleep
  staging with single-eeg channel,'' \emph{Methods}, vol. 204, pp. 84--91,
  2022.

\bibitem{Sleep65}
L.~Y. and et~al., ``Mtclss: Multi-task contrastive learning for semi-supervised
  pediatric sleep staging,'' \emph{IEEE Journal of Biomedical and Health
  Informatics}, vol. 27(6), pp. 2647--2655, 2022.

\bibitem{Sleep66}
Z.~C and et~al., ``Hybrid manifold-deep convolutional neural network for sleep
  staging,'' \emph{Methods}, vol. 202, pp. 164--172, 2022.

\bibitem{Sleep67}
Z.~Y and et~al., ``Shnn: A single-channel eeg sleep staging model based on
  semi-supervised learning,'' \emph{Expert Systems with Applications}, vol.
  213, p. 119288, 2023.

\bibitem{Sleep68}
A.~M. Munk and et~al., ``Semi-supervised sleep-stage scoring based on single
  channel eeg,'' in \emph{2018 IEEE International Conference on Acoustics,
  Speech and Signal Processing (ICASSP)}.\hskip 1em plus 0.5em minus
  0.4em\relax IEEE, 2018, pp. 2551--2555.

\bibitem{Sleep69}
B.~Haoran and L.~Guanze, ``Semi-supervised end-to-end automatic sleep stage
  classification based on pseudo-label,'' in \emph{2021 IEEE International
  Conference on Power Electronics, Computer Applications (ICPECA)}.\hskip 1em
  plus 0.5em minus 0.4em\relax IEEE, 2021, pp. 83--87.

\bibitem{Sleep70}
Z.~J and W.~Y, ``Competition convolutional neural network for sleep stage
  classification,'' \emph{Biomedical Signal Processing and Control}, vol.~64,
  p. 102318, 2021.

\bibitem{Sleep71}
Z.~J. and W.~Y., ``Complex-valued unsupervised convolutional neural networks
  for sleep stage classification,'' \emph{Computer methods and programs in
  biomedicine}, vol. 164, pp. 181--191, 2018.

\bibitem{Sleep72}
L.~Fraiwan and K.~Lweesy, ``Neonatal sleep state identification using deep
  learning autoencoders,'' in \emph{2017 IEEE 13th International Colloquium on
  Signal Processing \& its Applications (CSPA)}.\hskip 1em plus 0.5em minus
  0.4em\relax IEEE, 2017, pp. 228--231.

\bibitem{Sleep73}
Z.~J and et~al., ``Automatic sleep stage classification based on sparse deep
  belief net and combination of multiple classifiers,'' \emph{Transactions of
  the Institute of Measurement and Control}, vol. 38(4), pp. 435--451, 2016.

\bibitem{Sleep74}
T.~O and et~al., ``Automatic sleep stage scoring using time-frequency analysis
  and stacked sparse autoencoders,'' \emph{Annals of biomedical engineering},
  vol.~44, pp. 1587--1597, 2016.

\bibitem{Sleep75}
L.~M and et~al., ``Sleep stage classification using unsupervised feature
  learning,'' \emph{Advances in Artificial Neural Systems}, vol. 2012(1), p.
  107046, 2012.

\bibitem{MDD2}
H.-G. Wang and et~al., ``Amgcn-l: an adaptive multi-time-window graph
  convolutional network with long-short-term memory for depression detection,''
  \emph{Journal of Neural Engineering}, vol.~20, no.~5, p. 056038, 2023.

\bibitem{MDD3}
C.~Y and et~al., ``Dctnet: hybrid deep neural network-based eeg signal for
  detecting depression,'' \emph{Multimedia Tools and Applications}, vol.
  82(26), pp. 41\,307--41\,321, 2023.

\bibitem{MDD4}
S.~G and et~al., ``Depcap: a smart healthcare framework for eeg based
  depression detection using time-frequency response and deep neural network,''
  \emph{IEEE Access}, vol.~11, pp. 52\,327--52\,338, 2023.

\bibitem{MDD7}
Y.~Wang and et~al., ``Diffmdd: A diffusion-based deep learning framework for
  mdd diagnosis using eeg,'' \emph{IEEE Transactions on Neural Systems and
  Rehabilitation Engineering}, 2024.

\bibitem{MDD8}
Y.~M and et~al., ``Edt: An eeg-based attention model for feature learning and
  depression recognition,'' \emph{Biomedical Signal Processing and Control},
  vol.~93, p. 106182, 2024.

\bibitem{MDD9}
L.~Yang and et~al., ``A gated temporal-separable attention network for
  eeg-based depression recognition,'' \emph{Computers in Biology and Medicine},
  vol. 157, p. 106782, 2023.

\bibitem{MDD10}
W.~Z and et~al., ``Hybrideegnet: A convolutional neural network for eeg feature
  learning and depression discrimination,'' \emph{IEEE Access}, vol.~8, pp.
  30\,332--30\,342, 2020.

\bibitem{MDD11}
X.~Song and et~al., ``Lsdd-eegnet: An efficient end-to-end framework for
  eeg-based depression detection,'' \emph{Biomedical Signal Processing and
  Control}, vol.~75, p. 103612, 2022.

\bibitem{MDD13}
X.~Sun and et~al., ``Multi-granularity graph convolution network for major
  depressive disorder recognition,'' \emph{IEEE Transactions on Neural Systems
  and Rehabilitation Engineering}, vol.~32, pp. 559--569, 2023.

\bibitem{MDD14}
W.~Cui and et~al., ``A multiview sparse dynamic graph convolution-based
  region-attention feature fusion network for major depressive disorder
  detection,'' \emph{IEEE Transactions on Computational Social Systems},
  vol.~11, pp. 2691--2702, 2023.

\bibitem{MDD15}
C.~T and et~al., ``Exploring self-attention graph pooling with eeg-based
  topological structure and soft label for depression detection,'' \emph{IEEE
  transactions on affective computing}, vol. 13(4), pp. 2106--2118, 2022.

\bibitem{MDD16}
Z.~Z and et~al., ``A novel eeg-based graph convolution network for depression
  detection: incorporating secondary subject partitioning and attention
  mechanism,'' \emph{Expert Systems with Applications}, vol. 239, p. 122356,
  2024.

\bibitem{MDD17}
C.~Yang and et~al., ``Tsunet-cc: Temporal spectrogram unet embedding cross
  channel-wise attention mechanism for mdd identification,'' in \emph{2023 45th
  Annual International Conference of the IEEE Engineering in Medicine \&
  Biology Society (EMBC)}.\hskip 1em plus 0.5em minus 0.4em\relax IEEE, 2023,
  pp. 1--4.

\bibitem{MDD18}
X.~Y and et~al., ``Depressive disorder recognition based on frontal eeg signals
  and deep learning,'' \emph{Sensors}, vol. 23(20), p. 8639, 2023.

\bibitem{MDD19}
K.~H and et~al., ``Cloud‐aided online eeg classification system for brain
  healthcare: A case study of depression evaluation with a lightweight cnn,''
  \emph{Software: Practice and Experience}, vol. 50(5), pp. 596--610, 2020.

\bibitem{MDD20}
K.~M and et~al., ``Deep-asymmetry: Asymmetry matrix image for deep learning
  method in pre-screening depression,'' \emph{Sensors}, vol. 20(22), p. 6526,
  2020.

\bibitem{MDD21}
S.~A and et~al., ``Major depressive disorder diagnosis based on effective
  connectivity in eeg signals: a convolutional neural network and long
  short-term memory approach,'' \emph{Cognitive Neurodynamics}, vol. 15(2), pp.
  239--252, 2021.

\bibitem{MDD22}
L.~H. W and et~al., ``Decision support system for major depression detection
  using spectrogram and convolution neural network with eeg signals,''
  \emph{Expert Systems}, vol. 39(3), p. e12773, 2022.

\bibitem{MDD23}
M.~Kang and et~al., ``Low channel electroencephalogram based deep learning
  method to pre-screening depression,'' in \emph{2020 International Conference
  on Information and Communication Technology Convergence (ICTC)}.\hskip 1em
  plus 0.5em minus 0.4em\relax IEEE, 2020, pp. 449--451.

\bibitem{MDD24}
D.~W and et~al., ``Multilayer brain network combined with deep convolutional
  neural network for detecting major depressive disorder,'' \emph{Nonlinear
  Dynamics}, vol. 102(2), pp. 667--677, 2020.

\bibitem{MDD25}
M.~W and Q.~A, ``A deep learning framework for automatic diagnosis of unipolar
  depression,'' \emph{International journal of medical informatics}, vol. 132,
  p. 103983, 2019.

\bibitem{MDD26}
S.~X and et~al., ``A novel complex network-based graph convolutional network in
  major depressive disorder detection,'' \emph{IEEE Transactions on
  Instrumentation and Measurement}, vol.~71, pp. 1--8, 2022.

\bibitem{MDD27}
A.~A and et~al., ``Automated major depressive disorder diagnosis using a
  dual-input deep learning model and image generation from eeg signals,''
  \emph{Waves in Random and Complex Media}, 2023: 1-16.

\bibitem{MDD28}
X.~M and et~al., ``An end-to-end deep learning model for eeg-based major
  depressive disorder classification,'' \emph{IEEE Access}, vol.~11, pp.
  41\,337--41\,347, 2023.

\bibitem{MDD29}
A.~I. A and et~al., ``A robust deep-learning model to detect major depressive
  disorder utilising eeg signals,'' \emph{IEEE Transactions on Artificial
  Intelligence}, 2024.

\bibitem{MDD30}
A.~Rafiei and et~al., ``Automated detection of major depressive disorder with
  eeg signals: A time series classification using deep learning,'' \emph{IEEE
  Access}, vol.~10, pp. 73\,804--73\,817, 2022.

\bibitem{MDD31}
D.~M. Khan and et~al., ``Development of wavelet coherence eeg as a biomarker
  for diagnosis of major depressive disorder,'' \emph{IEEE Sensors Journal},
  vol.~22, pp. 4315--4325, 2022.

\bibitem{MDD32}
L.~Li and et~al., ``An eeg-based marker of functional connectivity: Detection
  of major depressive disorder,'' \emph{Cognitive Neurodynamics}, vol.~18, pp.
  1671--1687, 2024.

\bibitem{MDD34}
P.~Sandheep and et~al., ``Performance analysis of deep learning cnn in
  classification of depression eeg signals,'' in \emph{TENCON 2019-2019 IEEE
  Region 10 Conference}.\hskip 1em plus 0.5em minus 0.4em\relax IEEE, 2019, pp.
  1339--1344.

\bibitem{MDD35}
L.~Duan and et~al., ``Machine learning approaches for mdd detection and emotion
  decoding using eeg signals,'' \emph{Frontiers in Human Neuroscience},
  vol.~14, 2020.

\bibitem{MDD37}
X.~Zhang and et~al., ``Eeg-based depression detection using convolutional
  neural network with demographic attention mechanism,'' in \emph{2020 42nd
  annual international conference of the IEEE Engineering in Medicine \&
  Biology Society (EMBC)}.\hskip 1em plus 0.5em minus 0.4em\relax IEEE, 2020,
  pp. 128--133.

\bibitem{MDD38}
L.~X and et~al., ``A deep learning approach for mild depression recognition
  based on functional connectivity using electroencephalography,''
  \emph{Frontiers in neuroscience}, vol.~14, p. 192, 2020.

\bibitem{MDD39}
Y.~Xie and et~al., ``Anxiety and depression diagnosis method based on brain
  networks and convolutional neural networks,'' in \emph{2020 42nd Annual
  International Conference of the IEEE Engineering in Medicine \& Biology
  Society (EMBC)}.\hskip 1em plus 0.5em minus 0.4em\relax IEEE, 2020, pp.
  1503--1506.

\bibitem{MDD42}
A.~Qayyum and et~al., ``Hybrid deep shallow network for assessment of
  depression using electroencephalogram signals,'' in \emph{Neural Information
  Processing: 27th International Conference, ICONIP 2020, Bangkok, Thailand,
  November 23--27, 2020, Proceedings, Part III 27}.\hskip 1em plus 0.5em minus
  0.4em\relax Springer, 2020, pp. 245--257.

\bibitem{MDD43}
U.~C and et~al., ``Major depressive disorder classification based on different
  convolutional neural network models: deep learning approach,'' \emph{Clinical
  EEG and neuroscience}, vol. 52(1), pp. 38--51, 2021.

\bibitem{MDD44}
H.~D. S.~B. A and et~al., ``Integration of deep learning for improved diagnosis
  of depression using eeg and facial features,'' \emph{Materials Today:
  Proceedings}, vol.~80, pp. 1965--1969, 2023.

\bibitem{MDD45}
A.~O. Khadidos and et~al., ``Machine learning and electroencephalogram signal
  based diagnosis of dipression,'' \emph{Neuroscience Letters}, vol. 809, p.
  137313, 2023.

\bibitem{MDD46}
W.~Mao and et~al., ``Resting state eeg based depression recognition research
  using deep learning method,'' in \emph{Proceedings of the International
  Conference on Brain Informatics (BI)}, Arlington, TX, USA, December 2018, pp.
  329--338.

\bibitem{MDD47}
J.~Zhang and et~al., ``Depression screening using hybrid neural network,''
  \emph{Multimedia Tools and Applications}, vol.~82, pp. 26\,955--26\,970,
  2023.

\bibitem{MDD48}
W.~Wu and et~al., ``Few-electrode eeg from the wearable devices using domain
  adaptation for depression detection,'' \emph{Biosensors}, vol.~12, p. 1087,
  2022.

\bibitem{MDD49}
B.~Zhang and et~al., ``Spatial–temporal eeg fusion based on neural network
  for major depressive disorder detection,'' \emph{Interdisciplinary Science:
  Computational Life Sciences}, vol.~15, pp. 542--559, 2023.

\bibitem{MDD50}
D.~A and et~al., ``Deep learning in computer aided diagnosis of mdd,''
  \emph{Int J In-novat Technol Explor Eng}, vol. 8(6), pp. 464--468, 2019.

\bibitem{MDD51}
T.~P. P and et~al., ``Eeg-based deep learning model for the automatic detection
  of clinical depression,'' \emph{Physical and Engineering Sciences in
  Medicine}, vol.~43, pp. 1349--1360, 2020.

\bibitem{song2022lsdd}
X.~Song, D.~Yan, L.~Zhao, and L.~Yang, ``Lsdd-eegnet: An efficient end-to-end
  framework for eeg-based depression detection,'' \emph{Biomedical Signal
  Processing and Control}, vol.~75, p. 103612, 2022.

\bibitem{zhu2022eeg}
J.~Zhu and et~al., ``Eeg based depression recognition using improved graph
  convolutional neural network,'' \emph{Computers in Biology and Medicine},
  vol. 148, p. 105815, 2022.

\bibitem{MDD12}
S.~Zhang and et~al., ``Multi-view graph contrastive learning via adaptive
  channel optimization for depression detection in eeg signals,''
  \emph{International Journal of Neural Systems}, vol.~33, p. 2350055, 2023.

\bibitem{MDD1}
D.~Wang and et~al., ``Identification of depression with a semi-supervised gcn
  based on eeg data,'' in \emph{2021 IEEE International Conference on
  Bioinformatics and Biomedicine (BIBM)}.\hskip 1em plus 0.5em minus
  0.4em\relax IEEE, 2021, pp. 2338--2345.

\bibitem{li2023gcns}
W.~Li and et~al., ``Gcns--fsmi: Eeg recognition of mental illness based on
  fine-grained signal features and graph mutual information maximization,''
  \emph{Expert Systems With Applications}, vol. 228, p. 120227, 2023.

\bibitem{SZ2}
O.~S. L and et~al., ``Deep convolutional neural network model for automated
  diagnosis of schizophrenia using eeg signals,'' \emph{Applied Sciences}, vol.
  9(14), p. 2870, 2019.

\bibitem{SZ3}
K.~S. K and et~al., ``Schizonet: a robust and accurate margenau–hill
  time-frequency distribution based deep neural network model for schizophrenia
  detection using eeg signals,'' \emph{Physiological Measurement}, vol. 44(3),
  p. 035005, 2023.

\bibitem{SZ4}
K.~M. R and et~al., ``Weighted ordinal connection based functional network
  classification for schizophrenia disease detection using eeg signal,''
  \emph{Physical and Engineering Sciences in Medicine}, vol. 46(3), pp.
  1055--1070, 2023.

\bibitem{SZ5}
W.~Z and et~al., ``Automated rest eeg-based diagnosis of depression and
  schizophrenia using a deep convolutional neural network,'' \emph{IEEE
  Access}, vol.~10, pp. 104\,472--104\,485, 2022.

\bibitem{SZ6}
H.~F and et~al., ``Fusion of multivariate eeg signals for schizophrenia
  detection using cnn and machine learning techniques,'' \emph{Information
  Fusion}, vol.~92, pp. 466--478, 2023.

\bibitem{SZ8}
K.~S. K and et~al., ``Spwvd-cnn for automated detection of schizophrenia
  patients using eeg signals,'' \emph{IEEE Transactions on Instrumentation and
  Measurement}, vol.~70, pp. 1--9, 2021.

\bibitem{SZ9}
L.~I and et~al., ``Identification and diagnosis of schizophrenia based on
  multichannel eeg and cnn deep learning model,'' \emph{Schizophrenia
  Research}, vol. 271, pp. 28--35, 2024.

\bibitem{SZ10}
E.~Lillo and et~al., ``Automated diagnosis of schizophrenia using eeg
  microstates and deep convolutional neural network,'' \emph{Expert Systems
  with Applications}, vol. 209, p. 118236, 2022.

\bibitem{SZ11}
H.~Göker, ``1d-convolutional neural network approach and feature extraction
  methods for automatic detection of schizophrenia,'' \emph{Signal, Image and
  Video Processing}, vol.~17, no.~5, pp. 2627--2636, 2023.

\bibitem{SZ12}
A.~Khodabakhsh and et~al., ``U-net based estimation of functional connectivity
  from time series multi-channel eeg from schizophrenia patients,'' in
  \emph{2021 IEEE Nuclear Science Symposium and Medical Imaging Conference
  (NSS/MIC)}, 2021.

\bibitem{SZ13}
G.~Sahu and et~al., ``Scz-scan: an automated schizophrenia detection system
  from electroencephalogram signals,'' \emph{Biomedical Signal Processing and
  Control}, vol.~86, p. 105206, 2023.

\bibitem{SZ14}
L.~Chu and et~al., ``Individual recognition in schizophrenia using deep
  learning methods with random forest and voting classifiers: insights from
  resting state eeg streams,'' \emph{arXiv preprint arXiv:1707.03467}, 2017.

\bibitem{SZ15}
S.~K and et~al., ``Spectral features based convolutional neural network for
  accurate and prompt identification of schizophrenic patients,''
  \emph{Proceedings of the Institution of Mechanical Engineers}, vol. 2021,
  Part H: Journal of Engineering in Medicine.

\bibitem{SZ16}
Z.~Aslan and M.~Akin, ``Automatic detection of schizophrenia by applying deep
  learning over spectrogram images of eeg signals.'' \emph{Traitement du
  Signal}, vol.~37, no.~2, 2020.

\bibitem{SZ21}
S.~Siuly and et~al., ``Schizogooglenet: The googlenet-based deep feature
  extraction design for automatic detection of schizophrenia,''
  \emph{Computational Intelligence and Neuroscience}, vol. 2022, no.~1, p.
  1992596, 2022.

\bibitem{SZ23}
B.~C and et~al., ``From sound perception to automatic detection of
  schizophrenia: an eeg-based deep learning approach,'' \emph{Frontiers in
  Psychiatry}, vol.~12, p. 813460, 2022.

\bibitem{SZ24}
Z.~Guo and et~al., ``Deep neural network classification of eeg data in
  schizophrenia,'' in \emph{Proc 2021 IEEE 10th Data Driven Control Learn Syst
  Conf (DDCLS)}, 2021, pp. 1322--1327.

\bibitem{SZ25}
C.~A.~T. Naira and et~al., ``Classification of people who suffer schizophrenia
  and healthy people by eeg signals using deep learning,'' \emph{International
  Journal of Advanced Computer Science and Applications}, vol.~10, pp.
  511--516, 2019.

\bibitem{SZ26}
N.~Ilakiyaselvan and et~al., ``Reconstructed phase space portraits for
  detecting brain diseases using deep learning,'' \emph{Biomedical Signal
  Processing and Control}, vol.~71, p. 103278, 2022.

\bibitem{SZ28}
D.~Calhas and et~al., ``On the use of pairwise distance learning for brain
  signal classification with limited observations,'' \emph{Artificial
  Intelligence in Medicine}, vol. 105, p. 101852, 2020.

\bibitem{SZ29}
E.~Nsugbe and et~al., ``Intelligence combiner: a combination of deep learning
  and handcrafted features for an adolescent psychosis prediction using eeg
  signals,'' in \emph{2022 IEEE Int Work Metrol Ind 4.0 IoT (MetroInd4.0IoT)},
  2022, pp. 92--97.

\bibitem{SZ30}
V.~Divya and et~al., ``Signal conducting system with effective optimization
  using deep learning for schizophrenia classification,'' \emph{Computer
  Systems Science and Engineering}, vol.~45, pp. 1869--1886, 2023.

\bibitem{SZ31}
M.~Shen and et~al., ``Automatic identification of schizophrenia based on eeg
  signals using dynamic functional connectivity analysis and 3d convolutional
  neural network,'' \emph{Computers in Biology and Medicine}, vol. 160, p.
  107022, 2023.

\bibitem{SZ32}
M.~Saeedi and et~al., ``Schizophrenia diagnosis via fft and wavelet
  convolutional neural networks utilizing eeg signals,'' 2022.

\bibitem{SZ33}
C.~A. Ellis and et~al., ``Examining effects of schizophrenia on eeg with
  explainable deep learning models,'' in \emph{2022 IEEE 22nd Int Conf
  Bioinformatics Bioengineering (BIBE)}, 2022, pp. 301--304.

\bibitem{SZ34}
A.-A. D and et~al., ``Identification of children at risk of schizophrenia via
  deep learning and eeg responses,'' \emph{IEEE Journal of biomedical and
  health informatics}, vol. 25(1), pp. 69--76, 2020.

\bibitem{SZ36}
S.~G and J.~A. M, ``Szhnn: a novel and scalable deep convolution hybrid neural
  network framework for schizophrenia detection using multichannel eeg,''
  \emph{IEEE Transactions on Instrumentation and Measurement}, vol.~71, pp.
  1--9, 2022.

\bibitem{SZ37}
K.~Jindal and et~al., ``Bi-lstm-deep cnn for schizophrenia detection using
  msst-spectral images of eeg signals,'' in \emph{Artificial Intelligence-Based
  Brain-Computer Interface}.\hskip 1em plus 0.5em minus 0.4em\relax Elsevier,
  2022, pp. 145--162.

\bibitem{SZ38}
B.~S and et~al., ``Detection of schizophrenia using hybrid of deep learning and
  brain effective connectivity image from electroencephalogram signal,''
  \emph{Computers in Biology and Medicine}, vol. 146, p. 105570, 2022.

\bibitem{SZ39}
S.~S and J.~S. D, ``A novel approach to schizophrenia detection: Optimized
  preprocessing and deep learning analysis of multichannel eeg data,''
  \emph{Expert Systems with Applications}, vol. 246, p. 122937, 2024.

\bibitem{SZ40}
A.~Shoeibi and et~al., ``Automatic diagnosis of schizophrenia in eeg signals
  using cnn-lstm models,'' \emph{Frontiers in Neuroinformatics}, vol.~15, pp.
  1--16, 2021.

\bibitem{SZ41}
G.~Sharma and et~al., ``A smart healthcare framework for accurate detection of
  schizophrenia using multichannel eeg,'' \emph{IEEE Transactions on
  Instrumentation and Measurement}, vol.~72, pp. 1--9, 2023.

\bibitem{SZ42}
S.~Guhan and et~al., ``Eeg based classification of children with learning
  disabilities using shallow and deep neural network,'' \emph{Biomedical Signal
  Processing and Control}, vol.~82, p. 104553, 2023.

\bibitem{SZ43}
C.~R. Phang and et~al., ``Classification of eeg-based effective brain
  connectivity in schizophrenia using deep neural networks,'' in \emph{Int
  IEEE/EMBS Conf Neural Engineering (NER)}, 2019, pp. 401--406.

\bibitem{SZ44}
Q.~Chang and et~al., ``Classification of first-episode schizophrenia, chronic
  schizophrenia and healthy control based on brain network of mismatch
  negativity by graph neural network,'' \emph{IEEE Transactions on Neural
  Systems and Rehabilitation Engineering}, vol.~29, pp. 1784--1794, 2021.

\bibitem{SZ45}
A.~Nikhil~Chandran and et~al., ``Eeg-based automated detection of schizophrenia
  using long short-term memory (lstm) network,'' in \emph{Advances in Machine
  Learning and Computational Intelligence: Proceedings of ICMLCI 2019}.\hskip
  1em plus 0.5em minus 0.4em\relax Springer Singapore, 2021, pp. 229--236.

\bibitem{SZ46}
S.~R and et~al., ``A deep learning based model using rnn-lstm for the detection
  of schizophrenia from eeg data,'' \emph{Computers in Biology and Medicine},
  vol. 151, p. 106225, 2022.

\bibitem{SZ47}
L.~B and et~al., ``Automatic detection of schizophrenia based on
  spatial–temporal feature mapping and levit with eeg signals,'' \emph{Expert
  Systems with Applications}, vol. 224, p. 119969, 2023.

\bibitem{alves2022eeg}
C.~L. Alves and et~al., ``Eeg functional connectivity and deep learning for
  automatic diagnosis of brain disorders: Alzheimer’s disease and
  schizophrenia,'' \emph{Journal of Physics: complexity}, vol.~3, no.~2, p.
  025001, 2022.

\bibitem{SZ48}
Y.~Wu and et~al., ``Schizophrenia detection based on eeg using recurrent
  auto-encoder framework,'' in \emph{International Conference on Neural
  Information Processing}.\hskip 1em plus 0.5em minus 0.4em\relax Springer,
  2022, pp. 62--73.

\bibitem{SZ49}
P.~S. K and L.~S. W, ``Sasdl and rbatq: sparse autoencoder with swarm based
  deep learning and reinforcement based q-learning for eeg classification,''
  \emph{IEEE open journal of engineering in medicine and biology}, vol.~3, pp.
  58--68, 2022.

\bibitem{SZ50}
S.~Parija and et~al., ``Autoencoder-based improved deep learning approach for
  schizophrenic eeg signal classification,'' \emph{Pattern Analysis and
  Applications}, vol.~26, no.~2, pp. 403--435, 2023.

\bibitem{AD2}
N.~M and et~al., ``A novel hybrid model in the diagnosis and classification of
  alzheimer's disease using eeg signals: Deep ensemble learning (del)
  approach,'' \emph{Biomedical Signal Processing and Control}, vol.~89, p.
  105751, 2024.

\bibitem{AD4}
Ieracitano and et~al., ``A convolutional neural network based self-learning
  approach for classifying neurodegenerative states from eeg signals in
  dementia,'' in \emph{2020 International Joint Conference on Neural Networks
  (IJCNN)}.\hskip 1em plus 0.5em minus 0.4em\relax IEEE, 2020, pp. 1--8.

\bibitem{AD5}
F.~C. Morabito and et~al., ``Deep convolutional neural networks for
  classification of mild cognitive impaired and alzheimer's disease patients
  from scalp eeg recordings,'' in \emph{2016 IEEE 2nd International Forum on
  Research and Technologies for Society and Industry Leveraging a better
  tomorrow (RTSI)}.\hskip 1em plus 0.5em minus 0.4em\relax IEEE, 2016, pp.
  1--6.

\bibitem{AD6}
B.~X and W.~H, ``Early alzheimer’s disease diagnosis based on eeg spectral
  images using deep learning,'' \emph{Neural Networks}, vol. 114, pp. 119--135,
  2019.

\bibitem{AD7}
D.~L. D and et~al., ``An intelligent alzheimer’s disease prediction using
  convolutional neural network (cnn),'' \emph{International Journal of Advanced
  Research in Engineering and Technology (IJARET)}, vol. 11(4), pp. 12--22,
  2020.

\bibitem{AD8}
Huggins and et~al., ``Deep learning of resting-state electroencephalogram
  signals for three-class classification of alzheimer’s disease, mild
  cognitive impairment and healthy ageing,'' \emph{Journal of Neural
  Engineering}, vol.~18, no.~4, p. 046087, 2021.

\bibitem{AD9}
X.~W and et~al., ``A novel method for diagnosing alzheimer's disease using deep
  pyramid cnn based on eeg signals,'' \emph{Heliyon}, vol. 9(4), 2023.

\bibitem{AD10}
R.~K and Z.~M, ``Diagnose alzheimer’s disease and mild cognitive impairment
  using deep cascadenet and handcrafted features from eeg signals,''
  \emph{Biomedical Signal Processing and Control}, vol.~99, p. 106895, 2025.

\bibitem{AD12}
I.~C and et~al., ``A convolutional neural network approach for classification
  of dementia stages based on 2d-spectral representation of eeg recordings,''
  \emph{Neurocomputing}, vol. 323, pp. 96--107, 2019.

\bibitem{AD13}
A.~K and et~al., ``Eeg-based clinical decision support system for alzheimer's
  disorders diagnosis using emd and deep learning techniques,'' \emph{Frontiers
  in Human Neuroscience}, vol.~17, p. 1190203, 2023.

\bibitem{AD14}
D.~Kim and K.~Kim, ``Detection of early stage alzheimer’s disease using eeg
  relative power with deep neural network,'' in \emph{2018 40th Annual
  International Conference of the IEEE Engineering in Medicine and Biology
  Society (EMBC)}.\hskip 1em plus 0.5em minus 0.4em\relax IEEE, 2018, pp.
  352--355.

\bibitem{AD15}
Y.~Zhao and L.~He, ``Deep learning in the eeg diagnosis of alzheimer’s
  disease,'' in \emph{Computer Vision-ACCV 2014 Workshops: Singapore,
  Singapore, November 1-2, 2014, Revised Selected Papers, Part I 12}.\hskip 1em
  plus 0.5em minus 0.4em\relax Springer, 2015, pp. 340--353.

\bibitem{AD16}
L.~K and et~al., ``Feature extraction and identification of alzheimer’s
  disease based on latent factor of multi-channel eeg,'' \emph{IEEE
  Transactions on Neural Systems and Rehabilitation Engineering}, vol.~29, pp.
  1557--1567, 2021.

\bibitem{AD17}
M.~F. C and et~al., ``Deep learning representation from electroencephalography
  of early-stage creutzfeldt-jakob disease and features for differentiation
  from rapidly progressive dementia,'' \emph{International journal of neural
  systems}, vol. 27(02), p. 1650039, 2017.

\bibitem{PD1}
S.~S.~A. A and et~al., ``Dynamical system based compact deep hybrid network for
  classification of parkinson disease related eeg signals,'' \emph{Neural
  Networks}, vol. 130, pp. 75--84, 2020.

\bibitem{PD3}
K.~S. K and et~al., ``Pdcnnet: An automatic framework for the detection of
  parkinson’s disease using eeg signals,'' \emph{IEEE Sensors Journal}, vol.
  21(15), pp. 17\,017--17\,024, 2021.

\bibitem{PD5}
Z.~R and et~al., ``Eeg analysis of parkinson's disease using time–frequency
  analysis and deep learning,'' \emph{Biomedical Signal Processing and
  Control}, vol.~78, p. 103883, 2022.

\bibitem{PD6}
X.~Shi, T.~Wang, L.~Wang, H.~Liu, and N.~Yan, ``Hybrid convolutional recurrent
  neural networks outperform cnn and rnn in task-state eeg detection for
  parkinson's disease,'' in \emph{2019 Asia-Pacific signal and information
  processing association annual summit and conference (APSIPA ASC)}.\hskip 1em
  plus 0.5em minus 0.4em\relax IEEE, 2019, pp. 939--944.

\bibitem{PD8}
C.~Chu and et~al., ``Deep learning reveals personalized spatial spectral
  abnormalities of high delta and low alpha bands in eeg of patients with early
  parkinson’s disease,'' \emph{Journal of Neural Engineering}, vol.~18,
  no.~6, p. 066036, 2021.

\bibitem{PD9}
E.~Arasteh and et~al., ``Deep transfer learning for parkinson’s disease
  monitoring by image-based representation of resting-state eeg using
  directional connectivity,'' \emph{Algorithms}, vol.~15, no.~1, p.~5, 2021.

\bibitem{PD11}
P.~M and et~al., ``Deep-learning detection of mild cognitive impairment from
  sleep electroencephalography for patients with parkinson’s disease,''
  \emph{Plos one}, vol. 18(8), p. e0286506, 2023.

\bibitem{PD12}
M.~Shaban, ``Automated screening of parkinson's disease using deep learning
  based electroencephalography,'' in \emph{2021 10th international IEEE/EMBS
  conference on neural engineering (NER)}.\hskip 1em plus 0.5em minus
  0.4em\relax IEEE, 2021, pp. 158--161.

\bibitem{PD14}
S.~R. J and D.~P, ``Generalizable electroencephalographic classification of
  parkinson's disease using deep learning,'' \emph{Informatics in Medicine
  Unlocked}, vol.~42, p. 101352, 2023.

\bibitem{PD15}
L.~S and et~al., ``A convolutional-recurrent neural network approach to
  resting-state eeg classification in parkinson’s disease,'' \emph{Journal of
  neuroscience methods}, vol. 361, p. 109282, 2021.

\bibitem{PD16}
S.~Lee, R.~Hussein, and M.~J. McKeown, ``A deep convolutional-recurrent neural
  network architecture for parkinson’s disease eeg classification,'' in
  \emph{2019 IEEE global conference on signal and information processing
  (GlobalSIP)}.\hskip 1em plus 0.5em minus 0.4em\relax IEEE, 2019, pp. 1--4.

\bibitem{PD17}
L.~K and et~al., ``Parkinson’s disease detection and classification using eeg
  based on deep cnn-lstm model,'' \emph{Biotechnology and Genetic Engineering
  Reviews}, vol. 40(3), pp. 2577--2596, 2024.

\bibitem{PD18}
Z.~S and et~al., ``An interpretable model based on graph learning for diagnosis
  of parkinson’s disease with voice-related eeg,'' \emph{NPJ Digital
  Medicine}, vol. 7(1), p.~3, 2024.

\bibitem{dubreuil2020deep}
L.~Dubreuil-Vall and et~al., ``Deep learning convolutional neural networks
  discriminate adult adhd from healthy individuals on the basis of
  event-related spectral eeg,'' \emph{Frontiers in neuroscience}, vol.~14, p.
  251, 2020.

\bibitem{ADHD5}
A.~Ahmadi and et~al., ``Computer aided diagnosis system using deep
  convolutional neural networks for adhd subtypes,'' \emph{Biomedical Signal
  Processing and Control}, vol.~63, p. 102227, 2021.

\bibitem{ADHD6}
M.~Bakhtyari and S.~Mirzaei, ``Adhd detection using dynamic connectivity
  patterns of eeg data and convlstm with attention framework,''
  \emph{Biomedical Signal Processing and Control}, vol.~76, p. 103708, 2022.

\bibitem{ADHD8}
B.~Karakaş and et~al., ``Convmixer ve sdd kullanılarak dehb hastalığının
  eeg sinyalleri ile otomatik olarak tespit edilmesi,'' \emph{Türk Doğa ve
  Fen Dergisi}, vol.~13, no.~1, p. 19–25, 2024.

\end{thebibliography}

%





\end{document}